\documentclass{aa}  

\usepackage{graphicx}
\usepackage{lscape}
\usepackage{longtable}
\usepackage{txfonts}
\usepackage{xcolor}
\usepackage{array}
\begin{document} 

   \title{Metallicities in M dwarfs: \\
   Investigating different determination techniques}

   \author{V.\,M.~Passegger\inst{\ref{hs},\ref{ou}}
          \and
        A.~Bello-Garc\'ia\inst{\ref{ovi}}
        \and
        J.~Ordieres-Mer\'e\inst{\ref{upm}}
            \and
        A.~Antoniadis-Karnavas\inst{\ref{caup},\ref{fcup}}
        \and
        E.~Marfil\inst{\ref{ucm},\ref{cab2}}
        \and
        C.~Duque-Arribas\inst{\ref{ucm}}
        \and
            P.\,J.~Amado\inst{\ref{iaa}}
        \and
        E.~Delgado-Mena\inst{\ref{caup}}
        \and
        D.~Montes\inst{\ref{ucm}}
        \and
        B.~Rojas-Ayala\inst{\ref{chile}}
        \and
        A.~Schweitzer\inst{\ref{hs}}
        \and
        H.\,M.~Tabernero\inst{\ref{cab1}}
        \and 
        V.\,J.\,S.~B\'ejar\inst{\ref{iac},\ref{uiac}}
        \and
        J.\,A.~Caballero\inst{\ref{cab2}}
        \and
        A.\,P.~Hatzes\inst{\ref{tls}}
        \and
        Th.~Henning\inst{\ref{mpia}}
        \and
        S.~Pedraz\inst{\ref{caha}}
        \and
        A.~Quirrenbach\inst{\ref{lsw}}
        \and
        A.~Reiners\inst{\ref{iag}}
        \and
        I.~Ribas\inst{\ref{ice},\ref{ieec}}
        }

   \institute{Hamburger Sternwarte, Gojenbergsweg 112, D-21029 Hamburg, Germany         \newline
        \email{vpassegger@hs.uni-hamburg.de}\label{hs}
        \and
        Homer L. Dodge Department of Physics and Astronomy, University of Oklahoma, 440 West Brooks Street, Norman, OK 73019, United States of America \label{ou}
            \and
        Departamento de Construcci\'on e Ingenier\'ia de Fabricaci\'on, Universidad de Oviedo, c/ Pedro Puig Adam, Sede Departamental Oeste, M\'odulo 7, 1$^a$ planta, 33203 Gij\'on, Spain \label{ovi}
        \and
        Departamento de Ingenier\'ia de Organizaci\'on, Administraci\'on de Empresas y Estad\'istica, Universidad Polit\'ecnica de Madrid, c/~Jos\'e Guti\'errez Abascal 2, 28006 Madrid, Spain \label{upm}
        \and
        Instituto de Astrof\'isica e Ci\^encias do Espa\c{c}o, Universidade do Porto, CAUP, Rua das Estrelas, 4150-762 Porto, Portugal
        \label{caup}
        \and
        Departamento de F\'isica e Astronomia, Faculdade de Ci\^encias, Universidade do Porto, Rua do Campo Alegre, 4169-007 Porto, Portugal
        \label{fcup}
        \and
        Departamento de F\'{\i}sica de la Tierra y Astrof\'{\i}sica and IPARCOS-UCM (Instituto de F\'{\i}sica de Part\'{\i}culas y del Cosmos de la UCM), Facultad de Ciencias F\'{\i}sicas, Universidad Complutense de Madrid, 28040 Madrid, Spain \label{ucm}
        \and
        Centro de Astrobiolog\'{i}a (CSIC-INTA), ESAC, Camino Bajo del Castillo s/n, E-28691, Villanueva de la Ca\~{n}ada, Madrid, Spain
        \label{cab2}
        \and
        Instituto de Astrof\'isica de Andaluc\'ia (IAA-CSIC), Glorieta de la Astronom\'ia s/n, 18008 Granada, Spain \label{iaa}
        \and
        Instituto de Alta Investigación, Universidad de Tarapacá, Casilla 7D, Arica, Chile \label{chile}
        \and
        Centro de Astrobiolog\'ia (CSIC-INTA), Carretera de Ajalvir km 4, Torrej\'{o}n de Ardoz, 28850, Madrid, Spain \label{cab1}
        \and
        Instituto de Astrof\'{\i}sica de Canarias, c/ V\'ia L\'actea s/n, 38205 La Laguna, Tenerife, Spain \label{iac}
        \and
        Departamento de Astrof\'{\i}sica, Universidad de La Laguna,-38206 La Laguna, Tenerife, Spain \label{uiac}
        \and
        Th\"uringer Landessternwarte Tautenburg, Sternwarte 5, 07778 Tautenburg, Germany\label{tls}
        \and        
        Max-Planck-Institut f\"ur Astronomie, K\"onigstuhl 17, 69117 Heidelberg, Germany \label{mpia}
        \and
        Centro Astron\'omico Hispano-Alem\'an (CSIC-MPG), Observatorio Astron\'omico de Calar Alto,  Sierra de los Filabres, 04550 G\'ergal, Almer\'ia, Spain\label{caha}
        \and
        Landessternwarte, Zentrum f\"ur Astronomie der Universit\"at Heidelberg, K\"onigstuhl 12, 69117 Heidelberg, Germany\label{lsw}
        \and
        Institut f\"ur Astrophysik, Georg-August-Universit\"at, Friedrich-Hund-Platz 1, 37077 G\"ottingen, Germany \label{iag}
        \and
        Institut de Ci\`encies de l'Espai (CSIC-IEEC), Campus UAB, c/ de Can Magrans s/n, 08193 Bellaterra, Barcelona, Spain \label{ice}
        \and
        Institut d'Estudis Espacials de Catalunya (IEEC), 08034 Barcelona, Spain \label{ieec}
        }

   \date{Received 30 July 2021 /  Accepted 16 November 2021}

 
  \abstract
   {Deriving metallicities for solar-like stars follows well-established methods, but for cooler stars such as M dwarfs, the determination is much more complicated due to forests of molecular lines that are present. Several methods have been developed in recent years to determine accurate stellar parameters for these cool stars ($T_{\rm eff} \lesssim$ 4000\,K). However,  significant differences can be found at times when comparing metallicities for the same star derived using different methods. 
   In this work, we determine the effective temperatures, surface gravities, and metallicities of 18 well-studied M dwarfs observed with the CARMENES high-resolution spectrograph following different approaches, including synthetic spectral fitting, 
   analysis of pseudo-equivalent widths, and machine learning. We analyzed the discrepancies in the derived stellar parameters, including metallicity, in several analysis runs. Our goal is to minimize these discrepancies and find stellar parameters that are more consistent with the literature values. 
   We attempted to achieve this consistency by standardizing the most commonly used components, such as wavelength ranges, synthetic model spectra, continuum normalization methods, and stellar parameters. We conclude that although such modifications work quite well for hotter main-sequence stars, they do not improve the consistency in stellar parameters for M dwarfs, leading to mean deviations of around 50--200\,K in temperature and 0.1--0.3\,dex in metallicity. In particular, M dwarfs are much more complex and a standardization of the aforementioned components cannot be considered as a straightforward recipe for bringing consistency to the derived parameters. 
   Further in-depth investigations of the employed methods would be necessary in order to identify and correct for the  discrepancies that remain. }


   \maketitle


\section{Introduction}

Precise stellar metallicity determinations are an essential step to achieving a fuller understanding of the dynamical and chemical evolution of the Galaxy. Several methods have been developed to study element abundances of all kinds of stars. Among these, M dwarfs are the most prevalent type in our Galaxy \citep{Henry2016,Reyle2021} and therefore an accurate determination of their abundances is of utmost interest. In the fast-growing field of exoplanet detection and characterization, abundance determinations of the host star are also important to better understand the formation and evolution of planetary systems \citep[e.g.,][]{Burn2021}.

A popular method for deriving the metallicities of M dwarfs is based on the measurement of pseudo-equivalent widths (pEWs) of spectral lines. This method was used by  \cite{Neves2013,Neves2014}, \cite{Mann2013,Mann2014}, \cite{Newton2014}, \cite{Maldonado2015}, and \cite{Khata2020}, among others.
Another widely used approach is spectral synthesis, where the stellar spectrum is synthesized using stellar atmosphere models along with radiative transfer codes and atomic and molecular line lists. The PHOENIX stellar atmosphere code \citep{Hauschildt1992,Hauschildt1993} is the basis for stellar model grids such as the BT-Settl model atmospheres \citep{Allard2012,Allard2013} and the PHOENIX-ACES synthetic model grid \citep{Husser2013}. 

\cite{Marfil2021} used the BT-Settl model atmospheres and the radiative transfer code {\tt turbospectrum} \citep{Plez2012} to generate synthetic spectra around 75 \ion{Fe}{i} and \ion{Ti}{i} lines, along with the TiO $\gamma$ and $\epsilon$ bands, to determine $T_{\rm eff}$, $\log{g}$, and [Fe/H] for 342 M dwarfs from the CARMENES survey by means of the {\tt SteParSyn} code \citep{Tabernero2018,Tabernero2021a}. 
The {\tt turbospectrum} code was also employed by \cite{Souto2017, Souto2020} and \cite{Sarmento2021}, together with 1-D MARCS stellar atmospheres \citep{Gustafsson2008}, to derive stellar parameters and abundances for several M dwarfs observed with the Apache Point Observatory Galactic Evolution Experiment
\citep[APOGEE,][]{Majewski2017}. Operating in a wavelength range from 15000\,{\AA} to 17000\,{\AA} and with high-resolution \citep[$\mathcal{R} \approx$ 22,500;][]{Wilson2010}, APOGEE is dedicated to observing red giants, but it has additionally 
observed around 2000 M dwarfs. \cite{Onehag2012} and \cite{Lindgren2016} fitted synthetic spectra to high-resolution CRIRES $J$-band spectra of M dwarfs using the Spectroscopy Made Easy package \citep[{\tt SME},][]{ValentiPiskunov1996,ValentiFischer2005} with MARCS atmospheres. {\tt SME} computes synthetic spectra on the fly and determines the best fit stellar parameters by $\chi^2$-minimization with the observed spectra.  
\cite{Passegger2018} fitted the PHOENIX-ACES model spectra grid to high-resolution CARMENES spectra of 300 M dwarfs and derived $T_{\rm eff}$, $\log{g}$, and [Fe/H]. 

Over the last several years, machine learning has emerged as a valuable tool for predicting stellar parameters for large sets of stars. Several applications of neural networks in stellar parameter determination can be found in \cite{Fabbro2018}, \cite{Birky2020}, \cite{Antoniadis2020}, and \cite{Passegger2020}, among others.
For a more detailed overview on previous works on stellar parameter determinations in M dwarfs, we refer to the literature summaries in \cite{Passegger2020} and \cite{Marfil2021}.  

It is known from previous stellar parameter studies that different determination methods sometimes provide significantly different results for the same stars.
This is shown in the comparison plots of several parameter determination studies, for instance, 
Fig.\,13 in \cite{RojasAyala2012}, 
Figs.\,13--14 in \cite{Neves2014}, 
Figs.\,1 and\,5 in \cite{Lindgren2016}, 
Figs.\,5--7 in \cite{Passegger2019}, 
Fig.\,7 in \cite{Passegger2020}, 
Figs.\,10--12 in \cite{Marfil2020}, 
Figs.\,12 and 13 in \cite{Sarmento2021}, 
and Figs.\,9, 11, as well as\,A1--A6 in \cite{Marfil2021}. 
These inconsistencies challenge the reliability of the determined stellar parameters for the lowest-mass stars.

However, 
there are different types of inconsistencies. The most relevant cases in this context are inconsistencies between different methods and between different observations of the same star with different instruments.
Since there is no way yet to measure the absolute correct physical and atmospheric properties of a given star, we have to rely on the parameters that different methods and observations provide. Deriving consistent values for the same star with different methods (or different instruments) can therefore be considered as a proxy for the reliability of the value of the stellar parameters and of the methods themselves. 

Several studies have conducted such consistency analyses for FGK-type stars and examined the differences introduced when deriving abundances with different methods. 
For example, \cite{Hinkel2016} investigated four G-type stars with high-resolution MIKE spectra ($\mathcal{R} \approx$ 50,000) from the Magellan Planet Search Program, with an average signal-to-noise ratio (S/N) of 200, and covering the wavelength range of 5050--7100\,\AA. Six different teams participated in the analysis and determined abundances for ten elements (C, O, Na, Mg, Al, Si, Fe, Ni, Ba, and Eu), in four different runs. In Run\,1, each group used their individual techniques, while in Run 2 standard stellar parameters for $T_{\rm eff}$, $\log{g}$, and microturbulent velocity $\xi$ were provided. Run 3 included a standard line list, whereas Run 4 was a combination between Runs 2 and 3. 
The authors found that Run 2 gave consistently better results between the elements, followed by Run 4, which suggests that stellar parameters other than abundances or line lists should be standardized in order to produce similar results. 

A larger sample of 34 {\it Gaia} benchmark FGK-type stars was used by \cite{Jofre2014}. The spectra were collected with HARPS ($\mathcal{R} \approx 115,000$), NARVAL ($\mathcal{R} \approx 80,000$), and UVES ($\mathcal{R} \geq 70,000$), covering a spectral range from 4760\,{\AA} to 6840\,{\AA}. Seven different teams participated in this study and derived Fe abundances in three runs. Their main aim was to analyze the effects of instrumental resolution on the determination of metallicity when fixing $T_{\rm eff}$ and $\log{g}$ to independently derived values. Furthermore, all teams used a common line list and the same atomic data \citep[see][]{Heiter2021} and atmospheric models (MARCS). In the different runs, they used spectra with their original resolution and with resolution downgraded to $\mathcal{R} 70,000$ to study instrumental effects.
They found that different resolutions result in a metallicity difference of less than 0.05\,dex, and that metallicities agree when using different instruments. 
A comparison of the different methods showed larger standard deviations in metallicity for the coolest stars (0.1\,dex, $T_{\rm eff} <$ 5000\,K) than for the hottest stars (0.07\,dex, $T_{\rm eff} >$ 5000\,K).
A follow-up study by \cite{Jofre2015} analyzed ten different element abundances with eight methods taking into account non-local thermodynamical equilibrium (NLTE) corrections for Fe and errors of the fixed stellar parameters. They performed a detail analysis of systematic errors for differential and absolute abundances. For an extensive discussion on each element and NLTE effects, we refer to \cite{Jofre2015}.

\cite{Jofre2017} provided a detailed study of four {\it Gaia} benchmark stars, the Sun (G2\,V), Arcturus (K1.5\,III), 61 Cyg A (K5\,V), and HD\,22879 (G0\,V). Their high-resolution spectra from NARVAL and HARPS were convolved to a common resolution of 70,000. Also in this work, the stellar parameters $T_{\rm eff}$, $\log{g}$, micro-turbulence $v_{mic}$, and $\varv\sin{i}$ were fixed for each star. 
The analysis was performed by six different teams in eight different runs, including tests regarding continuum normalization, common line lists, hyperfine structure, $\alpha$-enhancement, and radiative transfer code. They concluded that the most important point for consistent metallicity values is a common continuum flux. 

Focusing on cooler stars, \cite{Slumstrup2019} conducted a similar study for red giant stars in the open clusters NGC 6819, M67, and NGC 188. They compared several combinations of line lists and methods to derive EWs, and analyzed the systematic uncertainties from a line-by-line spectroscopic analysis. As a result, they found scatter of around 170\,K in $T_{\rm eff}$, 0.4\,dex in $\log{g}$, and 0.25\,dex in metallicity, concluding that even for high-precision spectroscopic analyses, external constraints are necessary to obtain consistent results between different methods. 

Up to now, no such analysis has been performed for M dwarfs. In this work, we aim to follow the approach by \cite{Hinkel2016} to study the deviations in metallicity, as well as $T_{\rm eff}$ and $\log{g}$, coming from different determination methods, and to identify ways to derive more consistent results for stars at the cool end of the main sequence. This paper is structured as follows. Section\,\ref{methods} gives an overview on the methods we used in our analyses. Section\,\ref{analysis} explains our sample of benchmark stars and the different analysis runs we performed. In Sect.\,\ref{results}, we present the results of the investigation, followed by a discussion in Sect.\,\ref{discussion}. A short summary is given in Sect.\,\ref{summary}.

\section{Methods}
\label{methods}

In the following we describe the four different methods we use for deriving fundamental stellar parameters $T_{\rm eff}$, $\log{g}$, and [Fe/H]. 

\subsection{Synthetic spectra fitting}
\subsubsection{{\tt Pass19-code}}
\label{Pass19-code}

This method is fully described in \cite{Passegger2018,Passegger2019}, hereafter referred to as {\tt Pass19-code}. We used a downhill simplex method with a $\chi^2$ minimization to find the synthetic model spectrum that best fits the observed spectrum by fitting several wavelength ranges in the VIS and NIR simultaneously \citep[see Table~2 in][]{Passegger2019}.

The PHOENIX-ACES model spectra grid \citep{Husser2013} incorporated here is based on the PHOENIX code developed by \cite{Hauschildt1992,Hauschildt1993}. 
Improvements to the code are described in \citet{Hauschildt1997}, \citet{HauschildtBaron1999}, \citet{Claret2012}, and \citet{Husser2013}, for instance. 
The one-dimensional (1D) mode of the PHOENIX code computes spherically symmetric model atmospheres, which can be used to simulate main sequence stars and brown dwarfs, including L and T spectral types, as well as white dwarfs and giants. It also includes models for expanding envelopes of novae and supernovae, and accretion disks. PHOENIX can calculate synthetic spectra in 1D or 3D and can be executed in LTE or non-LTE radiative transfer mode.
Several model atmosphere grids for late-type stars are based on the PHOENIX code, for instance the NextGen models \citep{Hauschildt1999}, the AMES models \citep{Allard2001}, and the BT-Settl models \citep{Allard2011}. For the calculation of the aforementioned PHOENIX-ACES model spectra grid a new equation of state was used, which was especially designed for the formation of molecules in very cool stellar 
atmospheres. The grid takes into account solar chemical compositions from \cite{Asplund2009}, updated with meteoritic values from \cite{Lodders2009}. Since the PHOENIX-ACES grid we use has [$\alpha$/Fe]\,=\,0, our metallicity results of [M/H] directly translate into identical [Fe/H] values. However, for certain parameter ranges, an $\alpha$-enhanced PHOENIX-ACES grid is available \citep[see][]{Husser2013}.

To match the instrumental resolution and wavelength grid of observed spectra, the PHOENIX-ACES model spectra are convolved with a Gaussian and linearly interpolated in wavelength. The synthetic spectra are broadened to account for the rotational velocity $\varv\sin{i}$ of the star \citep{Reiners2018}. Therefore, a separate function estimates the effect on the line spread function and the synthetic spectrum is convolved with the resulting line 
spread function. The pseudo-continuum of both the observed and synthetic spectra is normalized with a linear fit within each small wavelength region that is analyzed. 

The surface gravity ($\log{g}$) is determined from evolutionary models as in \cite{Passegger2019}. This is done to break degeneracies between 
the parameters. The evolutionary models used in this work were taken from the PARSEC v1.2S library 
\citep{Bressan2012,Chen2014,Chen2015,Tang2014}, which provides $T_{\rm eff}$ and $\log{g}$ for metallicities in the range 
$-$2.2 < [M/H] < +0.7 and different stellar ages, among other parameters. To select the appropriate isochrone, we took 
the stellar ages from \cite{Passegger2019}. The $\log{g}$ is then calculated from this isochrone's $T_{\rm eff}$-$\log{g}$ relation 
depending on $T_{\rm eff}$ and [Fe/H] chosen by our algorithm. To get finer values we linearly interpolate for 
metallicities between $-$1.0 and $+$0.7. The PHOENIX-ACES model spectra grid is then interpolated according to these three 
parameters and the $\chi^2$ is calculated between the observed and synthetic spectrum. A downhill simplex finds the best fitting synthetic spectrum with the smallest $\chi^2$ by exploring the 2-D $T_{\rm eff}$-[Fe/H] parameter space and adjusting those parameters accordingly.

\subsubsection{{\tt SteParSyn}}
The {\tt SteParSyn} code is described in detail in \cite{Tabernero2021b}. It is a Bayesian implementation of the spectral synthesis technique that determines the probability distributions of the stellar atmospheric parameters ($T_{\rm eff}$, $\log{g}$, [Fe/H], $\varv\sin{i}$, and $\zeta$) from a Markov Chain Monte Carlo (MCMC) approach. In general terms, the code compares a grid of synthetic spectra pre-computed around certain spectral features of interest. Therefore, we used a selection of 75 magnetically insensitive Ti\,{\sc i} and Fe\,{\sc i} lines, as well as the TiO $\gamma$ and $\epsilon$ bands in a range between 5850--15800\,{\AA}. The assessment of any point in the parameter space is done in a computationally inexpensive way employing principal component analysis (PCA). The code finally returns the posterior probability distributions in the stellar atmospheric parameters along with the best synthetic fit for the input spectral features.

With the aim of avoiding any potential degeneracy in the M-dwarf parameter space, especially between $\log{g}$ and [Fe/H], we assumed Gaussian prior probability distributions in $T_{\rm eff}$ and $\log{g}$ for all individual targets, with standard deviations of 200\,K and 0.2\,dex, respectively. 
The prior distributions are centered following \cite{Cifuentes2020}, who determined $T_{\rm eff}$ from a multi-band photometric analysis by means of the Virtual Observatory Spectral energy distribution Analyser \cite[VOSA,][]{Bayo2008}, and derived stellar radii and masses from the Stefan-Boltzmann law and the mass-radius relation presented in \cite{Schweitzer2019}.

Even though any model atmosphere grid can be used along with {\tt SteParSyn}, in the present work we employed BT-Settl model atmospheres \citep{Allard2012}. Since the grid is alpha-enhanced, metallicities derived using this method are corrected using a simple interpolation scheme between the mass fraction $Z$ and [Fe/H] following the standard composition in the MARCS models \citep{Gustafsson2008}, as explained in \citet{Marfil2021}. 

{\tt SteParSyn} was  also used in \citet{Tabernero2018} for the study of cool supergiants in the Magellanic clouds, as well as in \citet{Tabernero2021a} for the analysis of the AGB-star candidate VX\,Sgr. \cite{Marfil2021} applied {\tt SteParSyn} to the CARMENES GTO sample. The exoplanet host WASP-121 was also analyzed with {\tt SteParSyn} using ESPRESSO spectra \citep{Borsa2021}.

\subsection{Machine learning}

\subsubsection{Deep learning (DL)}
\label{DL}
This method has been described in detail in \cite{Passegger2020}. 
Artificial neural networks are machine learning methods that are constructed from a collection of artificial neurons organized in different layers that are meant to learn structures from data in a similar way as the human brain does. 
In deep learning (DL), the neural network models consist of multiple processing layers that can learn relevant features by themselves without user interaction. 

For each stellar parameter, we built a convolutional deep neural network with several hidden layers. In order to learn features from the input spectrum, the networks were trained with PHOENIX-ACES synthetic models. We linearly interpolated the existing grid using {\tt pyterpol} \citep{Nemravov2016} to increase the number of training samples. We applied additional restrictions to our grid that are similar to those of the {\tt Pass19-code}. Based on the PARSEC v1.2S evolutionary models we excluded combinations of $T_{\rm eff}$, $\log{g}$, and [Fe/H] that are physically unrealistic for M dwarfs (i.e., they represent stellar objects far away from the main sequence). In the end, we created 449\,806 synthetic model spectra for the reference set in training process. 

We convolved the synthetic spectra with a Voigt profile to account for instrumental broadening using a function based on {\tt libcerf} \citep{libcerf}. The Gaussian and Lorentzian components  of the Voigt function for CARMENES were determined by \cite{Nagel2021}. 
We also took into account the $\varv\sin{i}$ by broadening the synthetic spectra with a Fortran translation of the {\tt rotational\_convolution} function of {\tt Eniric}, assuming a default limb darkening coefficient of 0.6 \citep[see][]{Figueira2016}.
For the continuum normalization of the synthetic as well as the observed CARMENES spectra, we employed the Gaussian Inflection Spline Interpolation Continuum ({\tt GISIC}) routine\footnote{\url{https://pypi.org/project/GISIC/}}, developed by D.\,D.~Whitten. The routine smoothens the spectrum with a Gaussian before identifying molecular bands with a numerical gradient. Then continuum points are selected and a cubic spline interpolation normalizes the continuum within the desired spectral range. 
The observed CARMENES spectra are corrected for the spatial motion of the stars by using a cross-correlation between the observed spectrum and a PHOENIX-ACES model spectrum. Because this results in shifts of the wavelength grid of the observations, we linearly interpolate this grid to match the wavelength grid of the synthetic spectra. 

In the training, the reference set is divided into a training set (95\,\%) and a validation set (5\,\%). After running the training set through the deep neural network, the training error is estimated from the difference between the output and the known input stellar parameters. 
Based on this error, the hyper-parameters of the DL model are adjusted through backward propagation. 
The validation set is used to determine the validation error, that is, the mean square error (MSE) after each training epoch to verify that the adjustment of the model hyper-parameters is heading in the right direction to improve the DL model and to make sure the error continues to decrease. It also helps  avoid overfitting the training set, which happens when the DL model learns to describe random variations and is unable to generalize based on new data. The training is complete once the minimum validation error is reached. At this point, a test set of 100 randomly generated synthetic spectra is sent through the DL model to measure the test error. This presents a final test to the DL model before it is applied to observed spectra. The model is assumed to be performing well when the average test error is below a certain threshold, which we define as between $5 \cdot 10^{-4}$ and $10^{-5}$ depending on the stellar parameter under investigation. 

As explained in \cite{Passegger2020}, the range 8800--8835\,{\AA} and an individual neural network model for each stellar parameter separately give the smallest validation errors. We therefore follow that approach in this work.

\subsubsection{Pseudo-EW approach ({\tt ODUSSEAS})}
\label{ODUSSEAS}

A detailed description of the machine learning tool {\tt ODUSSEAS} can be found in \cite{Antoniadis2020}. 
{\tt ODUSSEAS} receives 1D spectra and their resolutions as input. 
The method is based on measuring the pEWs of absorption lines and blended lines in the range between 5300\,{\AA} and 6900\,{\AA}. Spectral sections that include the activity-sensitive Na doublet, H$\alpha$ line, and strong telluric lines, have been excluded from the line list. The line list consists of 4104 absorption features, the same as used by \cite{Neves2014}. 

{\tt ODUSSEAS} contains a supervised machine learning algorithm based on the “scikit learn” package of Python, in order to determine the $T_{\rm eff}$ and [Fe/H] of the stars. In the training, it is provided with both input and expected output, in order to create the  machine learning models using ridge regression. The pEWs in 65 HARPS spectra are used together with their $T_{\rm eff}$ and [Fe/H] from \cite{Casagrande08} and \cite{Neves2012}, respectively, as reference for training and testing its models. 

Applied to new spectra, {\tt ODUSSEAS} measures the pEWs of the lines and compares them to the model generated from the HARPS spectra, convolved to the respective resolution of the new spectra. In this case, the HARPS reference spectra are convolved from their resolution of 115\,000 to the CARMENES resolution of 94\,600. 
For each new star, the resulting parameters are calculated from the mean values of 100 determinations obtained from randomly shuffling and splitting each time the training (70\,\% of the sample, i.e. 45 stars) and testing groups (30\,\% of the population, i.e., 20 stars). This iterative process of multiple runs minimizes the possible dependence of the resulting parameters on how the stars from the HARPS dataset are split for training and testing in a single measurement. 

We report parameter uncertainties derived by quadratically adding the dispersion of the resulting stellar parameters and the uncertainties of the machine learning models at this resolution after having taken into consideration the intrinsic uncertainties of the reference dataset parameters during the machine learning process.
Since {\tt ODUSSEAS} only relies on pEWs from HARPS spectra, this method is independent of synthetic spectra. 
The tool is publicly available on Github\footnote{\url{https://github.com/AlexandrosAntoniadis/ODUSSEAS}}.

\section{Analysis}
\label{analysis}

\subsection{Stellar sample}
\label{sample}
Our stellar sample of benchmark stars consists of 18 M dwarfs, listed in Table\,\ref{tab:sample}. All stars are part of the CARMENES GTO sample and were observed with the CARMENES\footnote{
\url{http://carmenes.caha.es}} instrument. They were chosen such that they have a high-S/N CARMENES spectrum with a S/N of at least 75 in the optical (VIS) and near-infrared (NIR), as stated in \cite{Passegger2018}. Their spectral types cover the range between M0.0\,V and M5.5\,V, following the typical CARMENES GTO distribution \citep[see][]{Marfil2021}. The mean S/N over all spectrograph orders in the VIS and NIR for each spectrum is listed in Table\,\ref{tab:sample}. There is one exception to the S/N > 75 limit in the NIR, which is J13005+056 (GJ\,493.1) due to its high rotational velocity. All sample stars, except for the two high-rotation stars, show only minimal to no stellar activity \citep[see e.g.,][]{TalOr2018,Schoefer2019}.
Each star has literature photospheric parameters determined from at least 3, and up to 12 other studies.

CARMENES operates with two highly stable fiber-fed spectrographs covering 5200--9600\,{\AA} in the VIS and 9600--17100\,{\AA} in the NIR wavelength ranges. The spectral resolutions are $R \approx$ 94\,600 and 80\,500, respectively \citep[][]{Quirrenbach2018,Reiners2018}. The spectrographs are mounted on the Zeiss 3.5\,m telescope at the Calar Alto Observatory in Spain. The prime goal of CARMENES is the search for Earth-sized planets in the habitable zones of M dwarfs. A detailed description of the whole CARMENES GTO sample can be  found in \cite{Caballero2016a}.

\cite{Zechmeister2014}, \cite{Caballero2016}, and \cite{Passegger2019} presented a detailed description of the data reduction. After spectral extraction, each single spectrum is corrected for telluric lines by modeling a telluric absorption spectrum with the tool {\tt Molecfit} \citep{Kausch2014,Smette2015}. The process is described in \cite{Nagel2021}. The absorption telluric spectrum is subtracted from the observed spectrum resulting in a telluric-free spectrum that is then fed into the CARMENES radial velocity pipeline {\tt serval} \citep[SpEctrum Radial Velocity AnaLyser;][]{Zechmeister2018}. There, a high-S/N template spectrum is constructed for each star having at least five single spectra. This is a byproduct of the radial velocity calculation, where the radial velocities of the single spectra are derived from a least-square fit against the template. 
In this work, we apply our methods to these high-S/N templates of our 18 benchmark stars. 

The stellar photospheric parameters we collected from literature for the benchmark stars are summarized in Table\,A.1. Although most benchmark stars have $\varv\sin{i} < 2$\,km\,s$^{-1}$ \citep{Reiners2018}, there are two stars with larger values: J07558+833 
(12.1\,km\,s$^{-1}$) and J13005+056 
(16.4\,km\,s$^{-1}$). These stars are useful to investigate the performance of the algorithms when dealing with higher rotational velocities.
The literature values were derived with different methods. 
These methods include: interferometry to estimate the stellar radius and $T_{\rm eff}$ \citep{Boyajian2012,Segransan2003,vonBraun2014,Berger2006,Newton2015}, synthetic model fitting using BT-Settl models to determine $T_{\rm eff}$ \citep{Gaidos2014,Lepine2013,GaidosMann2014,Mann2015} and $\log{g}$ \citep{Lepine2013}, empirical relations to derive stellar mass in the form of mass-luminosity relations  \citep{Mann2015,Khata2020,Boyajian2012,Berger2006,Segransan2003}, along with the mass-magnitude relations \citep{Maldonado2015}, mass-radius relations \citep{vonBraun2014}, mass-$T_{\rm eff}$ relations \citep{GaidosMann2014,Gaidos2014}, empirical relations to derive the stellar radius in the form of mass-radius relations \citep{Maldonado2015} and $T_{\rm eff}$-radius relations \citep{GaidosMann2014,Gaidos2014,Houdebine2019}, pEW measurements to determine $T_{\rm eff}$ \citep{Maldonado2015,Neves2014,Newton2015} and [Fe/H] \citep{Maldonado2015,Neves2014,Gaidos2014,Mann2015}, the definition of spectral indices such as the H2O-K2 index to estimate $T_{\rm eff}$ \citep{RojasAyala2012}, as well as the combination of the H2O-K2 index with pEWs to derive [Fe/H] \citep{RojasAyala2012,Khata2020}, the stellar radius and $T_{\rm eff}$ \citep{Khata2020}, and spectral curvature indices for the determination of $T_{\rm eff}$ \citep{GaidosMann2014}. Additionally, [Fe/H] was derived by using color-magnitude metallicity relations \citep{Dittmann2016}, atomic line strength relations \citep{GaidosMann2014}, and spectral feature relations \citep{Terrien2015}. \cite{Terrien2015} used $K$-band magnitudes and the Dartmouth Stellar Evolution Program \citep{Dotter2008} to derive the stellar radius, whereas \cite{Mann2015} employed the Boltzmann equation with $T_{\rm eff}$ determined from synthetic model fits. Last, but not least, \cite{Houdebine2019} derived $T_{\rm eff}$ from photometric colors. For more details on the individual methods, we refer to the descriptions in the corresponding works.

In this work, it is not our aim to analyze the variations from different techniques, data sets, and observations in the literature, however, we can compare the results of our methods to the literature as a whole. 
Therefore, we calculated the median over all literature values to reduce possible biases introduced by different data sets and methods. Thus, we presume the median to be to some extent more accurate than the individual literature values and we consider the similarity between our values and the literature median as our quality measurement. The errors for the literature median come from the root-mean-squared-errors (RMSE) of the single measurements. 
Further, the median can be effective in smoothing extreme outliers, in case of contradicting literature values. 

\begin{table*}
\caption[]{Selected sample of benchmark stars.}
\label{tab:sample}
\centering
\begin{tabular}{lllcccccc}
\hline 
\hline 
\noalign{\smallskip}
Karmn  & Name & GJ & $\alpha$ (J2000) & $\delta$ (J2000)&  Spectral type\,$^{(a)}$ & $\varv\sin{i}$\,$^{(b)}$ & \multicolumn{2}{c}{Mean S/N} \\
       &      &    & [hh:nmm:ss] & [hh:nmm:ss]  &                & [km\,s$^{-1}$] & VIS & NIR \\
\noalign{\smallskip}
\hline 
\noalign{\smallskip}
J00067$-$075 & GJ 1002        & 1002  & 00:06:42.35 & $-$07:32:46.4 & M5.5\,V & $\leq2.0$ & 226 & 318\\
J00183$+$440 & GX And         & 15A   & 00:18:27.04 & $+$44:01:29.0 & M1.0\,V & $\leq2.0$ & 993 & 1419\\
J04429$+$189 & HD 285968      & 176   & 04:42:56.49 & $+$18:57:12.1 & M2.0\,V & $\leq2.0$ & 230 & 171\\
J05314$-$036 & HD 36395       & 205   & 05:31:28.18 & $-$03:41:10.5 & M1.5\,V & $\leq2.0$ & 518 & 581\\
J07558$+$833 & GJ 1101        & 1101  & 07:55:51.31 & $+$83:22:55.7 & M4.5\,V &     12.1  & 92 & 94\\
J09143$+$526 & HD 79210       & 338A  & 09:14:20.14 & $+$52:41:03.0 & M0.0\,V & $\leq2.0$ & 459 & 597\\
J09144$+$526 & HD 79211       & 338B  & 09:14:22.00 & $+$52:41:00.7 & M0.0\,V &      2.3  & 770 & 796\\
J10508$+$068 & EE Leo         & 402   & 10:50:51.14 & $+$06:48:16.6 & M4.0\,V & $\leq2.0$ & 319 & 415\\
J11033$+$359 & Lalande 21185  & 411   & 11:03:19.44 & $+$35:56:52.8 & M1.5\,V & $\leq2.0$ & 1112 & 1553\\
J11054$+$435 & BD+44 2051A    & 412A  & 11:05:22.32 & $+$43:31:51.6 & M1.0\,V & $\leq2.0$ & 633 & 697\\
J11421$+$267 & Ross 905       & 436   & 11:42:12.13 & $+$26:42:11.0 & M2.5\,V & $\leq2.0$ & 506 & 1080\\
J13005$+$056 & FN Vir         & 493.1 & 13:00:32.55 & $+$05:41:11.5 & M4.5\,V &     16.4  & 89 & 62\\
J13457$+$148 & HD 119850      & 526   & 13:45:45.67 & $+$14:53:06.9 & M1.5\,V & $\leq2.0$ & 941 & 1053\\
J15194$-$077 & HO Lib         & 581   & 15:19:25.55 & $-$07:43:21.7 & M3.0\,V & $\leq2.0$ & 341 & 409\\
J16581$+$257 & BD+25\,3173     & 649   & 16:58:08.72 & $+$25:44:31.1 & M1.0\,V & $\leq2.0$ & 384 & 407\\
J17578$+$046 & Barnard's star & 699   & 17:57:47.67 & $+$04:44:16.7 & M3.5\,V & $\leq2.0$ & 976 & 1600\\
J22565$+$165 & HD 216899      & 880   & 22:56:33.69 & $+$16:33:08.0 & M1.5\,V & $\leq2.0$ & 1140 & 1338\\
J23419$+$441 & HH And         & 905   & 23:41:55.20 & $+$44:10:14.1 & M5.0\,V & $\leq2.0$ & 309 & 637 \\
\noalign{\smallskip}
\hline
\end{tabular}
\tablefoot{ $^{(a)}$Spectral types from \citet{AlonsoFloriano2015}. $^{(b)}$Projected rotational velocities from \citet{Reiners2018}.}
\end{table*}

\subsection{Different runs}

We analyzed our stellar sample with each method in three different runs. Each run is described thoroughly in the following. 

\subsubsection{Run A}

For the first run (Run A), each team derived the stellar parameters with their methods, as described in Sect.\,\ref{methods}, without any particular restrictions. In this way, we were able to directly compare the algorithms themselves and see how they perform compared to literature references. 

\begin{figure*}[!ht]
\centering
\includegraphics[width=0.8\textwidth]{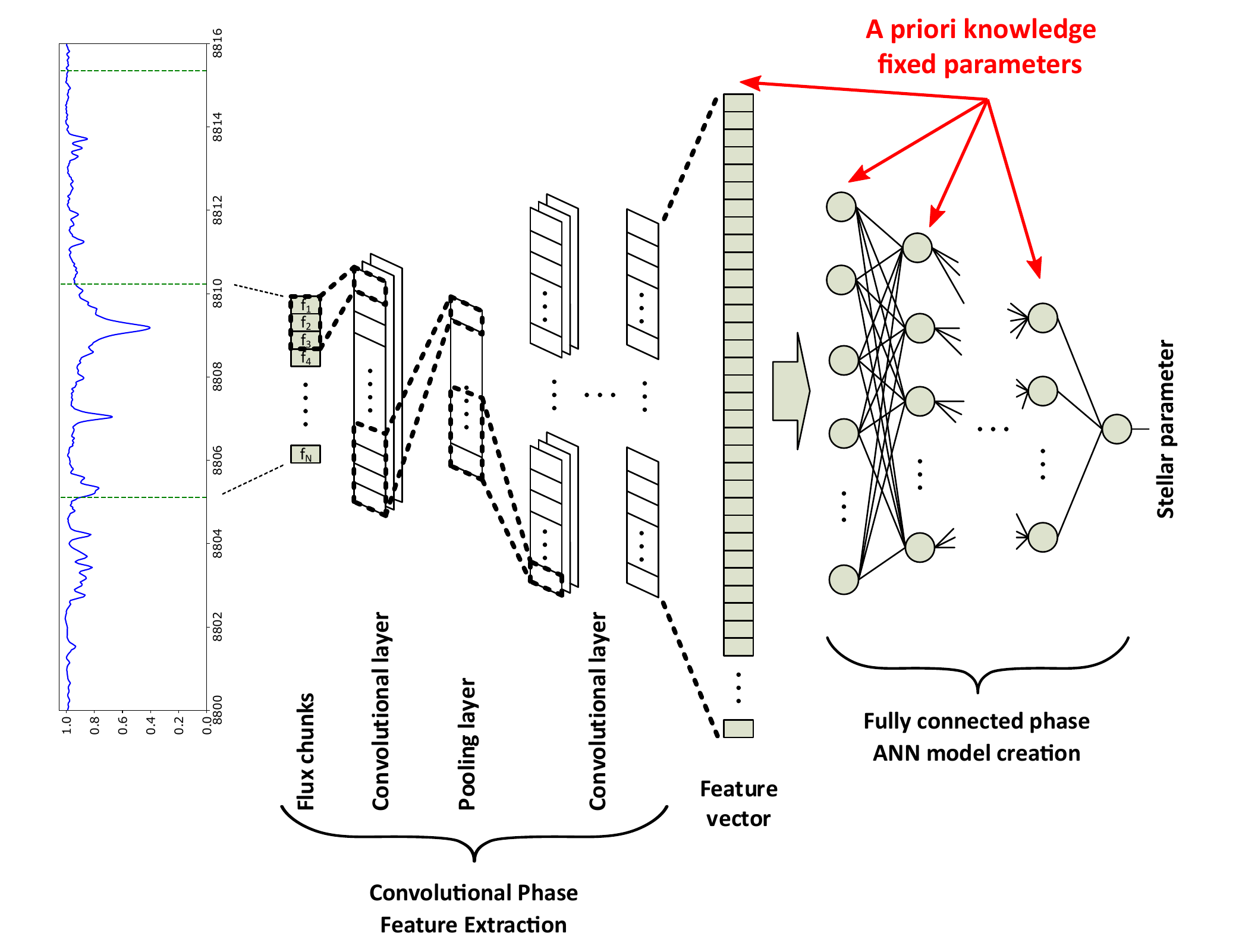}
\caption{Generic architecture for DL models in \cite{Passegger2020} and the different positions where we concatenated the values of the stellar parameters so that they could be fixed as needed in Run B.}
\label{fig:PriorKnowledge}
\end{figure*}

\subsubsection{Run B}
\label{methods:RunB}

In this run, all teams fixed the parameters $T_{\rm eff}$ and $\log{g}$ to the same values. They were calculated for each star as median values from the literature and the results from all teams from Run A (see Table\,A.1) and is hereafter referred to as the overall median. We did this in order to increase the amount of individual measurements for each star, especially when there are not many literature values available. This leaves metallicity as the only free parameter to be determined. With this setting, we are able to gain insight into how the algorithms perform if they focus on only one parameter and into whether this run gives any improvements compared to the previous run.

The implementation is straightforward in both the {\tt Pass19-code} and {\tt SteParSyn}, since $T_{\rm eff}$ and $\log{g}$ can be kept fixed so that the downhill simplex and the MCMC chains (respectively) only explore a 1-D parameter space for metallicity. Therefore, there is only one minimum and one best-fit metallicity value. 
To assess the uncertainty in metallicity for the {\tt Pass19-code} in the case of two fixed parameters, we follow the approach described in \cite{Passegger2016}. Thus, we produced a set of 1400 synthetic spectra with uniformly distributed random parameters for $T_{\rm eff}$, $\log{g}$, and [Fe/H], broadened to the resolution of the CARMENES spectrographs. To simulate a S/N of $\approx 100$, we added Poisson noise. The set of synthetic spectra was then sent through the {\tt Pass19-code,} keeping $T_{\rm eff}$ and $\log{g}$ fixed. This was done for the three different $\varv\sin{i}$ values of our stellar sample. The standard deviation of the mean deviation between input and derived output stellar parameter serves as an estimation for the uncertainty of the parameter. 

The DL approach presented in \cite{Passegger2020} was not initially targeted at fixing the stellar parameters as required in this run. Thus, we first constructed the models by restricting the training sample to synthetic spectra with fixed $T_{\rm eff}$ and $\log{g}$. This significantly reduced the training sample size and also indicated that new models be trained for every single star. Although it is always possible to apply the DL learning process to small datasets, the results were not as accurate or trustworthy as the predictions we obtained with a more extensive grid.
Instead, we tried different architectures to take into account this prior knowledge about $T_{\rm eff}$ and $\log{g}$ in our DL models (see Fig.\,\ref{fig:PriorKnowledge}). In this way, we were able to inject these conditions into the creation of DL models for predicting metallicity. The parameters that we fix are added at the end of the convolutional feature vector. 
We also consider the uncertainties of $T_{\rm eff}$ and $\log{g}$ from the overall median. For that purpose, we create two different sets of predictions. First, $T_{\rm eff}$ and $\log{g}$ are fixed without taking into account their uncertainties.
Second, we generate 50 copies of the original flux, but with $T_{\rm eff}$ and $\log{g}$ extracted from a binomial distribution with the overall median of $T_{\rm eff}$ and $\log{g}$ as the center and the uncertainties of each parameter as the corresponding standard deviations.
Finally, we aggregate the different predictions and create a probability density function using the Kernel Density Estimate \citep[KDE;][]{Rosenblatt1956,Parzen1962} for each benchmark star. The final result for metallicity is drawn from the maximum of the KDE, with its uncertainty derived from the $1\sigma$ threshold.

For {\tt ODUSSEAS,} it is not possible to fix any  parameters for technical reasons, therefore, this team cannot provide any metallicities for this run. Since the parameter determination process of {\tt ODUSSEAS} correlates the pEWs of new spectra with the pEWs and reference stellar parameters of the HARPS dataset of same resolution, fixing the $T_{\rm eff}$ of the new spectra, or even leaving out the $T_{\rm eff}$ prediction completely from the whole process, makes no difference to the derived [Fe/H] of new spectra.

\subsubsection{Run C}
\label{methods:RunC}

In the last run, we standardized our methods by using the same wavelength regions, the same synthetic model spectra, and the same continuum normalization method. The analyzed wavelength regions are provided by the {\tt SteParSyn} team and are summarized in Table\,\ref{tab:regions_C-C2}. Because some of these regions are a better fit for hotter M dwarfs but are shown to perform rather poorly for cooler spectral types, we manually selected 35 of them that yield good fits over the whole spectral type range and use those in an additional Run C2.
For the synthetic spectra, we used the PHOENIX-ACES model spectra grid as described in Sect.\,\ref{Pass19-code}. We incorporated the same normalization method as the DL team, the {\tt GISIC} routine (see Sect.\,\ref{DL}). 
In the end, all teams were provided with normalized CARMENES and PHOENIX-ACES synthetic spectra for all wavelength regions from Table\,\ref{tab:regions_C-C2} to then run their individual algorithms to derive the stellar parameters $T_{\rm eff}$, $\log{g}$, and [Fe/H]. 
The line list employed by {\tt ODUSSEAS} has a specific format of lower and upper wavelength boundaries for each absorption feature, which covers the range from 5300 to 6900\,{\AA}. Thus, {\tt ODUSSEAS} can only use those normalized CARMENES spectral regions within this range to measure the respective absorption lines and determine the stellar parameters based on them. Their modified Run C is designated as Run C* in the following.

\begin{table}
\small

\caption[]{Analyzed wavelength regions for Runs C and C2. Wavelengths are given in vacuum.}
\label{tab:regions_C-C2}
\centering
\begin{tabular}{cccccccc}
\hline 
\hline 
\noalign{\smallskip}
\multicolumn{2}{c}{Region}  & \multicolumn{2}{c}{Run} & \multicolumn{2}{c}{Region}  & \multicolumn{2}{c}{Run} \\
$\lambda_{\rm start}$\,[\AA] & $\lambda_{\rm end}$\,[\AA] & C & C2 & $\lambda_{\rm start}$\,[\AA] & $\lambda_{\rm end}$\,[\AA] & C & C2 \\
\noalign{\smallskip}
\hline 
\noalign{\smallskip}
5867.60 & 5868.55 & \textbullet &                &  8437.44 & 8438.51  & \textbullet & \textbullet\\
5923.32 & 5924.16 & \textbullet &                &  8452.76 & 8453.66  & \textbullet & \textbullet\\
5979.80 & 5980.62 & \textbullet &                &  8469.03 & 8469.97  & \textbullet &   \\
6065.88 & 6066.77 & \textbullet &                &  8470.22 & 8471.29  & \textbullet &   \\
6066.73 & 6067.62 & \textbullet &                &  8515.87 & 8516.95  & \textbullet & \textbullet\\
6086.46 & 6087.38 & \textbullet &                &  8516.99 & 8517.90  & \textbullet & \textbullet\\
6127.49 & 6128.36 & \textbullet &                &  8549.98 & 8550.92  & \textbullet &   \\
6137.88 & 6138.78 & \textbullet &                &  8584.16 & 8585.10  & \textbullet &   \\
6138.94 & 6139.86 & \textbullet &                &  8613.68 & 8614.66  & \textbullet &   \\
6394.95 & 6395.81 & \textbullet &                &  8676.66 & 8677.61  & \textbullet & \textbullet\\
6432.17 & 6433.08 & \textbullet &                &  8677.29 & 8678.23  & \textbullet & \textbullet\\
6477.00 & 6477.84 & \textbullet & \textbullet &  8684.87 & 8685.86  & \textbullet & \textbullet\\
6483.23 & 6484.09 & \textbullet & \textbullet &  8690.44 & 8691.63  & \textbullet & \textbullet\\
6557.43 & 6558.32 & \textbullet & \textbullet &  8694.27 & 8695.18  & \textbullet & \textbullet\\
6594.31 & 6595.20 & \textbullet &                &  8759.11 & 8760.09  & \textbullet &   \\
6600.51 & 6601.35 & \textbullet &                &  8826.06 & 8827.22  & \textbullet & \textbullet\\
7050.91 & 7061.52 & \textbullet & \textbullet &  8840.36 & 8841.35  & \textbullet & \textbullet\\
7084.05 & 7094.66 & \textbullet & \textbullet &  9012.65 & 9013.50  & \textbullet &       \\
7121.66 & 7132.27 & \textbullet & \textbullet &  9721.13 & 9722.16  & \textbullet & \textbullet\\
7390.98 & 7391.91 & \textbullet &                &  9730.48 & 9731.70  & \textbullet & \textbullet\\
7412.74 & 7413.67 & \textbullet &                &  9834.35 & 9835.38  & \textbullet &   \\
7491.21 & 7492.08 & \textbullet &                & 10343.25 & 10344.22 & \textbullet & \textbullet\\
7497.72 & 7498.62 & \textbullet &                & 10381.35 & 10382.32 & \textbullet &   \\
7585.42 & 7586.36 & \textbullet & \textbullet & 10398.17 & 10399.14 & \textbullet & \textbullet\\
7914.61 & 7915.50 & \textbullet & \textbullet & 10586.99 & 10588.10 & \textbullet & \textbullet\\
8000.61 & 8001.72 & \textbullet &                & 10663.98 & 10665.14 & \textbullet &   \\
8076.92 & 8077.84 & \textbullet & \textbullet & 10777.31 & 10778.38 & \textbullet &       \\
8206.78 & 8207.65 & \textbullet &                & 11799.85 & 11800.97 & \textbullet &   \\
8398.70 & 8399.73 & \textbullet & \textbullet & 11886.71 & 11887.97 & \textbullet & \textbullet\\
8403.31 & 8404.14 & \textbullet & \textbullet & 11952.26 & 11953.39 & \textbullet & \textbullet\\
8414.15 & 8415.22 & \textbullet & \textbullet & 12814.42 & 12815.54 & \textbullet & \textbullet\\
8418.80 & 8419.74 & \textbullet & \textbullet & 12922.87 & 12923.99 & \textbullet &       \\
8428.29 & 8429.39 & \textbullet & \textbullet & 15606.45 & 15607.80 & \textbullet &       \\
8436.73 & 8437.84 & \textbullet & \textbullet & 15719.19 & 15720.61 & \textbullet &       \\
8437.02 & 8447.62 & \textbullet & \textbullet &                   &                 &   \\
\noalign{\smallskip}
\hline 
\end{tabular}
\end{table}

\section{Results}
\label{results}

All the results for each star, run, and method are listed in Table\,\ref{tab:results1} and \ref{tab:results2}. In the following, we discuss the results for each run. As discussed in Sect.\,\ref{sample}, we compare our results to the literature median, assuming that the literature median represents accurate parameter values for each star, to investigate the consistency of our results over the different runs. 

\subsection{Run A}
In Run A, all teams determined the stellar parameters with their methods without any restrictions. Figure\,\ref{fig:A-comparison} shows the comparison of our results with the literature median for each star.
This gives a direct comparison of how each method performed. 

\subsection*{Effective temperature}
It can be seen in the top panel of Fig.\,\ref{fig:A-comparison} that all the methods are mostly consistent with the literature median (purple dot) within the errors and with only a few outliers. 
Overall, {\tt SteParSyn} reproduces the literature values best. Compared to the literature median, the mean difference $\overline{\Delta T_{\rm eff}} = {\rm mean} (T_{\rm eff}^{\rm our} - T_{\rm eff}^{\rm lit})$ is +7\,K, meaning that, on average, {\tt SteParSyn} derived $T_{\rm eff}$ 7\,K hotter. Their results fall only two times outside of the error range, which is defined from the combined error bars of the literature median and the respective method for each star. 
{\tt SteParSyn} is followed by the {\tt Pass19-code}, which is on average 50\,K hotter than the literature median and falls only once outside the error range. Results from DL lean on the hotter side as well, showing an average of 75\,K larger than the literature median, and also falling two times outside the error. 
In contrast to the previous methods, {\tt ODUSSEAS} consistently determines $T_{\rm eff}$ cooler than the other methods and, on average, 86\,K cooler compared to the literature median. There are only two exceptions when {\tt ODUSSEAS} derives hotter $T_{\rm eff}$, namely: J07558+833 
and J13005+056. 
Both stars have large $\varv\sin{i}$, which is the most likely reason for the larger $T_{\rm eff}$ values. Additionally, {\tt ODUSSEAS} falls outside the error ranges five times. 

Regarding large $\varv\sin{i}$, the other methods could determine values in good agreement with the literature median for these stars. 
In preparing  Run B, we calculated the median values between the literature and the results of Run A for each star. This overall median (red dot) is also plotted in Fig.\,\ref{fig:A-comparison} for comparison. As can be seen from the plot, the overall median is consistent with the literature median for all stars and sometimes differs only by a few K. Therefore, we considered the overall median as a benchmark value for each star and took it as a fixed value for our Run B. 

In order to provide a better visualisation of the agreement and spread of our methods, we present modified Bland-Altman plots \citep{Bland1986} for each parameter of Run A in the left column of Fig.\,\ref{fig:blandA}. These plots show the mean value of two measurements (in our case, e.g., the $T_{\rm eff}$ value from the literature median and one of our methods for each star) on the $x$-axis, and the difference on the $y$-axis. To achieve a more uniform distribution of the data points and increase the illustration of potential discrepancies the values can be plotted logarithmically or as ratio instead of the difference, as it is done here. The plot shows the same trends as described above, with DL deriving hotter $T_{\rm eff}$ and {\tt ODUSSEAS} cooler $T_{\rm eff}$ (see top left panel of Fig.\,\ref{fig:blandA}).

\subsection*{Surface gravity}
As described in Sect.\,\ref{ODUSSEAS}, {\tt ODUSSEAS} does not provide $\log{g}$. From the remaining methods, the {\tt Pass19-code} performs best. The differences between the results and the literature median are almost always within 0.1\,dex, and only once does their value fall outside the error range. On average, $\log{g}$ from the {\tt Pass19-code} are 0.01\,dex higher that the literature median, this difference is nearly negligible. The reason for this is likely the use of evolutionary models to constrain $\log{g}$. 
The DL method derives on average 0.06\,dex lower $\log{g}$ than the literature median and lies seven times outside of the error. 
For {\tt SteParSyn} the $\log{g}$ has values 11 times outside the error range, being, on average,  0.10\,dex higher than the literature median. In several cases, $\log{g}$ is significantly higher than the literature median and the other methods. The biggest outlier is GJ\,338B, where {\tt SteParSyn} derives a value of 0.53\,dex larger than the literature median.
A possible explanation for these high values could be either a still remaining degeneracy in the stellar parameter space or the synthetic gap (difference in feature distribution between synthetic and observed spectra). As shown in \cite{Marfil2021}, {\tt SteParSyn} retrieves tentatively higher $\log{g}$ values for the whole CARMENES GTO sample.

In the case of GJ\,1002 the overall median for $\log{g}$ represents the median of all our Run A results because there are no literature values for this star. In total, the literature and overall median differ less than 0.1\,dex in all cases, except for GJ\,493.1, which only has one literature value of 4.5\,dex. Therefore, we excluded this star from the analysis of $\log{g}$ in this section.

\subsection*{Metallicity}
\begin{figure*}[!ht]
  \centering
  \includegraphics[width=0.90\linewidth]{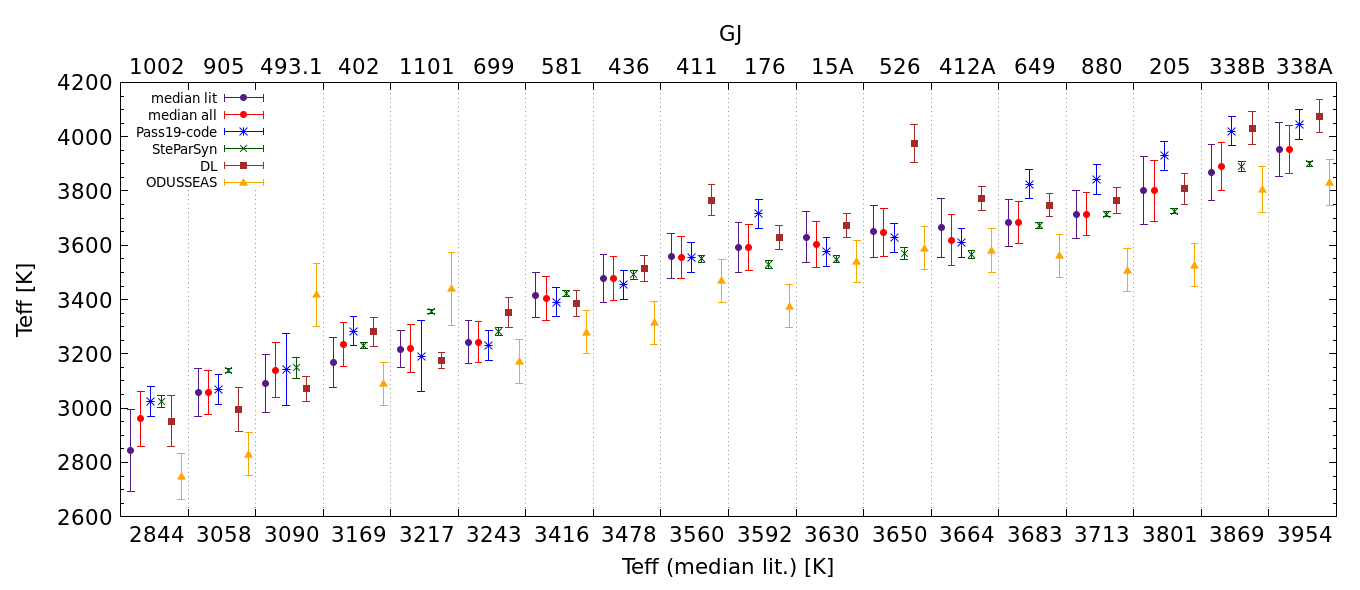}\hfil
  \includegraphics[width=0.90\linewidth]{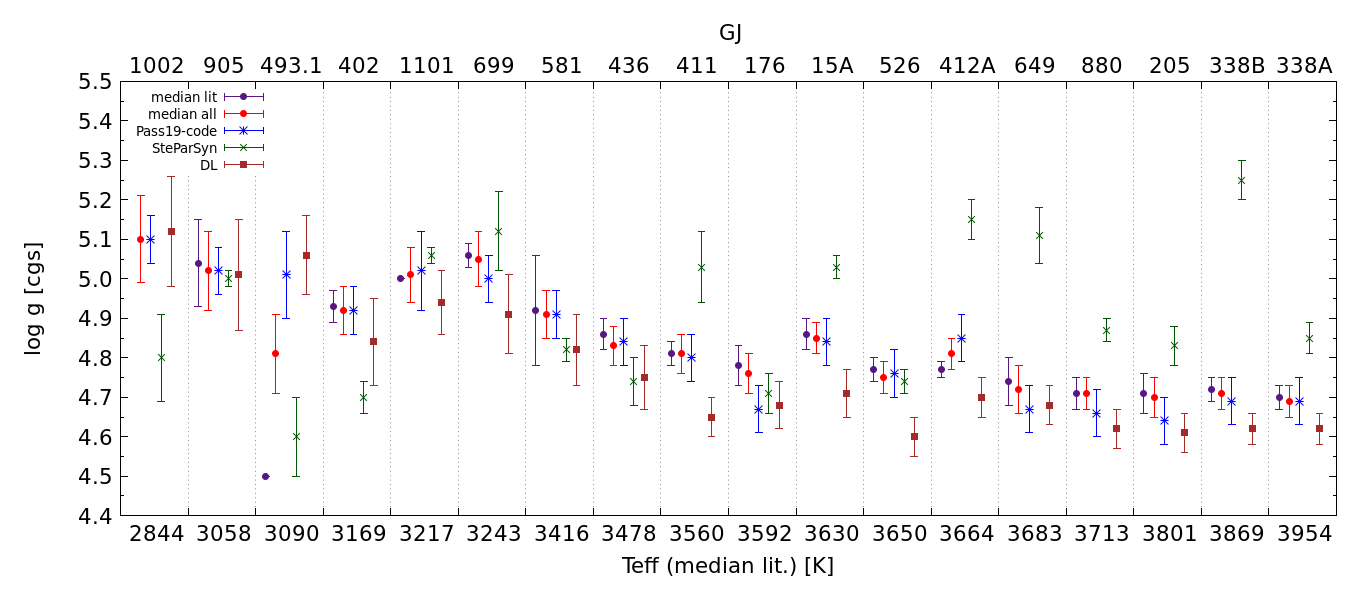}\par\medskip
  \includegraphics[width=0.90\linewidth]{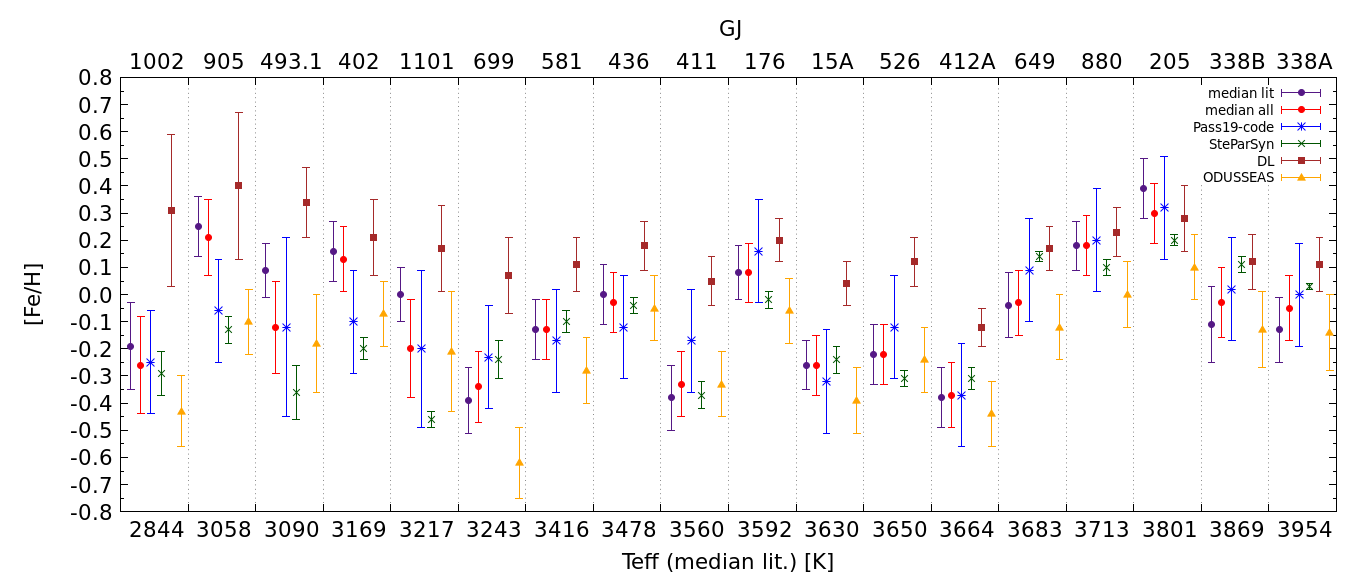}
\caption{Comparison of $T_{\rm eff}$ ({\it top}), $\log{g}$ ({\it middle}), and [Fe/H] ({\it bottom}) for the different methods in Run A. Each method is indicated with a different symbol and color. The median of all literature values and the median of literature+Run A are shown as purple and red dots, respectively. The $x$-axis indicates the $T_{\rm eff}$ from the literature median and the top axis shows the Gliese-Jahreiss (GJ) numbers for all sample stars, which are sorted by $T_{\rm eff}$ from the literature median to display any possible trends.}
\label{fig:A-comparison}
\end{figure*}

The bottom panel of Fig.\,\ref{fig:A-comparison} presents the results of all methods for [Fe/H]. Although this parameter is not fixed in Run B, we calculated and plotted an overall median for the purposes of comparison. 
The {\tt Pass19-code} performs best compared to the literature median and the other methods. On average, the metallicities are 0.02\,dex lower than the literature median. All values agree with each other within their errors, for eight stars the results are within 0.1\,dex difference to the literature median, and it is only for five stars that the difference is greater than 0.2\,dex. An explanation for this good performance can be the careful line selection of magnetically insensitive lines in the VIS and NIR. The use of multiple lines simultaneously also cancels out most of the effects coming from the synthetic gap, which impacts DL in particular. On this note,
DL performs worst when it comes to the metallicity determination. The results for 10 stars, which is more than half of our benchmark sample, lie outside the error range; for 11 stars, the values differ by more than 0.2\,dex from the literature median, while only two stars have differences less than 0.1\,dex. On average, DL provides metallicities 0.23\,dex higher than the literature median, tentatively deriving higher values for all but one star (GJ\,205). 

{\tt ODUSSEAS} and {\tt SteParSyn} determine tentatively lower values for metallicity, with $\overline{\Delta {\rm [Fe/H]}}$ of --0.14\,dex and --0.08\,dex, respectively. For {\tt ODUSSEAS}, two values fall outside the error range, while for {\tt SteParSyn} it is eight. Eight stars show differences of less than 0.1\,dex with {\tt ODUSSEAS}, and six stars differ by more than 0.2\,dex compared to the literature median. For {\tt SteParSyn}, eight stars fall within 0.1\,dex of the literature median and five stars outside of 0.2\,dex. All these numbers are summarized in Table\,\ref{tab:RunA_summary} for better readability.

As for $T_{\rm eff}$ and $\log{g}$, we provide a Bland-Altman plot for [Fe/H] in the bottom left panel of Fig.\,\ref{fig:blandA}. In order to avoid a possible division by zero on the $y$-axis due to values of solar metallicity, we transform [Fe/H] to the logarithmic number ratio of iron and hydrogen atoms via [Fe/H] = $\log(N_{\rm Fe}/N_{\rm H})_\star - \log(N_{\rm Fe}/N_{\rm H})_\odot$, with $\log(N_{\rm Fe}/N_{\rm H})_\odot = -4.5$ \citep[see Table 6 in][]{Lodders2009}.
Overall, the {\tt Pass19-code} performs best in $\log{g}$ and [Fe/H] compared to the literature median. For $T_{\rm eff}$, {\tt SteParSyn} would be the best choice. 

\begin{table}[!ht]
\caption[]{Analysis of Run A for $T_{\rm eff}$ ({\it top}), $\log{g}$ ({\it  middle}), and [Fe/H] ({\it bottom})$^{(a)}$. }
\label{tab:RunA_summary}
\centering
   \begin{tabular*}{\linewidth}{lcccc}
        \hline 
        \hline 
        \noalign{\smallskip}
         & {\tt Pass-19}& {\tt SteParSyn} & DL  & {\footnotesize {\tt ODUSSEAS}} \\
         & {\tt code} & & & \\
        \hline
        \noalign{\smallskip}
        $\overline{\Delta T_{\rm eff}}$ [K] & +50 & +7 & +75 & --86 \\
        \# o/s error & 1 & 2 & 2 & 6 \\
        \# <100\,K & 11 & 16 & 10 & 8 \\
        \# >200\,K & 0 & 0 & 2 & 6 \\
        \hline
        \noalign{\smallskip}
        $\overline{\Delta \log{g}}$ [cgs] & +0.01 & +0.10 & --0.06 & \ldots \\
        \# o/s error & 1 & 11 & 7 & \ldots \\
        \# <0.1\,dex & 15 & 6 & 8 & \ldots \\
        \# >0.2\,dex & 0 & 5 & 1 & \ldots \\
        \hline
        \noalign{\smallskip}
       $\overline{\Delta {\rm [Fe/H]}}$ [dex] & --0.02 & --0.08 & +0.23 & --0.14 \\
        \# o/s error & 0 & 8 & 10 & 2 \\
        \# <0.1\,dex & 8 & 8 & 2 & 7 \\
        \# >0.2\,dex & 5 & 5 & 11 & 7 \\
        \noalign{\smallskip}
            \hline 
        \noalign{\smallskip}
        
    \end{tabular*}
    \tablefoot{$^{(a)}$ The results for each team are provided in different columns, showing the mean difference to the literature median, the number of stars for which the results fall outside the error range, and the number of stars for which the results lie within 100\,K and 0.1\,dex, and outside 200\,K and 0.2\,dex of the literature median, for $T_{\rm eff}$, and $\log{g}$ and [Fe/H], respectively.}
 \end{table}

\subsection{Run B}

\begin{figure*}[!ht]
  \centering
  \includegraphics[width=0.95\linewidth]{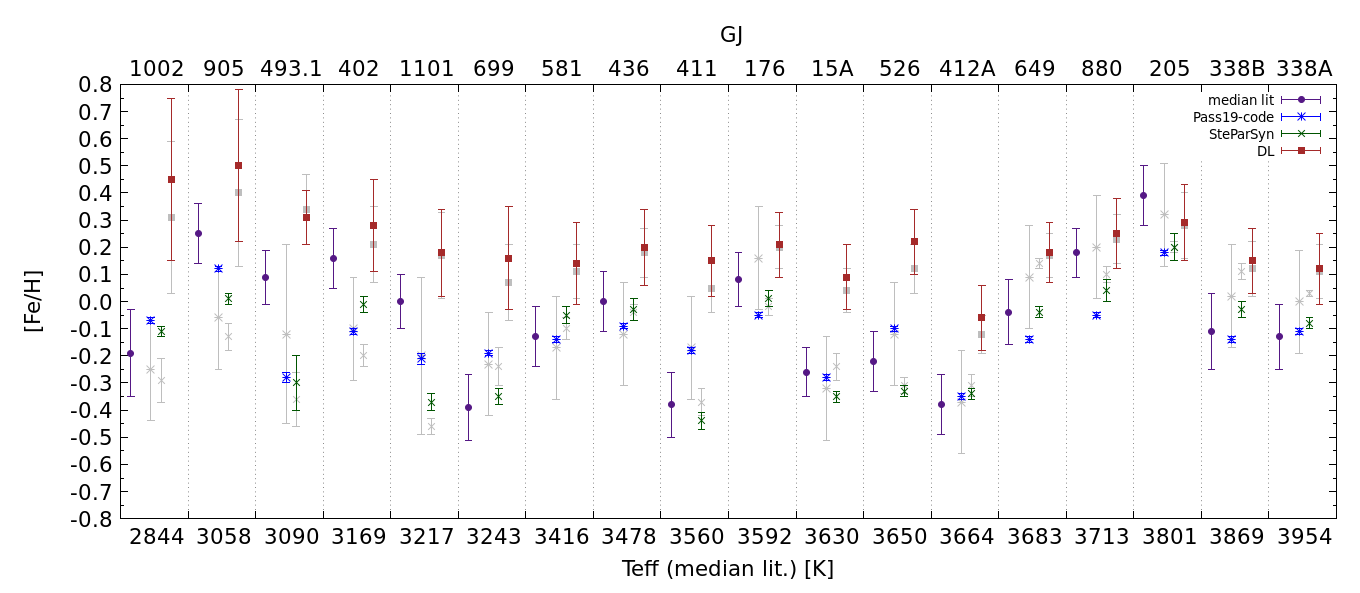}
\caption{Comparison of [Fe/H] for the different methods in Run B. Each method is indicated with a different symbol and color. The gray symbols indicate the results from Run A for comparison. The median of all literature values is shown as purple dots. The $x$-axis indicates $T_{\rm eff}$ from the median literature, the top axis shows the Gliese-Jahreiss (GJ) numbers for all sample stars.}
\label{fig:B-comparison}
\end{figure*}

For Run B, all teams (except {\tt ODUSSEAS}, see Sect.\,\ref{methods:RunB}) derived only [Fe/H], with $T_{\rm eff}$ and $\log{g}$ fixed to the median values determined from all teams in Run A and literature values. In Fig.\,\ref{fig:B-comparison} we show a comparison between our results and the literature median. Results from Run A are plotted in gray to illustrate the changes between the runs. We can see from this plot that fixing $T_{\rm eff}$ and $\log{g}$ does not improve the metallicities derived with the {\tt Pass19-code} and DL. For both methods, the discrepancies with regard to the literature median increased. This can be explained by looking at the results for $T_{\rm eff}$ from Run A. If the temperatures were further away from the overall median, which was used to fix this parameter in Run B, then the deviation in metallicity in Run B is larger than in Run A. Some correlation with $\log{g}$ can also be found in some cases. Since in Run B, the parameter determination is reduced to a 1-D problem, there are no longer any local minima  and there is only one best value for metallicity. If the fixed values $T_{\rm eff}$ and $\log{g}$ deviate from the best fitting values found in Run A, the deviation in metallicity will consequently increase as well. Therefore, there will be no improvement regarding metallicity, unless the other parameters $T_{\rm eff}$ and $\log{g}$ can be chosen freely as well. We performed the same analysis of the results here as we did for Run A, with the figures summarized in Table\,\ref{tab:RunB_summary}. The corresponding Bland-Altman plot is presented on the left in Fig.\,\ref{fig:blandB}. 

On the other hand, fixing parameters slightly improved the metallicities derived by {\tt SteParSyn}. The number of stars outside the error range decreased, whereas the number within 0.1\,dex increased. A good example here is GJ\,338B:\ from Run A, $T_{\rm eff}$ is close to the literature median, but $\log{g}$ is far off. By fixing $\log{g,}$ the metallicity improves and moves closer to the literature median. 
This run suggests that there is a dependency on the stellar synthetic spectra used in the analysis. The DL and the {\tt Pass19-code} both rely on the PHOENIX-ACES model spectra, but do not show any improvements towards literature values, whereas {\tt SteParSyn} incorporated BT-Settl model atmospheres. Therefore, the next step is for all methods to use the same synthetic models. 

\begin{table}
\caption[]{Analysis of Run B for [Fe/H].}
\label{tab:RunB_summary}
\centering
   \begin{tabular}{lccc}
        \hline 
        \hline 
        \noalign{\smallskip}
         & {\tt Pass-19}& {\tt SteParSyn} & DL \\
         & {\tt code} & & \\
        \hline
        \noalign{\smallskip}
        $\overline{\Delta {\rm [Fe/H]}}$ [dex] & --0.06 & --0.08 & +0.27\\
        \# o/s error & 8 & 5 & 7 \\
        \# <0.1\,dex & 6 & 11 & 1  \\
        \# >0.2\,dex & 6 & 3 & 12 \\
        \noalign{\smallskip}
            \hline 
        \noalign{\smallskip}
        
    \end{tabular}
 \end{table}

\subsection{Runs C and C2}
In Run C, all teams incorporated the same normalized PHOENIX-ACES model spectra and derived the parameters from the same wavelength regions. As mentioned in Sect.\,\ref{methods:RunC}, we carried out an additional Run C2, using a subset of 35 wavelength regions from Run C, but otherwise identical to Run C. Figure\,\ref{fig:C2-comparison} presents a comparison of stellar parameters between Run A and Run C2.
A comparison between Run C and C2 is shown in Fig.\,\ref{fig:C-comparison2} and as a Bland-Altman plot in the left column of Fig.\,\ref{fig:blandC} and Fig.\,\ref{fig:blandC2}. Table\,\ref{tab:RunC_summary} summarizes the statistics of Runs C and C2. 

\begin{table}
\caption[]{Analysis of Run C/C2 for $T_{\rm eff}$ ({\it top}), $\log{g}$ ({\it  middle}), and [Fe/H] ({\it bottom}).}
\label{tab:RunC_summary}
\centering
   \begin{tabular}{lccc}
        \hline 
        \hline 
        \noalign{\smallskip}
         & {\tt Pass-19}& {\tt SteParSyn} & DL \\
         & {\tt code} & & \\
        \hline
        \noalign{\smallskip}
        $\overline{\Delta T_{\rm eff}}$ [K] & +60/+101 & +46/+0.1 & +209/+140 \\
        \# o/s error & 0/2 & 8/7 & 5/2 \\
        \# <100\,K & 13/6 & 9/7 & 1/6 \\
        \# >200\,K & 0/3 & 3/4 & 5/3 \\
        \hline
        \noalign{\smallskip}
        $\overline{\Delta \log{g}}$ [cgs] & --0.08/+0.01 & --0.03/--0.03 & --0.02/+0.05 \\
        \# o/s error & 5/4 & 8/13 & 5/2 \\
        \# <0.1\,dex & 9/13 & 5/4 & 10/15 \\
        \# >0.2\,dex & 4/1 & 7/12 & 3/1 \\
        \hline
        \noalign{\smallskip}
        $\overline{\Delta {\rm [Fe/H]}}$ [dex] & +0.31/--0.17 & --0.07/--0.17 & --0.09/--0.07 \\
        \# o/s error & 8/4 & 8/11 & 1/2 \\
        \# <0.1\,dex & 4/1 & 4/3 & 8/8 \\
        \# >0.2\,dex & 9/9 & 6/11 & 7/5 \\
        \noalign{\smallskip}
            \hline 
        \noalign{\smallskip}
        
    \end{tabular}
 \end{table}

\subsection*{Effective temperature}
\begin{figure*}[!ht]
  \centering
  \includegraphics[width=0.90\linewidth]{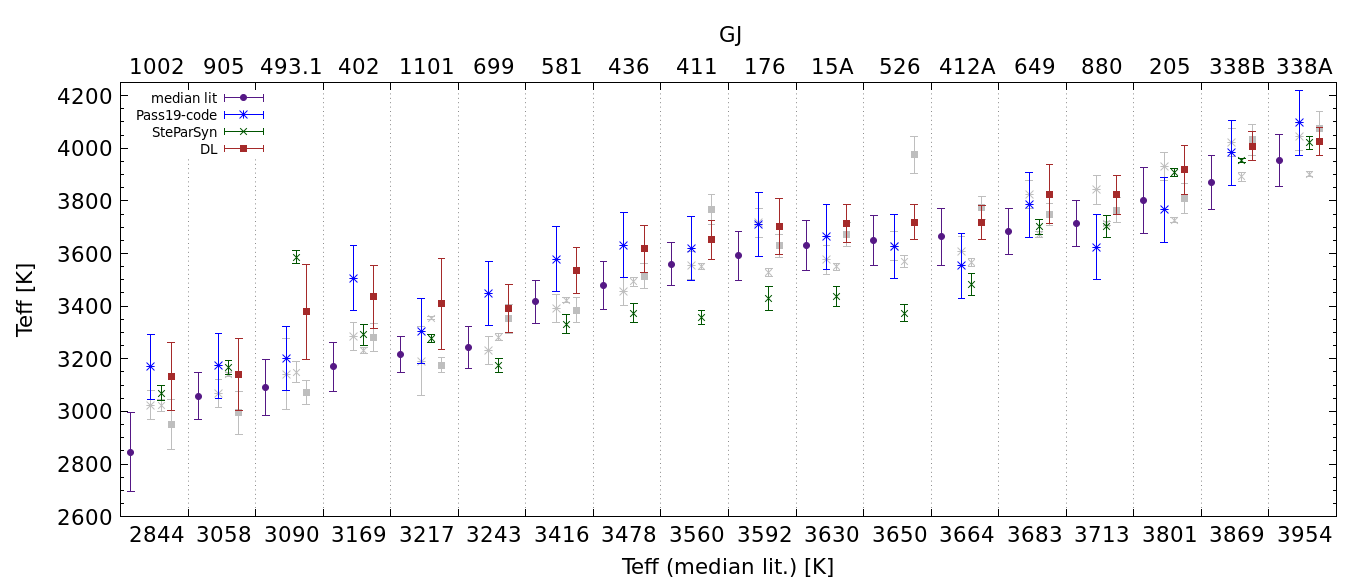} 
  \includegraphics[width=0.90\linewidth]{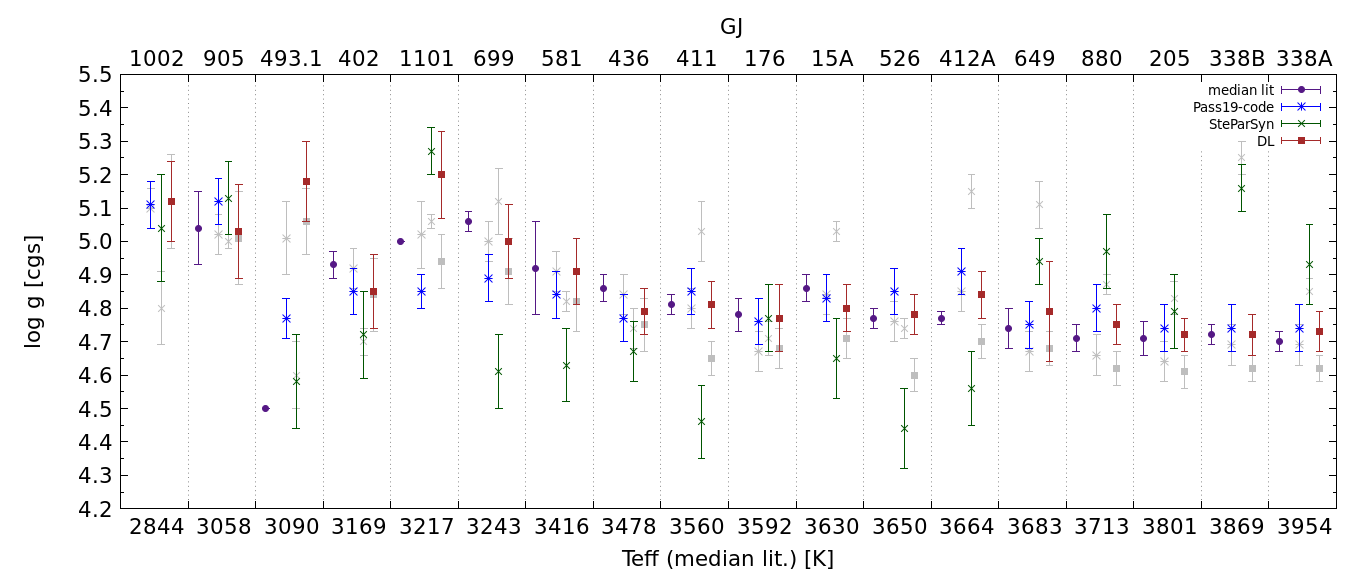} 
  \includegraphics[width=0.90\linewidth]{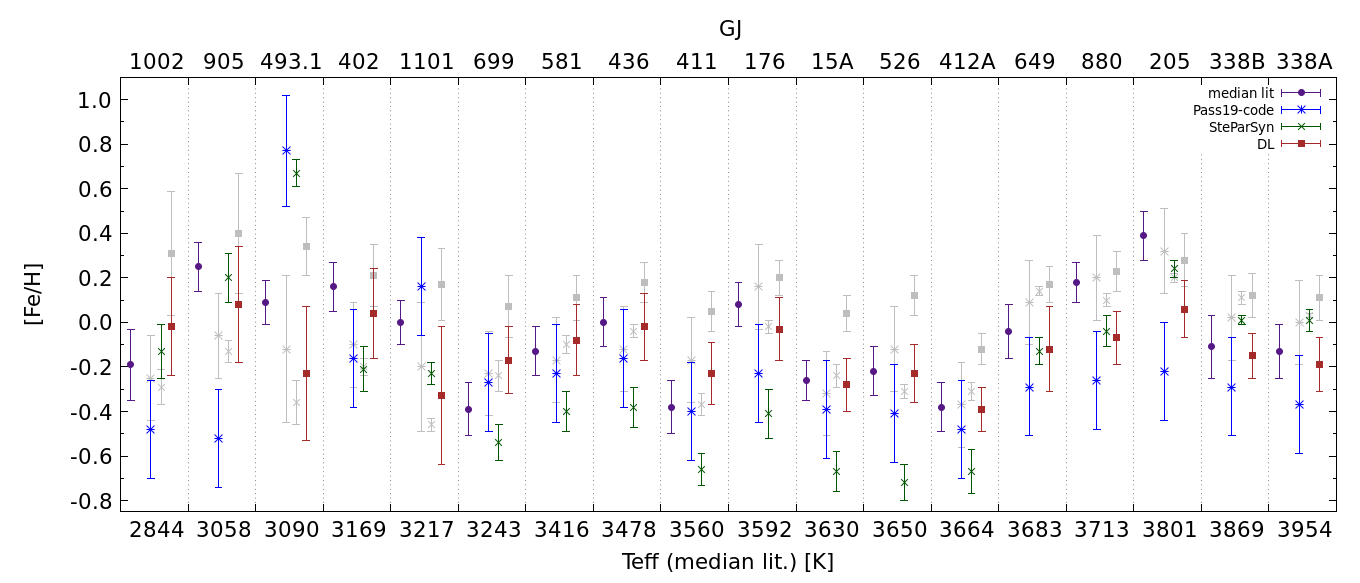}
\caption{Comparison of $T_{\rm eff}$ ({\it top}), $\log{g}$ ({\it middle}), and [Fe/H] ({\it bottom}) for the different methods in Run C2. Each method is indicated with a different symbol and color. The gray symbols indicate the results from Run A for comparison. The median of all literature values is shown as purple dots. The $x$-axis indicates $T_{\rm eff}$ from the literature median, the top axis shows the Gliese-Jahreiss (GJ) numbers for all sample stars.}
\label{fig:C2-comparison}
\end{figure*}

We compared our results from Run C2 to those derived in Run A. Figure\,\ref{fig:C2-comparison} shows that the stellar parameters do not improve from Run A to Run C2. This is most evident for $T_{\rm eff}$, where the 17 stars for {\tt SteParSyn}, 13 stars for DL, and 12 stars for the {\tt Pass19-code} show larger deviations to the literature median than in Run A. 
Analyzing Runs C and C2 shows that stellar parameters derived with DL agree better with the literature median in Run C2 than in Run C. This means an improvement towards Run C2, with all stars being closer to the literature median in $T_{\rm eff}$ compared to Run C.
For {\tt SteParSyn}, the results from Run C and C2 are a bit more ambiguous. Half of the sample stars are closer to the literature median in Run C for $T_{\rm eff}$, the other half in Run C2. Run C exhibits a larger $\overline{\Delta T_{\rm eff}}$ of +46\,K compared to Run C2 ($\overline{\Delta T_{\rm eff}}$ = {+0.1}\,K). 
Concerning the {\tt Pass19-code}, Run C is clearly better than Run C2, giving values closer to the literature median. The mean difference, $\overline{\Delta T_{\rm eff}}$, amounts to +60\,K, with all values being within the error range for all stars.

\subsection*{Surface gravity}
Compared to Run A, the {\tt Pass19-code} and {\tt SteParSyn} show similar results as for $T_{\rm eff}$, with no improvement from Run A to Run C2. On the other hand, 14 stars with $\log{g}$ derived from DL are closer to the literature median in Run C2 than in Run A. 
For Run C, the results from DL show an improvement towards Run C2 for eleven stars. 
Only five stars from {\tt SteParSyn} lie closer to the literature median in Run C2 than in Run C, which clearly favours the results from Run C in this case.  
For the {\tt Pass19-code}, the results derived in Run C are tentatively lower than for Run C2. Especially in the case of hotter stars, this means that Run C is closer to the literature median, as can be seen from Fig.\,\ref{fig:C-comparison2}. 

\subsection*{Metallicity}
Similarly to the cases of $T_{\rm eff}$ and $\log{g,}$ the results in metallicity for the {\tt Pass19-code} and {\tt SteParSyn} are closer to literature in Run A than in Run C2. However, DL presents a slight improvement, deriving values which are closer to the literature median for 12 stars in metallicity. Using multiple wavelength ranges instead of only one range, as done in Runs A and B, appears to trigger this improvement. Looking at Run C, eight of 18 stars have better values in Run C2 for DL. 
As for $\log{g}$, only a small number of five stars shows better results in Run C2 than in Run C for {\tt SteParSyn}. This could be explained by the fact that {\tt SteParSyn} is optimized for the wavelength ranges in Run C that are originally used by the method. Analyzing a subset of these ranges, as done in Run C2, does not improve the results. In Run C, the {\tt Pass19-code} consistently derives  too high metallicities, especially for cooler M-dwarfs (see Fig.\,\ref{fig:C-comparison2}, bottom panel). The selection of wavelength ranges for Run C2, however, clearly improves the metallicity determination.

\vspace{\baselineskip}
Comparing the numbers Table\,\ref{tab:RunC_summary} to those of Runs A and B also illustrates the differences and the better performance of Run A. It can be seen, for example, that the number of stars with differences larger than 200\,K and 0.2\,dex from the literature median significantly increased in Run C2. One exception here is DL, which is able to lower those numbers in Run C2 for [Fe/H]. 

\subsection{Comparison with interferometry}
\label{interferometry}

\begin{figure*}[!ht]
\centering
\includegraphics[width=0.89\textwidth]{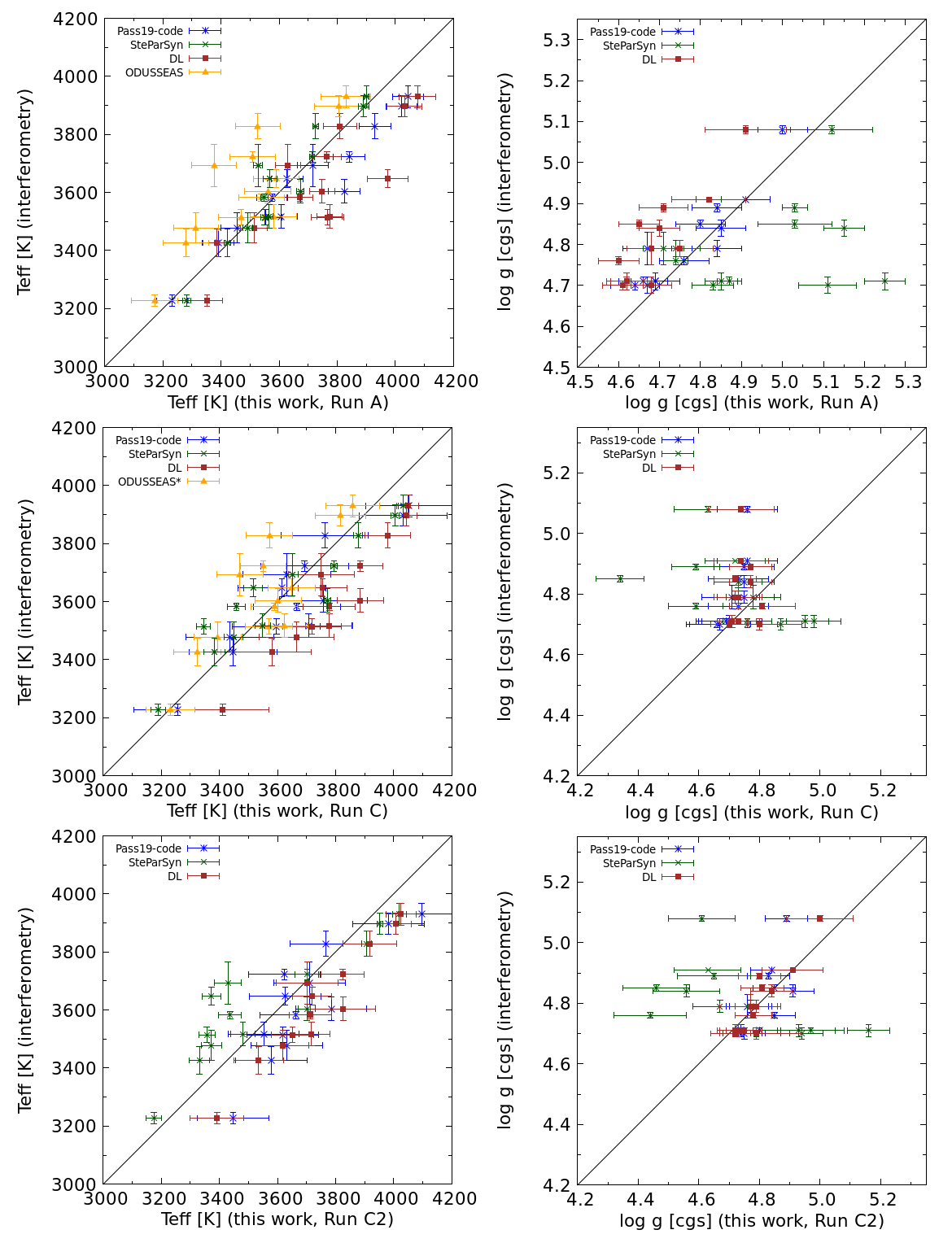}
\caption{Comparison of our results for $T_{\rm eff}$ ($left~column$) and $\log{g}$ ($right~column$) from Run A ($top$), Run C ($middle$), and Run C2 ($bottom$) with interferometric radii from \cite{Boyajian2012}, \cite{vonBraun2014}, \cite{Newton2015}, and \cite{Rabus2019}. If more than one value exists for a star in literature, we plot the mean with the RMSE for better readability. 
Results from {\tt ODUSSEAS} in Run C correspond to their Run C*. The black line indicates the 1:1 relationship. }
\label{fig:compare_interferometry}
\end{figure*}

\begin{table*}
\caption[]{Mean difference between our results and interferometric literature values, $\overline{\Delta T_{\rm eff}}$ / $\overline{\Delta \log{g}}$, standard deviation of the mean difference, std.\,dev., and Pearson correlation coefficients, $r_P$, for $T_{\rm eff}$ / $\log{g}$.}
\label{tab:pearson}
\centering
   \begin{tabular}{lcccc}
        \hline 
        \hline 
        \noalign{\smallskip}
         & {\tt Pass19-code} & {\tt SteParSyn} & DL & {\tt ODUSSEAS} \\
          
        \hline
        \noalign{\smallskip}
        Run A \\ 
        \noalign{\smallskip}
        $\overline{\Delta T_{\rm eff}}$ / $\overline{\Delta \log{g}}$ & +60 / $-0.04$ & $-18$ / +0.13 & +111 / $-0.11$ & $-123$ / ... \\
        std. dev. & 80 / 0.04 & 75 / 0.18 & 123 / 0.06 & 
           113 / ... \\ 
        $r_P$ & 0.950 / 0.928 & 0.923 / 0.300 & 0.817 / 0.881 & 0.815 / ... \\
        \\
        \hline
        \noalign{\smallskip}
        Run C \\
        \noalign{\smallskip}
        $\overline{\Delta T_{\rm eff}}$ / $\overline{\Delta \log{g}}$ &  +38 / $-0.08$ & $-5$ / $-0.09$ & +169 / $-0.06$ & $-67$ / ... \\
        std. dev & 90 / 0.09 & 106 / 0.25 & 66 / 0.12 & 112 / ... \\
        $r_P$ & 0.901 / 0.795 & 0.922 / $-0.480$ & 0.939 / $-0.021$ & 0.815 / ... \\
        \\
        \hline
        \noalign{\smallskip}
        Run C2 \\
        \noalign{\smallskip}
        $\overline{\Delta T_{\rm eff}}$ / $\overline{\Delta \log{g}}$ & +72 / $-0.01$ & $-68$ / $-0.07$ & +120 / +0.001 & ... / ... \\ 
        std. dev. & 105 / 0.09 & 130 / 0.29 & 62 / 0.05 & ... / ... \\
        $r_P$ & 0.839 / 0.741 & 0.875 / $-0.536$ & 0.948 / 0.929 & ... / ... \\
        \hline
        
        \noalign{\smallskip}
        
    \end{tabular}
 \end{table*}

For 11 stars in our study, we can compare our results for $T_{\rm eff}$ and $\log{g}$ to independent measurements coming from interferometry \citep{Boyajian2012,vonBraun2014,Newton2015,Rabus2019}. \cite{Boyajian2012} and \cite{vonBraun2014} used {\em Hipparcos} parallaxes \citep{vanLeeuwen2007} to convert the limb-darkened angular stellar diameter, $\Theta_{\rm LD}$, to stellar radius via $\Theta_{\rm LD} = 2 \cdot R/d$, whereas \cite{Rabus2019} used {\em Gaia} DR2 data \citep{Gaia2018}. \cite{Newton2015} collected interferometric radii from the literature. When there was more than one measurement per star, they calculated the weighted mean. Then, 
$T_{\rm eff}$ can be derived from the Stefan-Boltzmann law, 
$T_{\rm eff} = T_0 (F_{\rm bol}/\Theta_{\rm LD})^{1/4}$ (with $T_0$ = 2341\,K), 
when the bolometric flux $F_{\rm bol}$ is known. \cite{Boyajian2012} and \cite{vonBraun2014} produced spectral energy distributions (SEDs) using flux-calibrated photometry from the literature. \cite{Rabus2019} estimated $F_{\rm bol}$ by integrating the flux from synthetic photometric flux points using PHOENIX-ACES synthetic spectra. \cite{Newton2015} presented interferometric $T_{\rm eff}$ from \cite{Mann2013b} and updated $T_{\rm eff}$ for three stars following their approach. \cite{Mann2013b} determined $F_{\rm bol}$ by comparing the measured fluxes from observed visual and NIR spectra, incorporating BT-Settl synthetic models to cover wavelength gaps in the spectra, to photometric fluxes using a correction factor to adjust the overall flux level. 
From the stellar radius and mass, $\log{g}$ can be calculated via $g = GM/R^2$. This requires the stellar mass, which cannot be measured from interferometry. Therefore, \cite{Boyajian2012} and \cite{Rabus2019} used the $K$-band mass-luminosity relation from \cite{HenryMcCarthy1993}, and from \cite{Benedict2016} and \cite{Mann2019}, respectively. \cite{vonBraun2014} determined stellar mass by deriving a mass-radius relation from the results from \cite{Boyajian2012}. 
Although $T_{\rm eff}$ can be derived independently from interferometry, $\log{g}$ can be seen as semi-independent, since it involves interferometric radii, but also empirical mass-radius or mass-luminosity relations. Therefore, such quasi-interferometric $\log{g}$ tend to possess a tentatively higher level of accuracy than $\log{g}$ derived from, for instance, synthetic model fits alone; thus, the former can be used as a reliable comparison. 
A comparison plot is shown in Fig.\,\ref{fig:compare_interferometry}. We calculated the mean difference, standard deviation, and Pearson correlation coefficient ($r_P$) between our results and the literature, presented in Table\,\ref{tab:pearson}. A good consistency between the samples would result in a low mean difference and standard deviation, as well as a Pearson correlation coefficient close to 1. 

For Run A, {\tt SteParSyn} agrees quite well with interferometry in $T_{\rm eff}$ with $\overline{\Delta T_{\rm eff}} = -18$\,K, followed by the {\tt Pass19-code}, which gives slightly hotter values with $\overline{\Delta T_{\rm eff}} = +60$\,K compared to interferometry.  Also, DL is on the hotter side ($\overline{\Delta T_{\rm eff}} = +111$\,K), whereas {\tt ODUSSEAS}, as mentioned before, derived tentatively cooler temperatures ($\overline{\Delta T_{\rm eff}} = -123$\,K). In Run C (which corresponds to Run C* for {\tt ODUSSEAS}), temperatures from DL and {\tt ODUSSEAS} are shifted more towards the hotter side, bringing {\tt ODUSSEAS} closer to the interferometric values ($\overline{\Delta T_{\rm eff}} = -67$\,K). This is the same behavior seen in Fig.\,\ref{fig:compare_all}. In contrast, the {\tt Pass19-code} provides cooler temperatures, but still mostly consistent with those from interferometry ($\overline{\Delta T_{\rm eff}} = +38$\,K). {\tt SteParSyn} performs similar to Run A, however with some larger spread at low and high temperatures, which is represented in a higher standard deviation compared to Run\,A. This is similar for Run C2, where {\tt SteParSyn} again yields some cooler temperatures compared to Run C. Overall, {\tt SteParSyn} does best in Run A, where it shows the lowest standard deviation and highest $r_P$, similarly to the {\tt Pass19-code}. On the other hand, DL shows a better 1:1 relation in Run C2, represented by the larger $r_P$ (see Table\,\ref{tab:pearson}). This indicates that the selected wavelength ranges in Run C2 lead to an improvement in the results, although there seems to be a general offset towards hotter temperatures compared to interferometry. Again, this is also illustrated in Fig.\,\ref{fig:compare_all}.

For $\log{g}$, the {\tt Pass19-code} is closest to the quasi-interferometric $\log{g}$ for all runs, which is most likely due to the use of evolutionary models in the method. However, the values are slightly lower than those in the literature, on average. The smallest standard deviation and highest $r_P$ is presented by Run A, as for $T_{\rm eff}$. The results given by the {\tt Pass19-code} in Run C are systematically lower than interferometric ones, but they improve in Run C2.
In general, DL follows the relation, but those results are lower as well. A great improvement is shown for DL in Run C and even more in Run C2, where $\overline{\Delta \log{g}}$ decreases to as low as +0.001\,dex, which can be clearly attributed to the use of multiple wavelength ranges.
Overall, the values from {\tt SteParSyn} show a large spread with a high standard deviation and low $r_P$. 
The spread is persistent in Runs C and C2, although the mean difference of all {\tt SteParSyn} values moves closer to the 1:1 relation.  

Overall, this comparison is very similar to the comparison of the literature median and yet another indication that, for most stars, Run A gives very good results compared to the literature median, with the exception of DL, where the analysis of multiple wavelength ranges results in better measurements of $\log{g}$ and a higher correlation in $T_{\rm eff}$.
A similar analysis could be done for metallicity, should there be independent measurements from a hotter FGK-type binary companion available.

\section{Discussion}
\label{discussion}

\begin{figure}[] 
\centering
\includegraphics[width=0.45\textwidth]{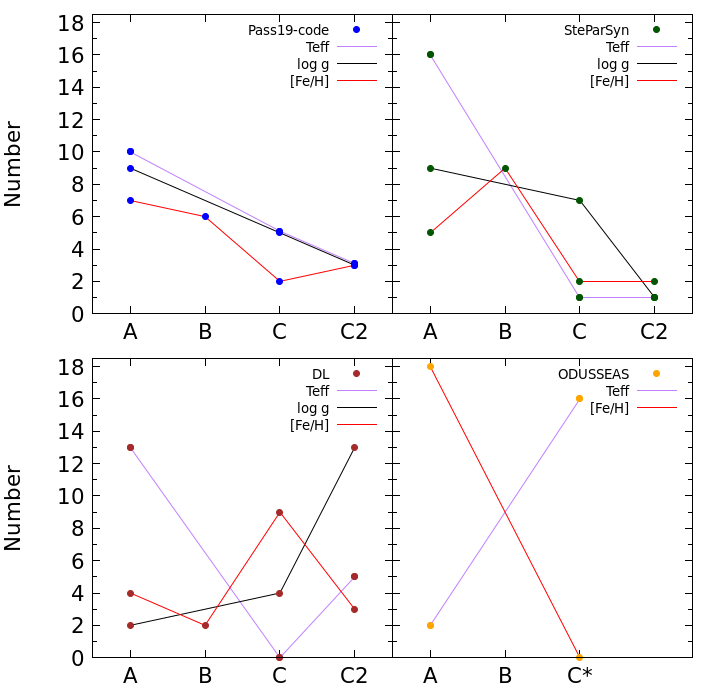}
\caption{Number of stars for each run where the stellar parameter lies closest to the literature median. Each method is shown in a separate panel. We note that {\tt ODUSSEAS} does not provide $\log{g}$, and their Run C* differs from our Runs C and C2 due to restrictions in the method itself.}
\label{fig:compare_all}
\end{figure}

We compare the results from all runs for each method in order to analyze which run gives the best results, namely, which run offers the most stars with values that are closest to the literature median. For example, in case of the {\tt Pass19-code}, each star has four determined [Fe/H] values, derived in Runs A, B, C, and C2. For each value, we calculated the difference from the literature median to find the minimum difference, for example, for Run A. This particular star then counts toward Run A. The procedure is repeated for all stars and for all three stellar parameters. As a consequence, the sum over all runs for each stellar parameter is always 18, except for $\log{g}$, where we excluded GJ\,1002 from the analysis. In this way, we can assess which run performed best for each stellar parameter.

Figure\,\ref{fig:compare_all} shows this number of stars for all parameters and methods.
From this, we can see that for $T_{\rm eff}$, all methods but {\tt ODUSSEAS} perform best in Run A. Generally, the {\tt Pass19-code} works better in Run A than in the other Runs. Runs C and C2 show a similar performance for [Fe/H]. 

The DL results get closer to the literature median for [Fe/H] and $\log{g}$ in Runs C and C2, respectively. Since the continuum normalization and synthetic spectra are the same as the ones used in Runs A and B, the only explanation for the improvement is the different wavelength regions. Then, DL can determine [Fe/H] and $\log{g}$ significantly better by taking into account more wavelength regions than just the one between 8800\,{\AA} and 8835\,{\AA}, although this region seems to work well for $T_{\rm eff}$ alone.

Similar to the {\tt Pass19-code}, {\tt SteParSyn} generally shows the best performance in Run A, with good results declining towards Runs C and C2. An exception is metallicity, which is best in Run B, directly followed by Run A. This indicates that the metallicity determinations with {\tt SteParSyn} could be improved by taking independent estimates and fixing $T_{\rm eff}$ and $\log{g}$. The stellar parameters derived in Runs C and C2 show tentatively larger deviations from the literature median than Run A, which is probably due to the different synthetic spectra used. This implies that {\tt SteParSyn} is optimized for the analysis of their selected wavelength ranges with BT-Settl models.

\subsection{{\tt ODUSSEAS} Run C*}

As described in Sect.\,\ref{methods:RunC}, {\tt ODUSSEAS} was only able to use the bluest wavelength ranges provided for Runs C and C2. Therefore, the results cannot be directly compared to the other methods. However, it is possible to assess the performance of the algorithm itself. Since {\tt ODUSSEAS} does not rely on synthetic model spectra, and a different continuum normalization does not affect the measurement of pEWs, the only difference between Runs A and C* is the choice of the wavelength ranges. In the bottom right panel of Fig.\,\ref{fig:compare_all}, it can be seen that {\tt ODUSSEAS} derives the best metallicities for all 18 stars in Run A. For $T_{\rm eff}$, Run A shows tentatively lower values compared to other methods and the literature. However, their modified Run C* gives significantly better results. From this, we can conclude that the wavelength ranges used in Run A are very good for deriving metallicity, but seem to be less sensitive to $T_{\rm eff}$. On the other hand, the ranges used in Run C* appear to be more appropriate when it comes to temperature determination.

\subsection{Consistency between methods}
\label{method_comparison}

\begin{table*}
\caption[]{Statistical analysis between our methods for Runs A, B, C, and C2. Mean difference, $\overline{\Delta}$, standard deviation of the mean difference (std. dev.), and Pearson correlation coefficients, $r_P$, for $T_{\rm eff}$, $\log{g}$, and [Fe/H] for all combinations of our methods.}
\label{tab:method-comparison}
\centering
   \begin{tabular}{lccccccccc}
        \hline 
        \hline 
        \noalign{\smallskip}
        Run A & \multicolumn{3}{c}{$T_{\rm eff}$}& \multicolumn{3}{c}{$\log{g}$} & \multicolumn{3}{c}{[Fe/H]} \\
         & $\overline{\Delta}$ & std. dev. & $r_P$ & $\overline{\Delta}$ & std. dev. & $r_P$ & $\overline{\Delta}$ & std. dev. & $r_P$ \\
          
        \hline
        \noalign{\smallskip}
        {\tt Pass-19} -- {\tt SteParSyn} & +43 & 96 & 0.974 & $-0.07$ & 0.24 & $-0.074$ & +0.06 & 0.12 & 0.804\\
        {\tt Pass-19} -- DL & $-24$ & 120 & 0.937 & +0.06 & 0.06 & 0.930 & $-0.25$ & 0.16 & 0.468 \\ 
        {\tt Pass-19} -- {\tt ODUSSEAS} & +136 & 178 & 0.829 & \ldots & \ldots & \ldots & +0.12 & 0.11 & 0.821 \\
        
        {\tt SteParSyn} -- DL &  $-67$ & 137 & 0.947 & +0.14 & 0.26 & $-0.189$ & $-0.30$ & 0.21 & 0.195 \\
        {\tt SteParSyn} -- {\tt ODUSSEAS} & +93 & 132 & 0.882 & \ldots & \ldots & \ldots & +0.07 & 0.16 & 0.639 \\
        
        DL -- {\tt ODUSSEAS} & +161 & 176 & 0.860 & \ldots & \ldots & \ldots & +0.37 & 0.15 & 0.530\\
        \\
        \hline\noalign{\smallskip}
        
        Run B \\
        \\
        \hline\noalign{\smallskip}
        {\tt Pass-19} -- {\tt SteParSyn} & \ldots & \ldots & \ldots & \ldots & \ldots & \ldots & +0.03 & 0.11 & 0.784  \\
        {\tt Pass-19} -- DL & \ldots & \ldots & \ldots & \ldots & \ldots & \ldots & $-0.33$ & 0.11 & 0.654 \\ 
        
        {\tt SteParSyn} -- DL & \ldots & \ldots & \ldots & \ldots & \ldots & \ldots & $-0.36$ & 0.17 & 0.418 \\
        \\
        \hline\noalign{\smallskip}
        
        Run C \\
        \\
        \hline\noalign{\smallskip}
        {\tt Pass-19} -- {\tt SteParSyn} & +15 & 123 & 0.900 & $-0.06$ & 0.22 & 0.177 & +0.38 & 0.38 & 0.346 \\
        {\tt Pass-19} -- DL & $-148$ & 86 & 0.954 & $-0.06$ & 0.13 & 0.339 & +0.40 & 0.30 & 0.359 \\ 
        {\tt Pass-19} -- {\tt ODUSSEAS}$^{*}$ & +74 & 152 & 0.844 & \ldots & \ldots & \ldots & +0.57 & 0.45 & $-0.222$ \\
        
        {\tt SteParSyn} -- DL &  $-163$ & 101 & 0.931 & +0.00 & 0.22 & 0.285 & +0.02 & 0.29 & 0.451 \\
        {\tt SteParSyn} -- {\tt ODUSSEAS}$^{*}$ & +59 & 180 & 0.777 & \ldots & \ldots & \ldots & +0.19 & 0.36 & 0.235\\
        
        DL -- {\tt ODUSSEAS}$^{*}$ & +222 & 114 & 0.901 & \ldots & \ldots & \ldots & +0.17 & 0.17 & 0.757 \\
        \\
        \hline\noalign{\smallskip}
        
        Run C2 \\
        \\
        \hline\noalign{\smallskip}
        {\tt Pass-19} -- {\tt SteParSyn} & +100 & 172 & 0.782 & +0.05 & 0.24 & 0.187 & $-0.01$ & 0.31 & 0.555 \\
        {\tt Pass-19} -- DL & $-39$ & 87 & 0.941 & $-0.03$ & 0.13 & 0.547 & $-0.10$ & 0.33 & $-0.162$ \\ 
        {\tt Pass-19} -- {\tt ODUSSEAS}$^{*}$ & +114 & 194 & 0.715 & \ldots & \ldots & \ldots & +0.09 & 0.46 & $-0.495$ \\
        
        {\tt SteParSyn} -- DL & $-140$ & 135 & 0.869 & $-0.08$ & 0.25 & 0.227 & $-0.09$ & 0.33 & 0.392 \\
        {\tt SteParSyn} -- {\tt ODUSSEAS}$^{*}$ & +14 & 181 & 0.772 & \ldots & \ldots & \ldots & +0.10 & 0.42 & 0.087  \\
        
        DL -- {\tt ODUSSEAS}$^{*}$ & +153 & 144 & 0.848 & \ldots & \ldots & \ldots & +0.19 & 0.18 & 0.704 \\
        \hline
        \noalign{\smallskip}
        
    \end{tabular}
\tablefoot{$^{(*)}$ Corresponding to Run C*. 
The results of Run C* from {\tt ODUSSEAS} technically cannot be compared to the other methods, but are shown here for completeness. For more details, see Sect.\,\ref{method_comparison}.}
 \end{table*}

As a last step, we analyzed the consistency between the methods we employed in this experiment. A statistical analysis similar to Table\,\ref{tab:pearson} is presented in Table\,\ref{tab:method-comparison}. We compare each method to each of the other methods to reveal trends. We plotted all these combination for further visualization in Figs.\,\ref{fig:Teff_all_A}--\ref{fig:metal_all_A} for $T_{\rm eff}$, $\log{g}$, and [Fe/H] for Run A; [Fe/H] for Run B is presented in Fig.\,\ref{fig:metal_all_B}, $T_{\rm eff}$, $\log{g}$, and [Fe/H] for Run C in Figs.\,\ref{fig:Teff_all_C}--\ref{fig:metal_all_C}, and $T_{\rm eff}$, $\log{g}$, and [Fe/H] for Run C2 in Figs.\,\ref{fig:Teff_all_C2}--\ref{fig:metal_all_C2}. The corresponding Bland-Altman plots similar to the literature comparison are shown in the right columns of Figs.\,\ref{fig:blandA}--\ref{fig:blandC2}.

\subsection*{Run A}
For $T_{\rm eff}$, {\tt SteParSyn} and the {\tt Pass19-code} show the best correlation with $r_P$ of 0.974, also being the only methods with a spread, namely, a standard deviation of less than 100\,K between them. Both methods compare well with DL, although DL shows higher deviations at higher $T_{\rm eff}$. As illustrated in previous comparisons (see Figs.\,\ref{fig:A-comparison} and \ref{fig:compare_interferometry}), {\tt ODUSSEAS} derives much lower $T_{\rm eff}$ values, on average, 130\,K cooler compared to the other methods. 
{\tt Pass19-code} and DL correlate quite well in $\log{g}$, whereas {\tt SteParSyn} exhibits a large spread compared to both other methods (see also Figs.\,\ref{fig:A-comparison} and \ref{fig:compare_interferometry}). 

{\tt SteParSyn}, the {\tt Pass19-code}, and {\tt ODUSSEAS} are in good agreement in [Fe/H], having small mean differences and a large $r_P$. The direct comparison between {\tt SteParSyn} and {\tt ODUSSEAS} displays a slightly larger spread and, therefore, a smaller $r_P$. 
For DL, it derives much higher [Fe/H] values, which are, on average, 0.3\,dex more metal-rich compared to the other methods. This behaviour can also be seen in Fig.\,\ref{fig:A-comparison}.

\subsection*{Run B}
As in Run A, {\tt SteParSyn} and the {\tt Pass19-code} are most consistent with a small mean difference of only 0.03\,dex, and a similar standard deviation. However, the values are not so well correlated, exhibiting a slightly smaller $r_P$ of 0.784 (compared to $r_P = 0.804$ in Run A). 
Here, DL performs even worse than it does in Run A, with $\overline{\Delta {\rm [Fe/H]}}$ being --0.33\,dex and --0.36\,dex compared to the {\tt Pass19-code} and {\tt SteParSyn}, respectively.

\subsection*{Runs C and C2}
As described in Sect.\,\ref{methods:RunC} and \ref{interferometry}, {\tt ODUSSEAS} was only able to use wavelength ranges between 5300 and 6900 {\AA}. Therefore, a direct comparison of results from this Run C* with the results from Runs C and C2 from the other methods is not meaningful. However, we included {\tt ODUSSEAS} in our analysis here for completeness and to visualize relative changes between Runs C and C2 for the other methods. 

A comparison of Run C with C2 reveals only minor differences for $T_{\rm eff}$. It can be seen from the numbers in Table\,\ref{tab:method-comparison} and the plot in Fig.\,\ref{fig:Teff_all_C} that DL performs a bit better in Run C2, where it derives slightly lower $T_{\rm eff}$ and therefore exhibits a smaller $\overline{\Delta T_{\rm eff}}$ compared to the other methods. {\tt SteParSyn} and the {\tt Pass19-code} show a somewhat smaller $\overline{\Delta T_{\rm eff}}$ and spread in Run C. This is also clearly shown by the comparison of the {\tt Pass19-code} and {\tt ODUSSEAS} in Runs C and C2. Since for both runs the {\tt Pass19-code} is compared to Run C* of {\tt ODUSSEAS}, relative improvements between the runs are revealed. Overall, it can be said that the correlation coefficients for $T_{\rm eff}$ are slightly greater in Run C compared to Run C2. 

On the other hand, there is almost no correlation in $\log{g}$ for any of the methods. The only notable improvement toward Run C2 is given between the {\tt Pass19-code} and DL, which present a little higher correlation and smaller $\overline{\Delta \log{g}}$ in Run C2. This can be attributed to an improvement of DL in Run C2, as already described in Sect.\,\ref{results} and \ref{interferometry}.

A clear difference can be seen for [Fe/H] between Runs C and C2 (see Figs.\,\ref{fig:metal_all_C} and \ref{fig:metal_all_C2}). In Run C, all methods appear more separated, also having higher mean differences. They determine, in general, higher [Fe/H] values, especially the {\tt Pass19-code}, which is depicted in Fig.\,\ref{fig:C-comparison2} as well. For Run C2, the derived values are more metal-poor, which causes the results to move closer together for all methods. This reduces the mean differences, although the spread and correlation coefficient are not necessarily improved  for all methods. 


Overall, the closest correlation between the methods for all stellar parameters is found in Run A, however we can see some trends. The determination of $\log{g}$ with {\tt SteParSyn} is generally not very well constrained: it has a large mean difference and spread compared to the other methods. The correlation increases toward Run C and C2, however the reason for this is not clear. In particular, {\tt ODUSSEAS} shows the best consistency in $T_{\rm eff}$ in Run C*, when compared to Run C using the other methods, with the smallest mean difference and a slightly better $r_P$ than in Run A. 
For [Fe/H], Run A as well as Run C2 show small $\overline{\Delta{\rm [Fe/H]}}$ in general; however, in Run C2 the spread is larger and $r_P$ is smaller, therefore, the consistency in [Fe/H] is better in Run A. Only DL is able to improve the consistency toward Runs C and C2, with negligible differences between C and C2. Consequently, $r_P$ increases and $\overline{\Delta{\rm [Fe/H]}}$ decreases. 

Possible improvements to increase the consistency are very specific to the method and there is no general recipe. For DL, the values in [Fe/H] are tentatively too high, using more wavelength ranges can improve [Fe/H] in Runs C and C2. However, for the determination of $T_{\rm eff}$, one wavelength range serves well. We see that {\tt ODUSSEAS} derives consistently lower $T_{\rm eff}$, and the use of different wavelength ranges, as done in Run C*, would increase the level of consistency with our other methods.
The analysis in this section confirms our findings in Sects.\,\ref{results} and \ref{interferometry}.

\section{Summary and conclusions}
\label{summary}

In this study, we applied four different methods, including synthetic spectral fitting, pEW measurements, and machine learning, to derive the stellar parameters $T_{\rm eff}$, $\log{g}$, and [Fe/H] for a sample of 18 M dwarfs from high-resolution and high-S/N CARMENES spectra. Our analysis consisted of four different runs: Run A allowed each team to use their method without restrictions, Run B fixed $T_{\rm eff}$ and $\log{g}$ to derive only [Fe/H], and for Runs C and C2, all the teams incorporated the same synthetic model spectra, continuum normalization method, and wavelength ranges. 

Although we provided several new stellar parameters for our sample, it was not our goal to measure more precise or accurate parameters for these stars in the context of a catalog, but to identify and understand discrepancies in the parameters between our groups, namely, the different parameter determination methods, with the aim to minimize these discrepancies in order to make a step forward to more consistent parameter determinations. 
At the beginning of this experiment, we expected that a standardization of underlying synthetic models and wavelength ranges, along with a reduction of the dimension of the parameter space by fixing stellar parameters would account for the inconsistencies between our results and literature medians. Therefore, we assumed to find the best agreement between our methods and with comparisons to the literature in Runs C and C2. However, we found that this is not necessarily the case as it is for FGK stars \citep[e.g.,][]{Jofre2014,Jofre2017}, and that our methods generally show the greatest consistency with the literature values when used in their original setting without any standardization. In general, the mean differences to the literature median are below 100\,K in $T_{\rm eff}$ for all methods, and also below 0.1\,dex in [Fe/H] for the {\tt Pass19-code} and {\tt SteParSyn}. In Runs C and C2, these differences increase significantly, up to over 200\,K and 0.3\,dex for some of our methods. 

This consistency in Run A is an indication that each team successfully calibrated their methods and optimized them to the use of certain wavelength ranges and synthetic spectra. It also implies that there might be other components responsible for the remaining differences that we see in the stellar parameters, which requires a more thorough and in-depth investigation of the methods themselves and the underlying concepts. One example is stellar atmosphere models and their corresponding spectra, where, despite constant improvements, they still suffer from some deficiencies. Various sets of synthetic spectra also show discrepancies when comparing the same spectral lines, due to the use of different equations of state, line lists, and other hyper-parameters. It cannot be excluded that these deviations contribute to the disagreements in derived stellar parameters found in this work. 
However, we were able to shed some light on deficiencies of some methods, for instance, showing that the DL method would benefit from the use of multiple wavelength ranges and that {\tt ODUSSEAS} could improve the $T_{\rm eff}$ determination by using different sets of lines or, most importantly, by using a new reference $T_{\rm eff}$ scale based on interferometry \citep[][]{Khata2021}. This possibility is currently being explored and is expected to be implemented as an option in an upgraded version of the tool.

%

\begin{acknowledgements}
We thank an anonymous referee for helpful comments that improved the quality of this paper.
CARMENES is an instrument for the Centro Astron\'omico Hispano-Alem\'an de Calar Alto (CAHA, Almer\'ia, Spain). 
CARMENES is funded by the German Max-Planck-Gesellschaft (MPG), the Spanish Consejo Superior de Investigaciones Cient\'ificas (CSIC), European Regional Development Fund (ERDF) through projects FICTS-2011-02, ICTS-2017-07-CAHA-4, and CAHA16-CE-3978, 
and the members of the CARMENES Consortium (Max-Planck-Institut f\"ur Astronomie, Instituto de Astrof\'isica de Andaluc\'ia, Landessternwarte K\"onigstuhl, Institut de Ci\`encies de l'Espai, Insitut f\"ur Astrophysik G\"ottingen, Universidad Complutense de Madrid, Th\"uringer Landessternwarte Tautenburg, Instituto de Astrof\'isica de Canarias, Hamburger Sternwarte, Centro de Astrobiolog\'ia and Centro Astron\'omico Hispano-Alem\'an), with additional contributions by the Spanish Ministry of Economy, the German Science Foundation through the Major Research Instrumentation Programme and DFG Research Unit FOR2544 ``Blue Planets around Red Stars'', the Klaus Tschira Stiftung, the states of Baden-W\"urttemberg and Niedersachsen, and by the Junta de Andaluc\'ia.
E.D.M. and A.A.K. acknowledge the support by the Investigador FCT contract IF/00849/2015/CP1273/CT0003. We acknowledge financial support from NASA through grant NNX17AG24G, 
the Agencia Estatal de Investigaci\'on of the Ministerio de Ciencia through fellowship FPU15/01476, Innovaci\'on y Universidades and the ERDF through projects
PID2019-109522GB-C51/2/3/4, 
AYA2016-79425-C3-1/2/3-P        
and AYA2018-84089,              
the Funda\c{c}\~{a}o para a Ci\^{e}ncia e a Tecnologia through and ERDF through grants UID/FIS/04434/2019, UIDB/04434/2020 and UIDP/04434/2020, PTDC/FIS-AST/28953/2017, and PTDC/FIS-AST/32113/2017, and COMPETE2020 - Programa Operacional Competitividade e Internacionaliza\c{c}\~{a}o POCI-01-0145-FEDER-028953, and POCI-01-0145-FEDER-032113. This research has been funded by the Spanish State Research Agency (AEI) Project No. MDM-2017-0737 Unidad de Excelencia "María de Maeztu"- Centro de Astrobiología (CSIC/INTA).
\end{acknowledgements}

\bibliographystyle{aa}  
\bibliography{HandH.bib}

\newpage
\appendix
\section{Literature summary of sample stars}

\onecolumn
\begin{longtable}{llccccc}
\label{tab:literature}\\
\caption{Collection of stellar parameters from the literature for the selected sample of benchmark stars.}\\
   \hline
   \hline
   \noalign{\smallskip}
  Karmn  & Author & $T_{\rm eff}$ [K] & $\log{g}$ [dex] & [Fe/H] [dex] & $R$ [R$_{\odot}$] & $M$ [M$_{\odot}$] \\ 
\noalign{\smallskip}
    \hline
    \noalign{\smallskip}                
 \endfirsthead
\caption{Collection of stellar parameters from the literature (cont.)}\\ 
  \hline
  \hline
  \noalign{\smallskip}
Karmn  & Author & $T_{\rm eff}$ [K] & $\log{g}$ [dex] & [Fe/H] [dex] & $R$ [R$_{\odot}$] & $M$ [M$_{\odot}$] \\ 
  \noalign{\smallskip}
  \hline
  \noalign{\smallskip}
  \endhead
  \noalign{\smallskip}
  \hline
  \endfoot
 
J00067$-$075 & \cite{Neves2014}         & $2718\pm150$    & \ldots            & $-0.27\pm0.20$ 
                                        & \ldots          & \ldots \\
             & \cite{Terrien2015}       & \ldots          & \ldots        & $-0.11\pm0.10$ 
                                        & \ldots          & \ldots \\
             & \cite{Houdebine2019}     & $2970\pm149$          & \ldots        & \ldots 
                                        & $0.127\pm0.011$ & \ldots \\
             \cline{2-7}\noalign{\smallskip}
             & Literature median          & $2844\pm149$    &   \ldots        & $-0.19\pm0.16$ 
                                        & \ldots          & \ldots\\
             & Literature \& Run A median & $2960\pm103$     & $5.10\pm0.11$ & $-0.26\pm0.18$ 
                                        & \ldots          & \ldots \\

\hline\noalign{\smallskip}

J00183$+$440 & \cite{Berger2006}  & 3747 $\pm$ 112   & 4.89 $\pm$ 0.07   & \ldots   & 0.379 $\pm$ 0.006 & 0.404 $\pm$ 0.040 \\
             & \cite{Boyajian2012}  & 3563 $\pm$ 11 & 4.89 $\pm$ 0.04$^{c}$ & \ldots   & 0.387 $\pm$ 0.002 & 0.423 $\pm$ 0.042   \\
             & \cite{Gaidos2014}  & 3669 $\pm$ 67 & 4.79 $\pm$ 0.02$^{c}$ & $-$0.29 $\pm$ 0.11 & 0.470 $\pm$ 0.040 & 0.500 $\pm$ 0.060 \\
             & \cite{GaidosMann2014}  & 3693 $\pm$ 91 & 4.77 $\pm$ 0.03$^{c}$ & $-$0.26 $\pm$ 0.08 & 0.490 $\pm$ 0.050 & 0.520 $\pm$ 0.070 \\
             & \cite{Houdebine2019}  & 3656 $\pm$ 183   & \ldots   & \ldots   & 0.365 $\pm$ 0.014 & \ldots   \\
             & \cite{Khata2020}  & 3493 $\pm$ 103 & 4.82 $\pm$ 0.04$^{c}$ & $-$0.19 $\pm$ 0.08 & 0.385 $\pm$ 0.027 & 0.357 $\pm$ 0.017 \\
             & \cite{Mann2015}  & 3603 $\pm$ 60 & 4.86 $\pm$ 0.01$^{c}$ & $-$0.30 $\pm$ 0.08 & 0.388 $\pm$ 0.013 & 0.398 $\pm$ 0.040 \\
             & \cite{Newton2015}  & 3534 $\pm$ 79 & \ldots   & \ldots   & 0.388 $\pm$ 0.028 & \ldots   \\
             &  & 3602 $\pm$ 13$^{int}$ & \ldots   & \ldots   & 0.386 $\pm$ 0.002$^{int}$ & \ldots   \\
             & \cite{Segransan2003}  & 3698 $\pm$ 95 & 4.89 $\pm$ 0.02$^{c}$ & \ldots   & 0.383 $\pm$ 0.020 & 0.414 $\pm$ 0.021 \\
             & \cite{Terrien2015}  & \ldots   & \ldots   & $-$0.26 $\pm$ 0.10 & 0.395 $\pm$ 0.004 & \ldots  \\
             \cline{2-7}\noalign{\smallskip}
             & Literature median  & 3630 $\pm$ 94 & 4.86 $\pm$ 0.04 & $-$0.26 $\pm$ 0.09 &  \ldots  & \ldots\\
             & Literature \& Run A median  & 3603 $\pm$ 84 & 4.85 $\pm$ 0.04 & $-$0.26 $\pm$ 0.11 &  \ldots  &  \ldots  \\

\hline\noalign{\smallskip}

J04429$+$189 & \cite{Gaidos2014} & 3680 $\pm$ 99 & 4.78 $\pm$ 0.03$^{c}$ & +0.04 $\pm$ 0.11 & 0.480 $\pm$ 0.050 & 0.510 $\pm$ 0.070 \\
             & \cite{GaidosMann2014} & 3721 $\pm$ 82 & 4.76 $\pm$ 0.03$^{c}$ & +0.14 $\pm$ 0.08 & 0.500 $\pm$ 0.050 & 0.530 $\pm$ 0.070 \\
             & \cite{Houdebine2019} & 3542 $\pm$ 177 & \ldots    & \ldots   & 0.453 $\pm$ 0.027 & \ldots   \\
             & \cite{Khata2020} & 3377 $\pm$ 110 & 4.87 $\pm$ 0.06$^{c}$ & $-$0.01 $\pm$ 0.09 & 0.338 $\pm$ 0.032 & 0.312 $\pm$ 0.015 \\
             & \cite{Lepine2013} & 3550 $\pm$ 57 & 4.50    & \ldots   & \ldots   & \ldots   \\
             & \cite{Maldonado2015} & 3603 $\pm$ 68 & 4.75 $\pm$ 0.04  & +0.03 $\pm$ 0.09 & 0.510 $\pm$ 0.047 & 0.520 $\pm$ 0.052 \\
             & \cite{Mann2015} & 3680 $\pm$ 60 & 4.82 $\pm$ 0.01$^{c}$ & +0.14 $\pm$ 0.08 & 0.452 $\pm$ 0.019 & 0.492 $\pm$ 0.049 \\
             & \cite{Neves2014} & 3355 $\pm$ 110 & \ldots    & $-$0.01 $\pm$ 0.09 & \ldots   & 0.500 $\pm$ 0.030 \\
             & \cite{Newton2015} & 3574 $\pm$ 78 & \ldots    & \ldots   & 0.514 $\pm$ 0.029 & \ldots   \\
             &  & 3701 $\pm$ 90$^{int}$ & \ldots    & \ldots   & 0.453 $\pm$ 0.022$^{int}$ & \ldots   \\
             & \cite{RojasAyala2012} & 3581 $\pm$ 20 & \ldots    & +0.15 $\pm$ 0.17 & \ldots   & \ldots   \\
             & \cite{Terrien2015} & \ldots   & \ldots    & +0.12 $\pm$ 0.10  & 0.478 $\pm$ 0.010 & \ldots   \\
             & \cite{vonBraun2014} & 3679 $\pm$ 77 & 4.78 $\pm$ 0.09$^{c}$ & \ldots   & 0.453 $\pm$ 0.022 & 0.450 $\pm$ 0.135   \\
             \cline{2-7}\noalign{\smallskip}
             & Literature median & 3592 $\pm$ 93 & 4.78 $\pm$ 0.05  & +0.08 $\pm$ 0.10 & \ldots   & \ldots   \\
             & Literature \& Run A median & 3592 $\pm$ 85 & 4.76 $\pm$ 0.05  & +0.08 $\pm$ 0.11 & \ldots   & \ldots   \\

\hline\noalign{\smallskip}

J05314$-$036 & \cite{Boyajian2012} & 3801 $\pm$ 9 & 4.71 $\pm$ 0.04$^{c}$ & \ldots   & 0.574 $\pm$ 0.004 & 0.615 $\pm$ 0.062 \\
             & \cite{Gaidos2014} & 3701 $\pm$ 61 & 4.77 $\pm$ 0.02$^{c}$ & \ldots   & 0.490 $\pm$ 0.040 & 0.520 $\pm$ 0.060 \\
             & \cite{GaidosMann2014} & 3895 $\pm$ 84 & 4.72 $\pm$ 0.01$^{c}$ & +0.43 $\pm$ 0.08 & 0.560 $\pm$ 0.040 & 0.600 $\pm$ 0.070 \\
             & \cite{Houdebine2019} & 3696 $\pm$ 185 & \ldots    & \ldots   & 0.588 $\pm$ 0.019 & \ldots   \\
             & \cite{Khata2020} & 3849 $\pm$ 275 & 4.69 $\pm$ 0.08$^{c}$ & \ldots   & 0.553 $\pm$ 0.149 & 0.554 $\pm$ 0.193 \\
             & \cite{Maldonado2015} & 3800 $\pm$ 68 & 4.68 $\pm$ 0.05  & +0.00 $\pm$ 0.09 & 0.580 $\pm$ 0.052 & 0.600 $\pm$ 0.056 \\
             & \cite{Mann2015} & 3801 $\pm$ 60 & 4.71 $\pm$ 0.01$^{c}$ & +0.49 $\pm$ 0.08 & 0.581 $\pm$ 0.019 & 0.633 $\pm$ 0.063 \\
             & \cite{Neves2014} & 3670 $\pm$ 110 & \ldots    & +0.19 $\pm$ 0.09 & \ldots   & 0.600 $\pm$ 0.070 \\
             & \cite{Newton2015} & 3872 $\pm$ 75 & \ldots    & \ldots   & 0.597 $\pm$ 0.027 & \ldots   \\
             &  & 3850 $\pm$ 22$^{int}$ & \ldots    & \ldots   & 0.574 $\pm$ 0.004$^{int}$ & \ldots   \\
             & \cite{RojasAyala2012} & 4012 $\pm$ 106 & \ldots    & +0.35 $\pm$ 0.17 & \ldots   & \ldots   \\
             & \cite{Segransan2003} & 3520 $\pm$ 170 & 4.54 $\pm$ 0.06$^{c}$ & \ldots   & 0.702 $\pm$ 0.063 & 0.631 $\pm$ 0.031 \\
             & \cite{Terrien2015} & \ldots   & \ldots    & +0.69 $\pm$ 0.10 & 0.587 $\pm$ 0.040 & \ldots   \\
             \cline{2-7}\noalign{\smallskip}
             & Literature median & 3801 $\pm$ 125 & 4.71 $\pm$ 0.05  & +0.39 $\pm$ 0.11 &  \ldots  &  \ldots  \\
             & Literature \& Run A median & 3801 $\pm$ 112 & 4.70 $\pm$ 0.05  & +0.30 $\pm$ 0.11 &  \ldots  &  \ldots  \\

\hline\noalign{\smallskip}

J07558$+$833 & \cite{Dittmann2016} & \ldots   & \ldots   & +0.00 $\pm$ 0.10 & \ldots & \ldots \\
             & \cite{Gaidos2014} & 3183 $\pm$ 60 & \ldots   & \ldots   & <0.19 & <0.14 \\
             & \cite{Lepine2013} & 3250 $\pm$ 76 & 5.00   & \ldots   & \ldots & \ldots \\
             \cline{2-7}\noalign{\smallskip}
             & Literature median & 3217 $\pm$ 68 & 5.00   & +0.00 $\pm$ 0.10 & \ldots & \ldots \\
             & Literature \& Run A median & 3220 $\pm$ 87 & 5.01 $\pm$ 0.07 &  $-$0.20 $\pm$ 0.18 & \ldots & \ldots \\

\hline\noalign{\smallskip}

J09143$+$526 & \cite{Boyajian2012} & 3907 $\pm$ 35 & 4.71 $\pm$ 0.02$^{c}$ & \ldots   & 0.577 $\pm$ 0.013 & 0.622 $\pm$ 0.062\\
             & \cite{Gaidos2014} & 3991 $\pm$ 66 & 4.69 $\pm$ 0.01$^{c}$ & $-$0.26 $\pm$ 0.11 & 0.590 $\pm$ 0.040 & 0.630 $\pm$ 0.070 \\
             & \cite{Houdebine2019} & 3921 $\pm$ 196 & \ldots    & \ldots   & 0.602 $\pm$ 0.020 & \ldots   \\
             & \cite{Khata2020} & 4002 $\pm$ 125 & 4.48 $\pm$ 0.05$^{c}$ & $-$0.08 $\pm$ 0.09 & 0.617 $\pm$ 0.051 & 0.424 $\pm$ 0.024 \\
             & \cite{Mann2015} & 3920 $\pm$ 60 & 4.74 $\pm$ 0.00$^{c}$ & $-$0.01 $\pm$ 0.08 & 0.550 $\pm$ 0.026 & 0.607 $\pm$ 0.061 \\
             & \cite{Newton2015} & 3955 $\pm$ 106 & \ldots    & \ldots   & 0.571 $\pm$ 0.029 & \ldots   \\
             &  & 3953 $\pm$ 41$^{int}$ & \ldots    & \ldots   & 0.577 $\pm$ 0.013$^{int}$ & \ldots   \\
             & \cite{RojasAyala2012} & 4031 $\pm$ 56 & \ldots    & $-0.18$ $\pm$ 0.17 & \ldots   & \ldots   \\
             \cline{2-7}\noalign{\smallskip}
             & Literature median & 3954 $\pm$ 100 & 4.70 $\pm$ 0.03  & $-$0.13 $\pm$ 0.12 & \ldots   & \ldots   \\
             & Literature \& Run A median & 3954 $\pm$ 88 & 4.69 $\pm$ 0.04  & $-$0.05 $\pm$ 0.12 & \ldots   & \ldots \\

\hline\noalign{\smallskip}

J09144$+$526 & \cite{Boyajian2012} & 3867 $\pm$ 37 & 4.71 $\pm$ 0.02 $^{c}$ & \ldots   & 0.567 $\pm$ 0.014 & 0.600 $\pm$ 0.060 \\
             & \cite{Gaidos2014} & 3770 $\pm$ 87 & 4.74 $\pm$ 0.03$^{c}$ & \ldots   & 0.520 $\pm$ 0.050 & 0.550 $\pm$ 0.07 \\
             & \cite{Houdebine2019} & 3921 $\pm$ 196 & \ldots    & \ldots   & 0.600 $\pm$ 0.040 & \ldots   \\
             & \cite{Khata2020} & 3844 $\pm$ 127 & \ldots    & $-$0.07 $\pm$ 0.09 & 0.582 $\pm$ 0.047 & \ldots   \\
             & \cite{Newton2015} & 3892 $\pm$ 92 & \ldots    & \ldots   & 0.562 $\pm$ 0.028 & \ldots   \\
             & & 3926 $\pm$ 37$^{int}$ & \ldots    & \ldots   & 0.567 $\pm$ 0.014$^{int}$ & \ldots   \\
             & \cite{RojasAyala2012} & 3869 $\pm$ 15 & \ldots    & $-0.15$ $\pm$ 0.17 & \ldots   & \ldots   \\
             \cline{2-7}\noalign{\smallskip}
             & Literature median & 3869 $\pm$ 103 & 4.72 $\pm$ 0.03  & $-$0.11 $\pm$ 0.14 & \ldots   & \ldots   \\
             & Literature \& Run A median & 3891 $\pm$ 89 & 4.71 $\pm$ 0.04  & $-$0.03 $\pm$ 0.13 & \ldots   & \ldots \\
             
\hline\noalign{\smallskip}

J10508$+$068 & \cite{Gaidos2014} & 3238 $\pm$ 81 & 5.02 $\pm$ 0.06$^{c}$ & +0.13 $\pm$ 0.11 & 0.190 $\pm$ 0.080 & 0.140 $\pm$ 0.100 \\
             & \cite{GaidosMann2014} & 3400 $\pm$ 63 & 4.92 $\pm$ 0.05$^{c}$ & +0.20 $\pm$ 0.08 & 0.320 $\pm$ 0.050 & 0.310 $\pm$ 0.060 \\
             & \cite{Houdebine2019} & 3099 $\pm$ 155 & \ldots    & \ldots   & 0.334 $\pm$ 0.031 & \ldots   \\
             & \cite{Khata2020} & 2388 $\pm$ 113 & \ldots    & +0.18 $\pm$ 0.10 & \ldots   & 0.155 $\pm$ 0.007 \\
             & \cite{Lepine2013} & 3100 $\pm$ 76 & 4.50    & \ldots   & \ldots   & \ldots   \\
             & \cite{Mann2015} & 3238 $\pm$ 60 & 4.94 $\pm$ 0.01$^{c}$ & +0.16 $\pm$ 0.08 & 0.276 $\pm$ 0.012 & 0.246 $\pm$ 0.025 \\
             & \cite{Neves2014} & 2943 $\pm$ 110 & \ldots    & +0.03 $\pm$ 0.09 & \ldots   & \ldots   \\
             & \cite{RojasAyala2012} & 3334 $\pm$ 23 & \ldots    & +0.20 $\pm$ 0.17 & \ldots   & \ldots   \\
             & \cite{Terrien2015} & \ldots   & \ldots    & +0.11 $\pm$ 0.10 & \ldots   & \ldots   \\
             \cline{2-7}\noalign{\smallskip}
             & Literature median & 3169 $\pm$ 93 & 4.93 $\pm$ 0.04  & +0.16 $\pm$ 0.11 & \ldots   & \ldots   \\
             & Literature \& Run A median & 3235 $\pm$ 82 & 4.92 $\pm$ 0.06  & +0.13 $\pm$ 0.12 & \ldots   & \ldots   \\

\hline\noalign{\smallskip}

J11033$+$359 & \cite{Boyajian2012} & 3465 $\pm$ 17 & 4.85 $\pm$ 0.04$^{c}$ & \ldots   & 0.392 $\pm$ 0.004 & 0.403 $\pm$ 0.040  \\
             & \cite{Gaidos2014} & 3593 $\pm$ 66 & 4.81 $\pm$ 0.02$^{c}$ & \ldots   & 0.440 $\pm$ 0.040 & 0.460 $\pm$ 0.060 \\
             & \cite{GaidosMann2014} & 3679 $\pm$ 110 & 4.78 $\pm$ 0.04$^{c}$ & $-$0.30 $\pm$ 0.08 & 0.480 $\pm$ 0.060 & 0.510 $\pm$ 0.080 \\
             & \cite{Houdebine2019} & 3602 $\pm$ 180 & \ldots    & \ldots   & 0.362 $\pm$ 0.013 & \ldots   \\
             & \cite{Khata2020} & 3560 $\pm$ 104 & 4.74 $\pm$ 0.04$^{c}$ & $-$0.18 $\pm$ 0.13 & 0.419 $\pm$ 0.028 & 0.355 $\pm$ 0.016 \\
             & \cite{Lepine2013} & 3530 $\pm$ 39 & 4.50    & \ldots   & \ldots   & \ldots   \\
             & \cite{Mann2015} & 3563 $\pm$ 60 & 4.84 $\pm$ 0.01$^{c}$ & $-$0.38 $\pm$ 0.08 & 0.389 $\pm$ 0.013 & 0.386 $\pm$ 0.039 \\
             & \cite{Newton2015} & 3532 $\pm$ 85 & \ldots    & \ldots   & 0.401 $\pm$ 0.029 & \ldots   \\
             &  & 3532 $\pm$ 17$^{int}$ & \ldots    & \ldots   & 0.392 $\pm$ 0.003$^{int}$ & \ldots   \\
             & \cite{RojasAyala2012} & 3526 $\pm$ 18 & \ldots    & $-$0.41 $\pm$ 0.17 & \ldots   & \ldots   \\
             & \cite{Segransan2003} & 3570 $\pm$ 42 & 4.85 $\pm$ 0.00$^{c}$ & \ldots   & 0.393 $\pm$ 0.008 & 0.403 $\pm$ 0.020 \\
             & \cite{Terrien2015} & \ldots   & \ldots    & $-$0.41 $\pm$ 0.10 & 0.378 $\pm$ 0.042 & \ldots   \\
             \cline{2-7}\noalign{\smallskip}
             & Literature median & 3560 $\pm$ 82 & 4.81 $\pm$ 0.03  & $-$0.38 $\pm$ 0.12 & \ldots   & \ldots   \\
             & Literature \& Run A median & 3555 $\pm$ 76 & 4.81 $\pm$ 0.05  & $-$0.33 $\pm$ 0.12 & \ldots   & \ldots   \\

\hline\noalign{\smallskip}

J11054$+$435 & \cite{Boyajian2012} & 3497 $\pm$ 39 & 4.84 $\pm$ 0.02$^{c}$ & \ldots   & 0.398 $\pm$ 0.009 & 0.403 $\pm$ 0.040 \\
             & \cite{Gaidos2014} & 3702 $\pm$ 65 & 4.77 $\pm$ 0.02$^{c}$ & $-$0.41 $\pm$ 0.11 & 0.490 $\pm$ 0.040 & 0.520 $\pm$ 0.060 \\
             & \cite{GaidosMann2014} & 3743 $\pm$ 84 & 4.75 $\pm$ 0.03$^{c}$ & $-$0.32 $\pm$ 0.08 & 0.510 $\pm$ 0.050 & 0.540 $\pm$ 0.070 \\
             & \cite{Houdebine2019} & 3692 $\pm$ 185 & \ldots    & \ldots   & 0.370 $\pm$ 0.030 & \ldots   \\
             & \cite{Lepine2013} & 3560 $\pm$ 44 & 4.50    & \ldots   & \ldots   & \ldots   \\
             & \cite{Mann2015} & 3619 $\pm$ 60 & 4.86 $\pm$ 0.01$^{c}$ & $-$0.37 $\pm$ 0.08 & 0.383 $\pm$ 0.013 & 0.390 $\pm$ 0.039 \\
             & \cite{Newton2015} & 3664 $\pm$ 227 & \ldots    & \ldots   & 0.425 $\pm$ 0.041 & \ldots   \\
             &  & 3537 $\pm$ 41$^{int}$ & \ldots    & \ldots   & 0.398 $\pm$ 0.009$^{int}$ & \ldots   \\
             & \cite{RojasAyala2012} & 3684 $\pm$ 20 & \ldots    & $-$0.40 $\pm$ 0.17 & \ldots   & \ldots   \\
             & \cite{Terrien2015} & \ldots   & \ldots    & $-$0.38 $\pm$ 0.10 & 0.378 $\pm$ 0.004 & \ldots   \\
             \cline{2-7}\noalign{\smallskip}
             & Literature median & 3664 $\pm$ 109 & 4.77 $\pm$ 0.02  & $-$0.38 $\pm$ 0.11 & \ldots   & \ldots   \\
             & Literature \& Run A median & 3619 $\pm$ 95 & 4.81 $\pm$ 0.04  & $-$0.37 $\pm$ 0.12 & \ldots   & \ldots   \\

\hline\noalign{\smallskip}

J11421+267 & \cite{Gaidos2014} & 3479 $\pm$ 61 & 4.88 $\pm$ 0.05$^{c}$ & +0.07 $\pm$ 0.11 & 0.370 $\pm$ 0.050 & 0.380 $\pm$ 0.060 \\
             & \cite{GaidosMann2014} & 3606 $\pm$ 72 & 4.82 $\pm$ 0.04$^{c}$ & +0.00 $\pm$ 0.08 & 0.440 $\pm$ 0.050 & 0.470 $\pm$ 0.060 \\
             & \cite{Houdebine2019} & 3464 $\pm$ 173 & \ldots    & \ldots   & 0.403 $\pm$ 0.012 & \ldots   \\
             & \cite{Khata2020} & 3534 $\pm$ 106 & 4.86 $\pm$ 0.04$^{c}$ & $-$0.06 $\pm$ 0.08 & 0.418 $\pm$ 0.029 & 0.460 $\pm$ 0.021 \\
             & \cite{Lepine2013} & 3400 $\pm$ 62 & 5.00    & \ldots   & \ldots   & \ldots   \\
             & \cite{Mann2015} & 3479 $\pm$ 60 & 4.78 $\pm$ 0.01$^{c}$ & +0.01 $\pm$ 0.08 & 0.449 $\pm$ 0.019 & 0.445 $\pm$ 0.044 \\
             & \cite{Neves2014} & 3354 $\pm$ 110 & \ldots    & $-$0.03 $\pm$ 0.09 & \ldots   & \ldots   \\
             & \cite{Newton2015} & 3477 $\pm$ 81 & \ldots    & \ldots   & 0.400 $\pm$ 0.028 & \ldots   \\
             &  & 3520 $\pm$ 66$^{int}$  & \ldots    & \ldots   & 0.455 $\pm$ 0.018$^{int}$  & \ldots   \\
             & \cite{RojasAyala2012} & 3469 $\pm$ 17 & \ldots    & +0.04 $\pm$ 0.17 & \ldots   & \ldots   \\
             & \cite{Terrien2015} & \ldots   & \ldots    & $-$0.06 $\pm$ 0.10 & 0.431 $\pm$ 0.008 & \ldots   \\
             \cline{2-7}\noalign{\smallskip}
             & Literature median & 3478 $\pm$ 90 & 4.86 $\pm$ 0.04  & $-$0.00 $\pm$ 0.11 & \ldots   & \ldots   \\
             & Literature \& Run A median & 3478 $\pm$ 81 & 4.83 $\pm$ 0.05  & $-$0.03 $\pm$ 0.11 & \ldots   & \ldots   \\

\hline\noalign{\smallskip}

J13005$+$056 & \cite{Dittmann2016} & \ldots   & \ldots   & +0.07 $\pm$ 0.10 & \ldots   & \ldots \\
             & \cite{Gaidos2014} & 3090 $\pm$ 77 & \ldots   & +0.28 $\pm$ 0.11 & <0.19   & <0.14 \\
             & \cite{Houdebine2019} & 3140 $\pm$ 157 & \ldots   & \ldots   & 0.191 $\pm$ 0.013 & \ldots \\
             & \cite{Lepine2013} & 2950 $\pm$ 61 & 4.50   & \ldots   & \ldots   & \ldots \\
             & \cite{Newton2015} & \ldots   & \ldots   & \ldots   & 0.170 $\pm$ 0.043 & \ldots \\
             & \cite{Terrien2015} & \ldots   & \ldots   & +0.09 $\pm$ 0.10 & \ldots   & \ldots \\
             \cline{2-7}\noalign{\smallskip}
             & Literature median & 3090 $\pm$ 107 & 4.50   & +0.09 $\pm$ 0.10 & \ldots   & \ldots \\
             & Literature \& Run A median & 3140 $\pm$ 100 & 4.81 $\pm$ 0.10 & --0.12 $\pm$ 0.17 & \ldots   & \ldots \\

\hline\noalign{\smallskip}

J13457$+$148 & \cite{Berger2006}        & $3662\pm110$ & $4.75\pm0.07$ & \ldots 
                                        & $0.493\pm0.033$ & $0.502\pm0.050$ \\
             & \cite{Boyajian2012}      & $3618\pm31$ & $4.78\pm0.03^{c}$ & \ldots 
                                        & $0.484\pm0.008$ & $0.520\pm0.052$ \\
             & \cite{Gaidos2014}        & $3703\pm73$ & $4.77\pm0.02^{c}$ & $-0.45\pm0.11$ 
                                        & $0.490\pm0.040$ & $0.520\pm0.060$ \\
             & \cite{GaidosMann2014}    & 3792 $\pm$ 92 & 4.75 $\pm$ 0.03$^{c}$ & $-$0.18 $\pm$ 0.09 
                                        & 0.520 $\pm$ 0.050 & 0.560 $\pm$ 0.070 \\
             & \cite{Houdebine2019}     & 3650 $\pm$ 183 & \ldots & \ldots 
                                        & 0.494 $\pm$ 0.033 & \ldots   \\
             & \cite{Khata2020}         & 3707 $\pm$ 103 & 4.78 $\pm$ 0.01$^{c}$ & $-$0.15 $\pm$ 0.10 
                                        & 0.502 $\pm$ 0.027 & 0.554 $\pm$ 0.044 \\
             & \cite{Maldonado2015}     & 3609 $\pm$ 68 & 4.79 $\pm$ 0.04 & $-$0.10 $\pm$ 0.09 
                                        & 0.470 $\pm$ 0.047 & 0.470 $\pm$ 0.052 \\
             & \cite{Mann2015}          & 3649 $\pm$ 60 & 4.74 $\pm$ 0.01$^{c}$ & $-$0.31 $\pm$ 0.08 
                                        & 0.478 $\pm$ 0.016 & 0.465 $\pm$ 0.046 \\
             & \cite{Neves2014}         & 3515 $\pm$ 110 & \ldots & $-$0.22 $\pm$ 0.09 
                                        & \ldots & 0.500 $\pm$ 0.030 \\
             & \cite{Newton2015}        & 3716 $\pm$ 125 & \ldots & \ldots 
                                        & 0.450 $\pm$ 0.033 & \ldots   \\
             &         & 3646 $\pm$ 34$^{int}$ & \ldots & \ldots 
                                        & 0.484 $\pm$ 0.008$^{int}$ & \ldots   \\
             & \cite{RojasAyala2012}    & 3642 $\pm$ 17 & \ldots & $-$0.30 $\pm$ 0.17 
                                        & \ldots & \ldots \\
             \cline{2-7}\noalign{\smallskip}
             & Literature median          & 3650 $\pm$ 95 & 4.77 $\pm$ 0.03  & $-$0.22 $\pm$ 0.11 
                                        & \ldots   & \ldots   \\
             & Literature \& Run A median & 3648 $\pm$ 88 & 4.75 $\pm$ 0.04  & $-$0.22 $\pm$ 0.11 
                                        & \ldots   & \ldots   \\

\hline\noalign{\smallskip}

J15194$-$077 & \cite{Gaidos2014} & 3413 $\pm$ 61 & 4.90 $\pm$ 0.05$^{c}$ & $-$0.21 $\pm$ 0.11 & 0.330 $\pm$ 0.050 & 0.320 $\pm$ 0.060 \\
             & \cite{GaidosMann2014} & 3357 $\pm$ 73 & 4.94 $\pm$ 0.05$^{c}$ & $-$0.10 $\pm$ 0.08 & 0.290 $\pm$ 0.060 & 0.270 $\pm$ 0.080 \\
             & \cite{Houdebine2019} & 3423 $\pm$ 171 & \ldots    & \ldots   & 0.285 $\pm$ 0.008 & \ldots   \\
             & \cite{Khata2020} & 3475 $\pm$ 119 & 4.71 $\pm$ 0.08$^{c}$ & $-$0.11 $\pm$ 0.12 & 0.364 $\pm$ 0.045 & 0.251 $\pm$ 0.015 \\
             & \cite{Maldonado2015} & 3419 $\pm$ 68 & 4.95 $\pm$ 0.08  & $-$0.20 $\pm$ 0.09 & 0.300 $\pm$ 0.078 & 0.290 $\pm$ 0.086 \\
             & \cite{Mann2015} & 3395 $\pm$ 60 & 4.92 $\pm$ 0.01$^{c}$ & $-$0.15 $\pm$ 0.08 & 0.311 $\pm$ 0.012 & 0.292 $\pm$ 0.029 \\
             & \cite{Neves2014} & 3248 $\pm$ 110 & \ldots    & $-$0.20 $\pm$ 0.09 & \ldots   & 0.300 $\pm$ 0.020 \\
             & \cite{Newton2015} & 3354 $\pm$ 74 & \ldots    & \ldots   & 0.329 $\pm$ 0.027 & \ldots   \\
             &  & 3487 $\pm$ 62$^{int}$ & \ldots    & \ldots   & 0.299 $\pm$ 0.010$^{int}$ & \ldots   \\
             & \cite{RojasAyala2012} & 3534 $\pm$ 18 & \ldots    & $-$0.10 $\pm$ 0.17 & \ldots   & \ldots   \\
             & \cite{Terrien2015} & \ldots   & \ldots    & $-$0.06 $\pm$ 0.10 & 0.322 $\pm$ 0.050 & \ldots   \\
             \cline{2-7}\noalign{\smallskip}
             & Literature median & 3416 $\pm$ 83 & 4.92 $\pm$ 0.14  & $-$0.13 $\pm$ 0.11 & \ldots   & \ldots   \\
             & Literature \& Run A median & 3404 $\pm$ 82 & 4.91 $\pm$ 0.06  & $-$0.13 $\pm$ 0.11 & \ldots   & \ldots   \\

\hline\noalign{\smallskip}

J16581$+$257 & \cite{Gaidos2014} & 3744 $\pm$ 65 & 4.75 $\pm$ 0.02$^{c}$ & $-$0.08 $\pm$ 0.11 & 0.510 $\pm$ 0.040 & 0.540 $\pm$ 0.060 \\
             & \cite{Houdebine2019} & 3705 $\pm$ 185 & \ldots    & \ldots   & 0.497 $\pm$ 0.020 & \ldots   \\
             & \cite{Khata2020} & 3654 $\pm$ 117 & 4.74 $\pm$ 0.05$^{c}$ & $-$0.03 $\pm$ 0.12 & 0.466 $\pm$ 0.036 & 0.438 $\pm$ 0.019 \\
             & \cite{Lepine2013} & 3590 $\pm$ 39 & 4.50    & \ldots   & \ldots   & \ldots   \\
             & \cite{Mann2015} & 3700 $\pm$ 60 & 4.75 $\pm$ 0.01$^{c}$ & +0.03 $\pm$ 0.08 & 0.507 $\pm$ 0.018 & 0.534 $\pm$ 0.053 \\
             & \cite{Newton2015} & 3683 $\pm$ 79 & \ldots    & \ldots   & 0.497 $\pm$ 0.028 & \ldots   \\
             & & 3604 $\pm$ 46$^{int}$ & \ldots    & \ldots   & 0.539 $\pm$ 0.016$^{int}$ & \ldots   \\
             & \cite{RojasAyala2012} & 3733 $\pm$ 20 & \ldots    & $-$0.04 $\pm$ 0.17 & \ldots   & \ldots   \\
             & \cite{Terrien2015} & \ldots   & \ldots    & $-$0.04 $\pm$ 0.10 & 0.505 $\pm$ 0.006 & \ldots   \\
             & \cite{vonBraun2014} & 3590 $\pm$ 45 & 4.70 $\pm$ 0.10$^{c}$ & \ldots   & 0.539 $\pm$ 0.016 & 0.540 $\pm$ 0.162 \\
             \cline{2-7}\noalign{\smallskip}
             & Literature median & 3683 $\pm$ 87 & 4.74 $\pm$ 0.06  & $-$0.04 $\pm$ 0.12 & \ldots   & \ldots   \\
             & Literature \& Run A median & 3683 $\pm$ 78 & 4.72 $\pm$ 0.06  & --0.03 $\pm$ 0.12 & \ldots   & \ldots   \\

\hline\noalign{\smallskip}

J17578$+$046 & \cite{Boyajian2012} & 3224 $\pm$ 10 & 5.06 $\pm$ 0.04$^{c}$ & \ldots   & 0.187 $\pm$ 0.001 & 0.146 $\pm$ 0.015 \\
             & \cite{Dittmann2016} & \ldots   & \ldots    & $-$0.44 $\pm$ 0.10 & \ldots   & \ldots   \\
             & \cite{Gaidos2014} & 3237 $\pm$ 60 & \ldots    & $-$0.32 $\pm$ 0.11 & <0.19   & <0.14   \\
             & \cite{GaidosMann2014} & 3247 $\pm$ 61 & 5.05 $\pm$ 0.04$^{c}$ & $-$0.32 $\pm$ 0.08 & 0.190 $\pm$ 0.060 & 0.150 $\pm$ 0.080 \\
             & \cite{Houdebine2019} & 3266 $\pm$ 163 & \ldots    & \ldots   & 0.186 $\pm$ 0.010 & \ldots   \\
             & \cite{Mann2015} & 3228 $\pm$ 60 & 5.09 $\pm$ 0.01$^{c}$ & $-$0.40 $\pm$ 0.08 & 0.186 $\pm$ 0.007 & 0.155 $\pm$ 0.015 \\
             & \cite{Neves2014} & 3338 $\pm$ 110 & \ldots    & $-$0.51 $\pm$ 0.09 & \ldots   & 0.160 $\pm$ 0.010 \\
             & \cite{Newton2015} & 3248 $\pm$ 81 & \ldots    & \ldots   & 0.188 $\pm$ 0.029 & \ldots   \\
             &  & 3238 $\pm$ 11$^{int}$ & \ldots    & \ldots   & 0.187 $\pm$ 0.001$^{int}$ & \ldots   \\
             & \cite{RojasAyala2012} & 3266 $\pm$ 29 & \ldots    & $-$0.39 $\pm$ 0.17 & \ldots   & \ldots   \\
             & \cite{Segransan2003} & 3163 $\pm$ 65 & 5.05 $\pm$ 0.01$^{c}$ & \ldots   & 0.196 $\pm$ 0.008 & 0.158 $\pm$ 0.008 \\
             & \cite{Terrien2015} & \ldots   & \ldots    & $-$0.34 $\pm$ 0.10 & 0.183 $\pm$ 0.002 & \ldots   \\
             \cline{2-7}\noalign{\smallskip}
             & Literature median & 3243 $\pm$ 78 & 5.06 $\pm$ 0.03  & $-$0.39 $\pm$ 0.12 & \ldots   & \ldots   \\
             & Literature \& Run A median & 3243 $\pm$ 75 & 5.05 $\pm$ 0.07  & $-$0.34 $\pm$ 0.13 & \ldots   & \ldots   \\

\hline\noalign{\smallskip}

J22565$+$165 & \cite{Berger2006} & 3373 $\pm$ 101  & 4.53 $\pm$ 0.07  & \ldots   & 0.689 $\pm$ 0.044 & 0.586 $\pm$ 0.059 \\
             & \cite{Boyajian2012} & 3713 $\pm$ 11 & 4.71 $\pm$ 0.04$^{c}$ & \ldots   & 0.548 $\pm$ 0.005 & 0.569 $\pm$ 0.057 \\
             & \cite{Gaidos2014} & 3673 $\pm$ 60 & 4.78 $\pm$ 0.02$^{c}$ & +0.18 $\pm$ 0.11 & 0.480 $\pm$ 0.040 & 0.510 $\pm$ 0.060 \\
             & \cite{GaidosMann2014} & 3786 $\pm$ 87 & 4.75 $\pm$ 0.03$^{c}$ & +0.17 $\pm$ 0.08 & 0.520 $\pm$ 0.050 & 0.560 $\pm$ 0.070 \\
             & \cite{Houdebine2019} & 3661 $\pm$ 183 & \ldots    & \ldots   & 0.567 $\pm$ 0.019 & \ldots   \\
             & \cite{Lepine2013} & 3520 $\pm$ 39 & 4.50    & \ldots   & \ldots   & \ldots   \\
             & \cite{Maldonado2015} & 3736 $\pm$ 68 & 4.71 $\pm$ 0.04  & $-$0.01 $\pm$ 0.09 & 0.550 $\pm$ 0.047 & 0.570 $\pm$ 0.052 \\
             & \cite{Mann2015} & 3720 $\pm$ 60 & 4.71 $\pm$ 0.01$^{c}$ & +0.21 $\pm$ 0.08 & 0.549 $\pm$ 0.018 & 0.574 $\pm$ 0.057 \\
             & \cite{Neves2014} & 3602 $\pm$ 110 & \ldots    & +0.03 $\pm$ 0.09 & \ldots   & 0.580 $\pm$ 0.030 \\
             & \cite{Newton2015} & 3749 $\pm$ 76 & \ldots    & \ldots   & 0.555 $\pm$ 0.028 & \ldots   \\
             &  & 3731 $\pm$ 16$^{int}$ & \ldots    & \ldots   & 0.548 $\pm$ 0.005$^{int}$ & \ldots   \\
             & \cite{Terrien2015} & \ldots   & \ldots    & +0.26 $\pm$ 0.10 & 0.545 $\pm$ 0.003 & \ldots   \\
             \cline{2-7}\noalign{\smallskip}
             & Literature median & 3713 $\pm$ 87 & 4.71 $\pm$ 0.04  & +0.18 $\pm$ 0.09 & \ldots   & \ldots   \\
             & Literature \& Run A median & 3714 $\pm$ 79 & 4.71 $\pm$ 0.04  & +0.18 $\pm$ 0.11 & \ldots   & \ldots \\

\hline\noalign{\smallskip}

J23419$+$441 & \cite{Dittmann2016} & \ldots   & \ldots    & +0.17 $\pm$ 0.10 & \ldots   & \ldots   \\
             & \cite{Gaidos2014} & 3005 $\pm$ 62 & \ldots    & \ldots   & <0.19   & <0.14   \\
             & \cite{GaidosMann2014} & 3067 $\pm$ 60 & \ldots    & +0.29 $\pm$ 0.08 & <0.19   & <0.14   \\
             & \cite{Houdebine2019} & 3032 $\pm$ 152 & \ldots    & \ldots   & 0.098 $\pm$ 0.003 & \ldots   \\
             & \cite{Khata2020} & 3104 $\pm$ 117 & 5.05 $\pm$ 0.15$^{c}$ & +0.26 $\pm$ 0.11 & 0.197 $\pm$ 0.038 & 0.161 $\pm$ 0.007 \\
             & \cite{Lepine2013} & 3110 $\pm$ 43 & 5.00    & \ldots   & \ldots   & \ldots   \\
             & \cite{Mann2015} & 2930 $\pm$ 60 & 5.04 $\pm$ 0.01$^{c}$ & +0.23 $\pm$ 0.08 & 0.189 $\pm$ 0.008 & 0.145 $\pm$ 0.015 \\
             & \cite{RojasAyala2012} & 3058 $\pm$ 65 & \ldots    & +0.19 $\pm$ 0.17 & \ldots   & \ldots   \\
             & \cite{Terrien2015} & \ldots   & \ldots    & +0.32 $\pm$ 0.10 & \ldots   & \ldots   \\
             \cline{2-7}\noalign{\smallskip}
             & Literature median & 3058 $\pm$ 88 & 5.04 $\pm$ 0.11  & +0.25 $\pm$ 0.11 & \ldots   & \ldots   \\
             & Literature \& Run A median & 3058 $\pm$ 80 & 5.02 $\pm$ 0.10  & +0.21 $\pm$ 0.14 & \ldots   & \ldots   \\

\end{longtable}
\tablefoot{$^{(c)}$ $\log{g}$ calculated from $M$ and $R$, $^{(int)}$ interferometric measurement.}

\begin{flushleft}
\end{flushleft}

\twocolumn
\clearpage

\section{Additional plots}

\begin{figure*}[!ht]
  \centering
  \includegraphics[width=0.95\linewidth]{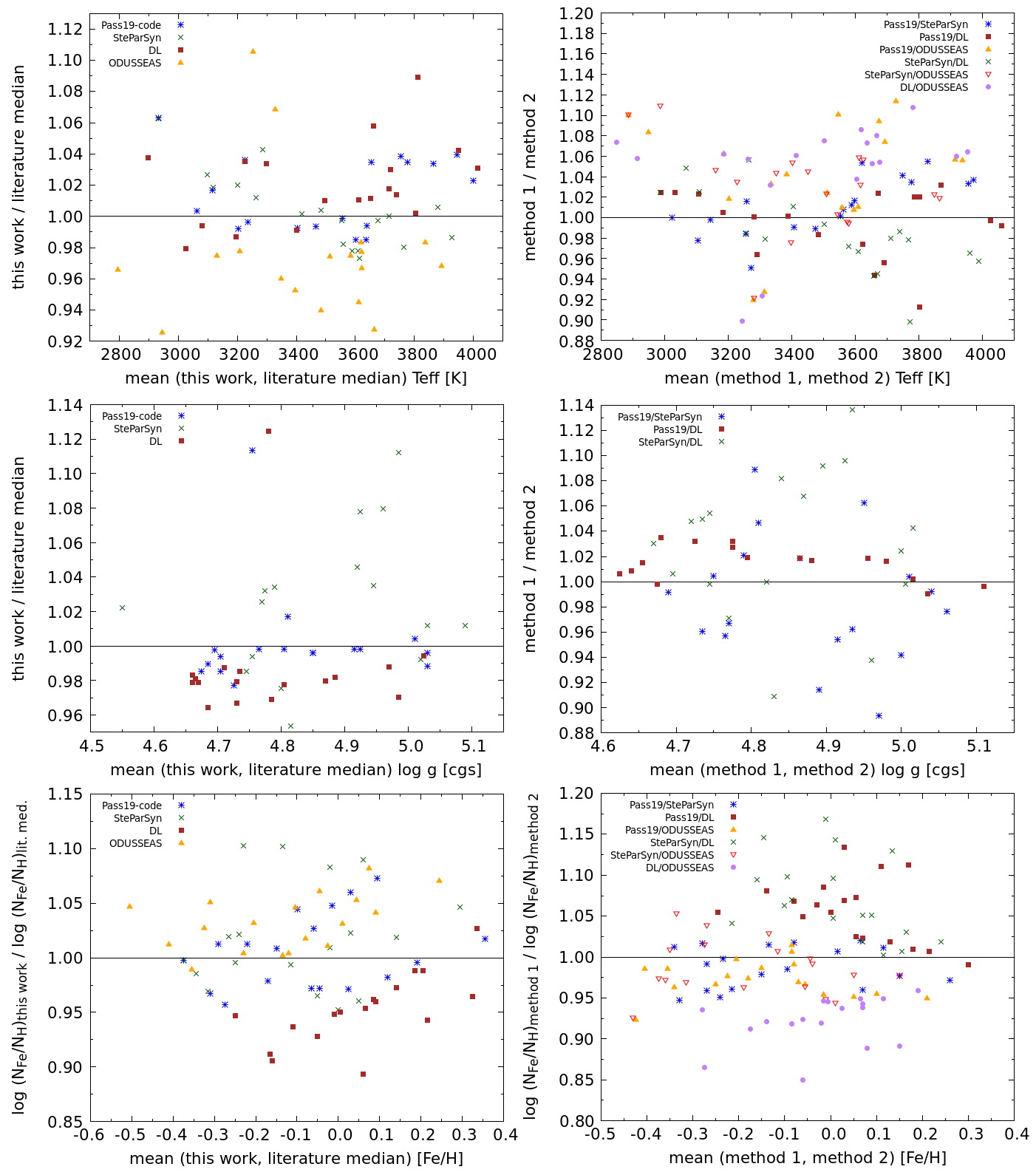}
\caption{Modified Bland-Altman plot for Run A, showing the mean of two methods on the $x$-axis and the ratio on the $y$-axis. The left column presents the comparison between the literature median and our methods (color-coded) for $T_{\rm eff}$ (top), $\log{g}$ (middle), and [Fe/H] (bottom). The right column presents the comparison between our methods. Each symbol represents a different combination of our methods, as shown in the legend. 
For [Fe/H] the $y$-axis shows the ratio of the number of Fe and H atoms in order to avoid division by zero. See text for details. }
\label{fig:blandA}
\end{figure*}
\clearpage

\begin{figure*}[!ht]
  \centering
  \includegraphics[width=0.95\linewidth]{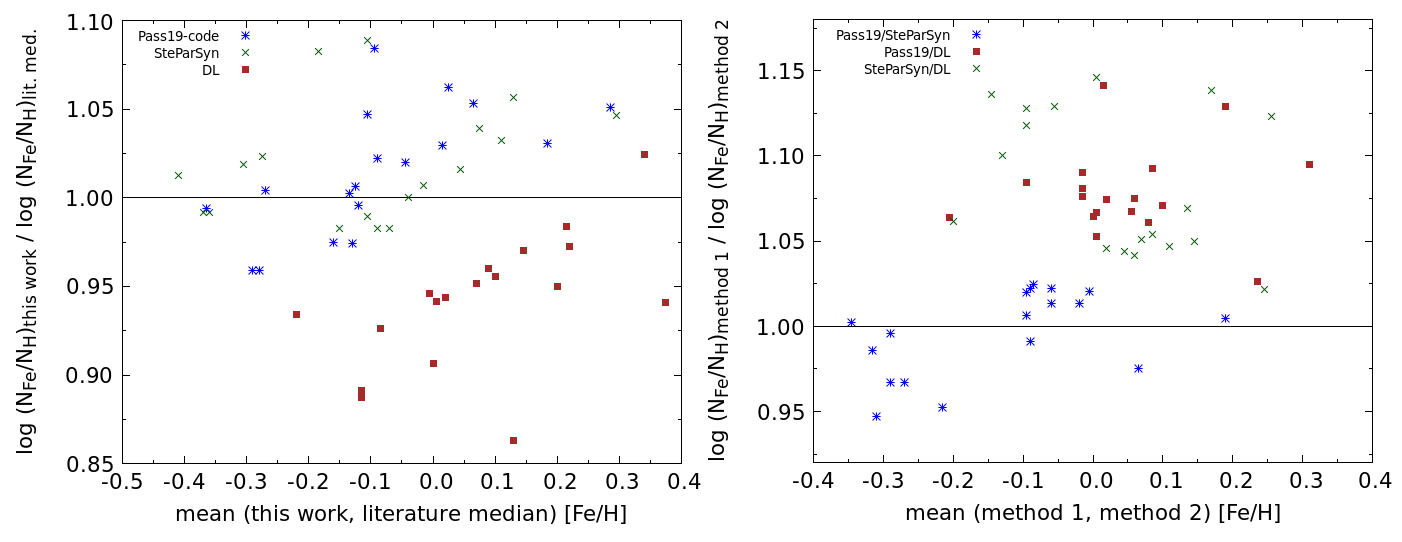}
\caption{Modified Bland-Altman plot for [Fe/H] in Run B, showing the mean of two methods on the $x$-axis and on the $y$-axis the ratio of the number of Fe and H atoms in order to avoid division by zero. The left plot presents the comparison between the literature median and our methods (color-coded). The right plot presents the comparison between our methods. Each symbol represents a different combination of our methods, as shown in the legend.}
\label{fig:blandB}
\end{figure*}
\clearpage

\begin{figure*}[!ht]
  \centering
  \includegraphics[width=0.95\linewidth]{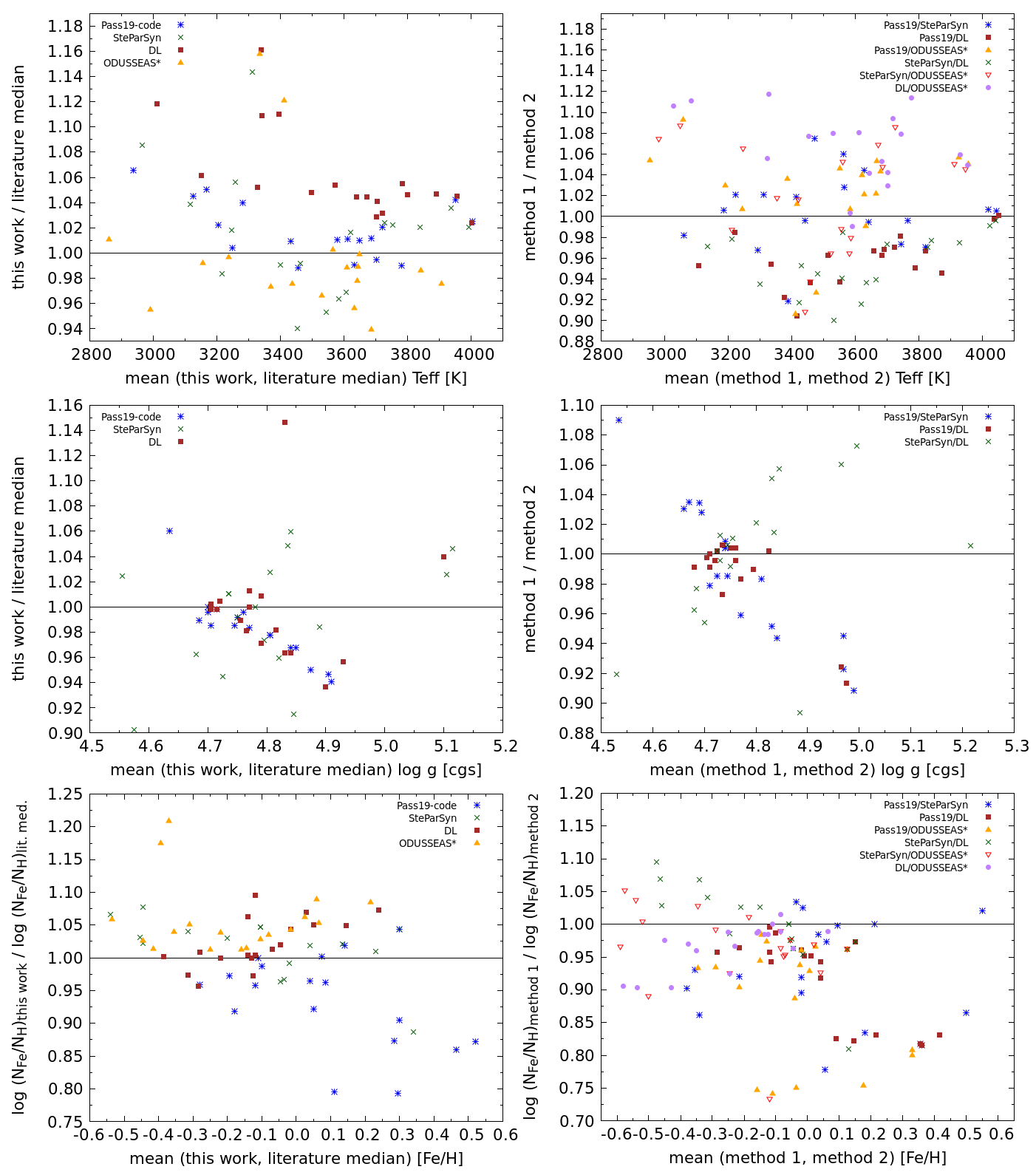}
\caption{Modified Bland-Altman plot for Run C, showing the mean of two methods on the $x$-axis and the ratio on the $y$-axis. The left column presents the comparison between the literature median and our methods (color-coded) for $T_{\rm eff}$ (top), $\log{g}$ (middle), and [Fe/H] (bottom). The right column presents the comparison between our methods. Each symbol represents a different combination of our methods, as shown in the legend. 
For [Fe/H] the $y$-axis shows the ratio of the number of Fe and H atoms in order to avoid division by zero. See text for details. }
\label{fig:blandC}
\end{figure*}
\clearpage

\begin{figure*}[!ht]
  \centering
  \includegraphics[width=0.95\linewidth]{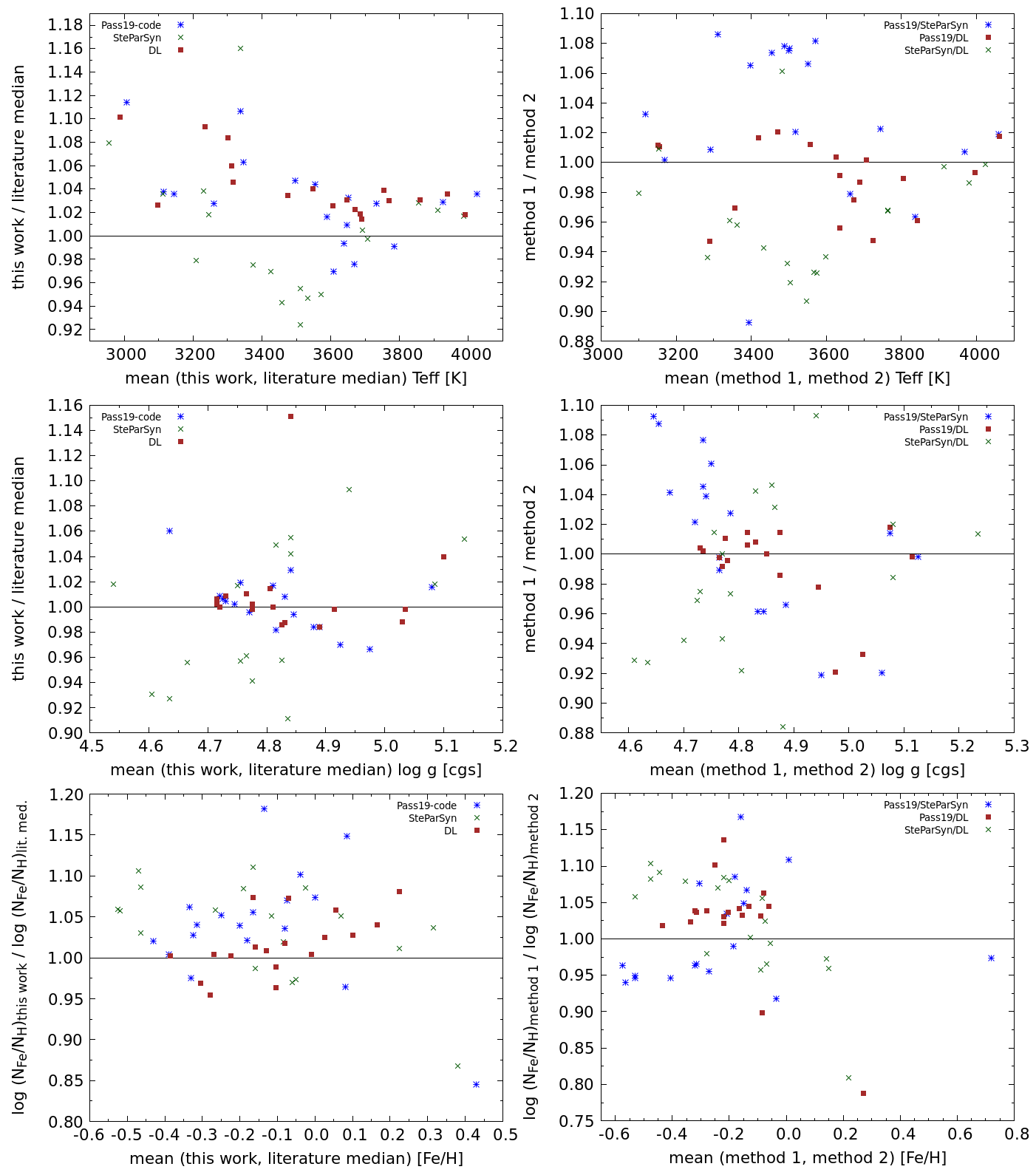}
\caption{Modified Bland-Altman plot for Run C2, showing the mean of two methods on the $x$-axis and the ratio on the $y$-axis. The left column presents the comparison between the literature median and our methods (color-coded) for $T_{\rm eff}$ (top), $\log{g}$ (middle), and [Fe/H] (bottom). The right column presents the comparison between our methods. Each symbol represents a different combination of our methods, as shown in the legend. 
For [Fe/H] the $y$-axis shows the ratio of the number of Fe and H atoms in order to avoid division by zero. See text for details. }
\label{fig:blandC2}
\end{figure*}
\clearpage

\begin{figure*}[!ht]
  \centering
  \includegraphics[width=0.95\linewidth]{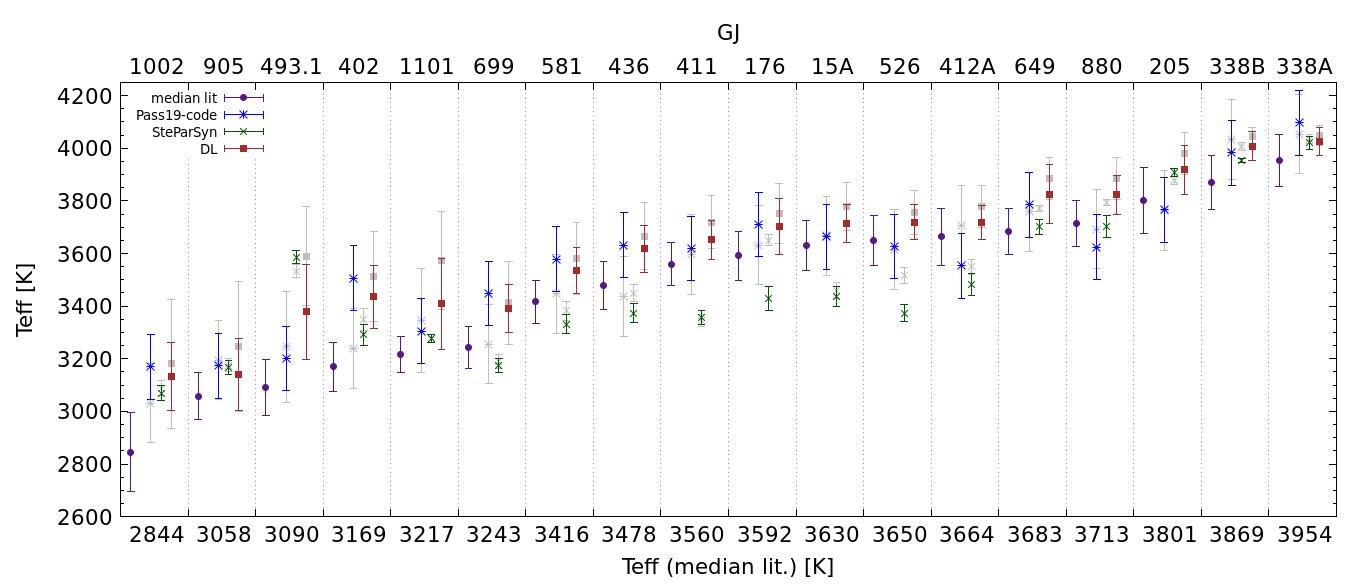}
  \includegraphics[width=0.95\linewidth]{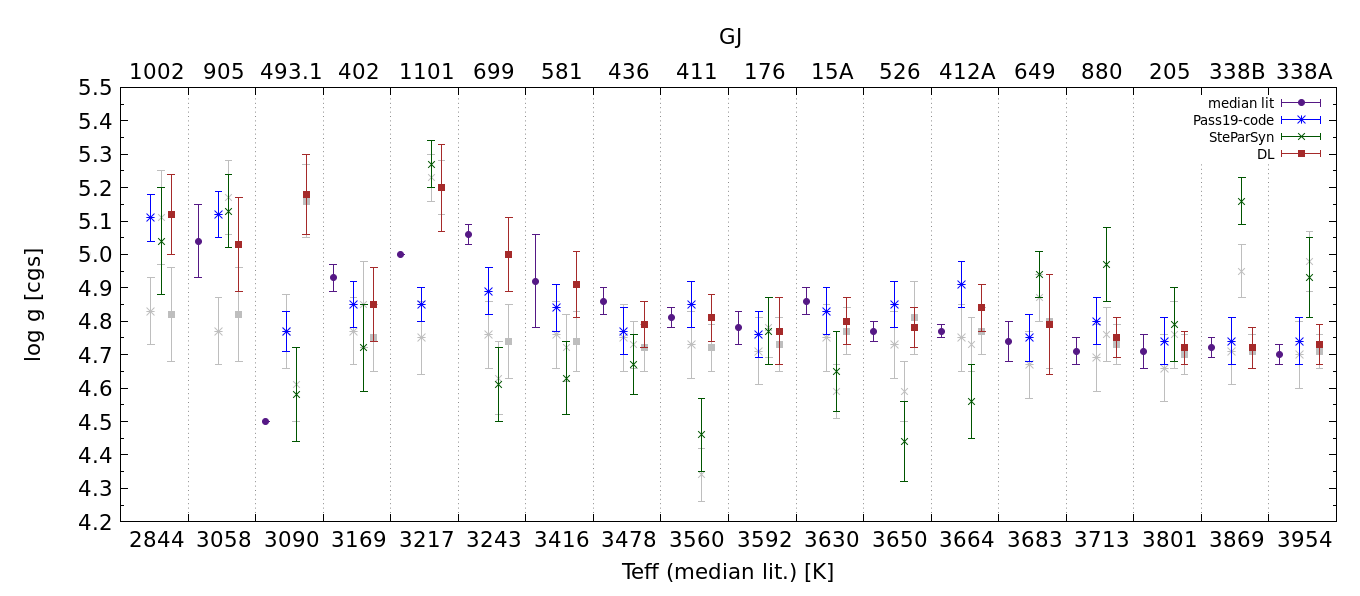}\par\medskip
  \includegraphics[width=0.95\linewidth]{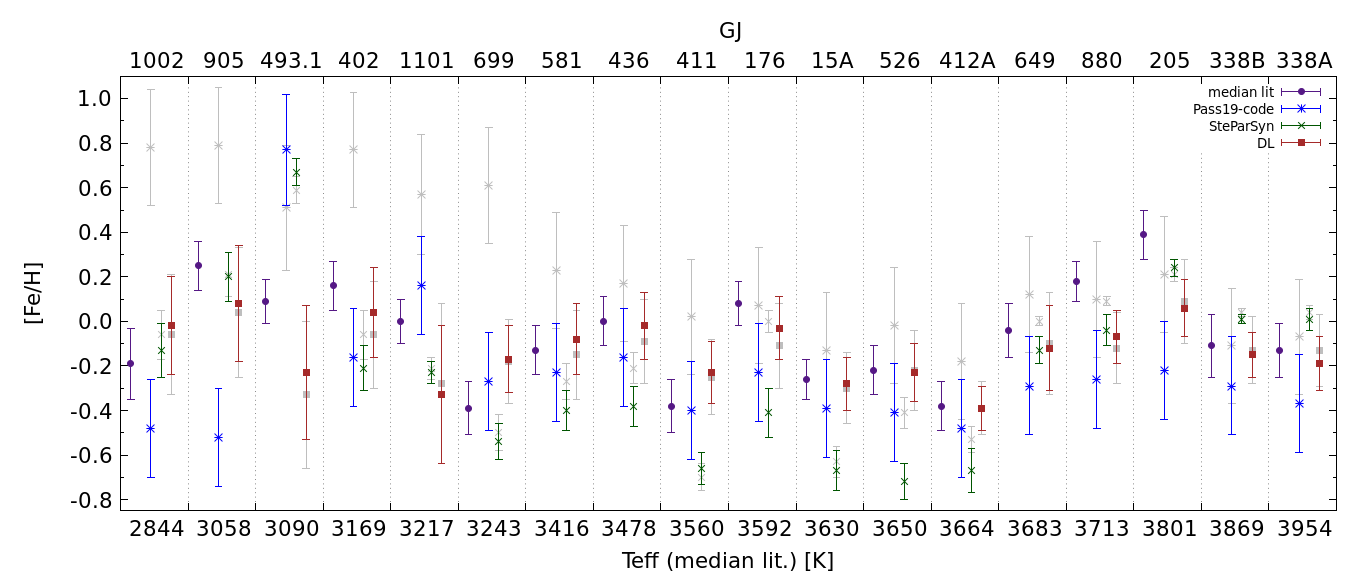}
\caption{Comparison of $T_{\rm eff}$ ({\it top}), $\log{g}$ ({\it middle}), and [Fe/H] ({\it bottom}) for the different methods in Run C2. Each method is indicated with a different symbol and color. The gray symbols indicate the results from Run C for comparison. The median of all literature values is shown as purple dots. The $x$-axis indicates $T_{\rm eff}$ from the literature median, the top axis shows the Gliese-Jahreiss (GJ) numbers for all sample stars.}
\label{fig:C-comparison2}
\end{figure*}
\clearpage

\begin{figure*}[!ht]
  \centering
  \includegraphics[width=0.60\linewidth]{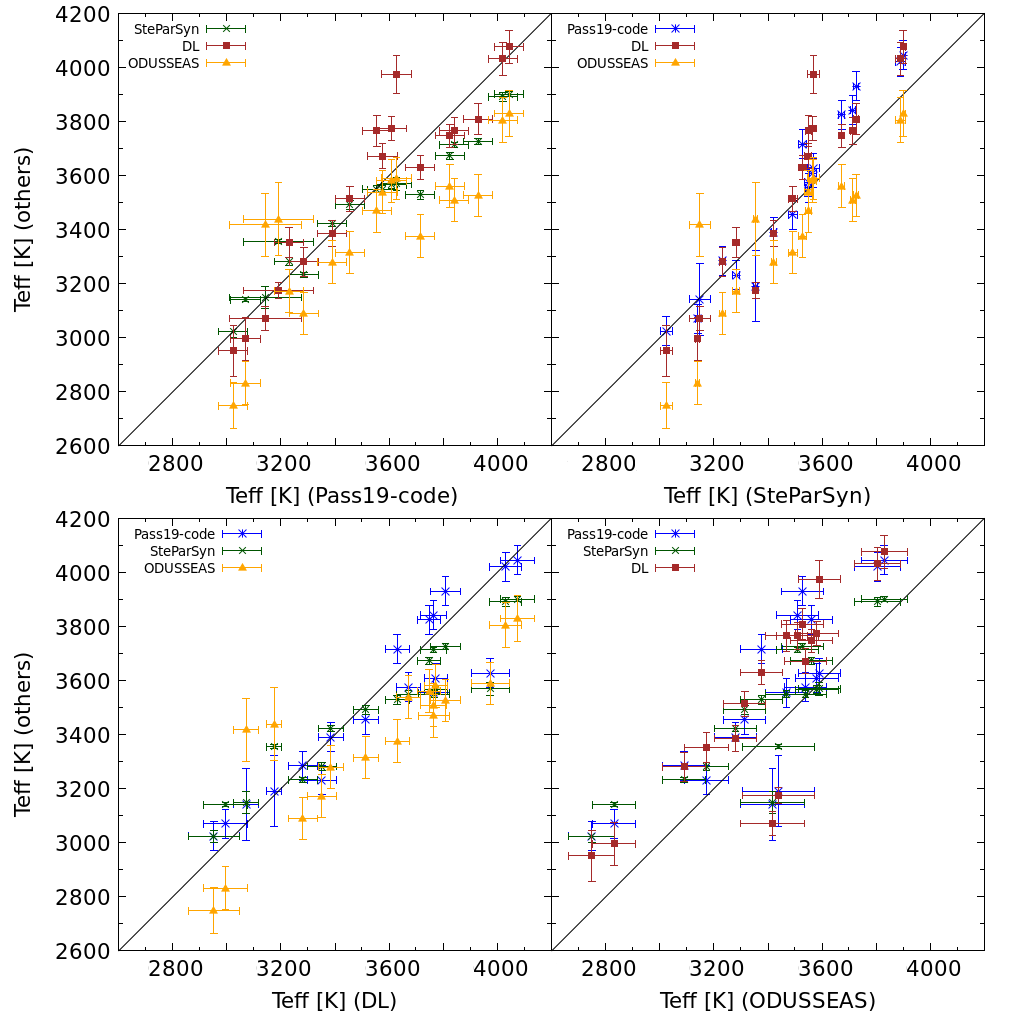}
\caption{Comparison between our methods, showing the derived $T_{\rm eff}$ in Run A. Each method is indicated by a different color and symbol. Each panel compares one method (denoted by the $x$-axis label) to all other methods. }
\label{fig:Teff_all_A}
\end{figure*}

\begin{figure*}[!ht]
  \centering
  \includegraphics[width=0.60\linewidth]{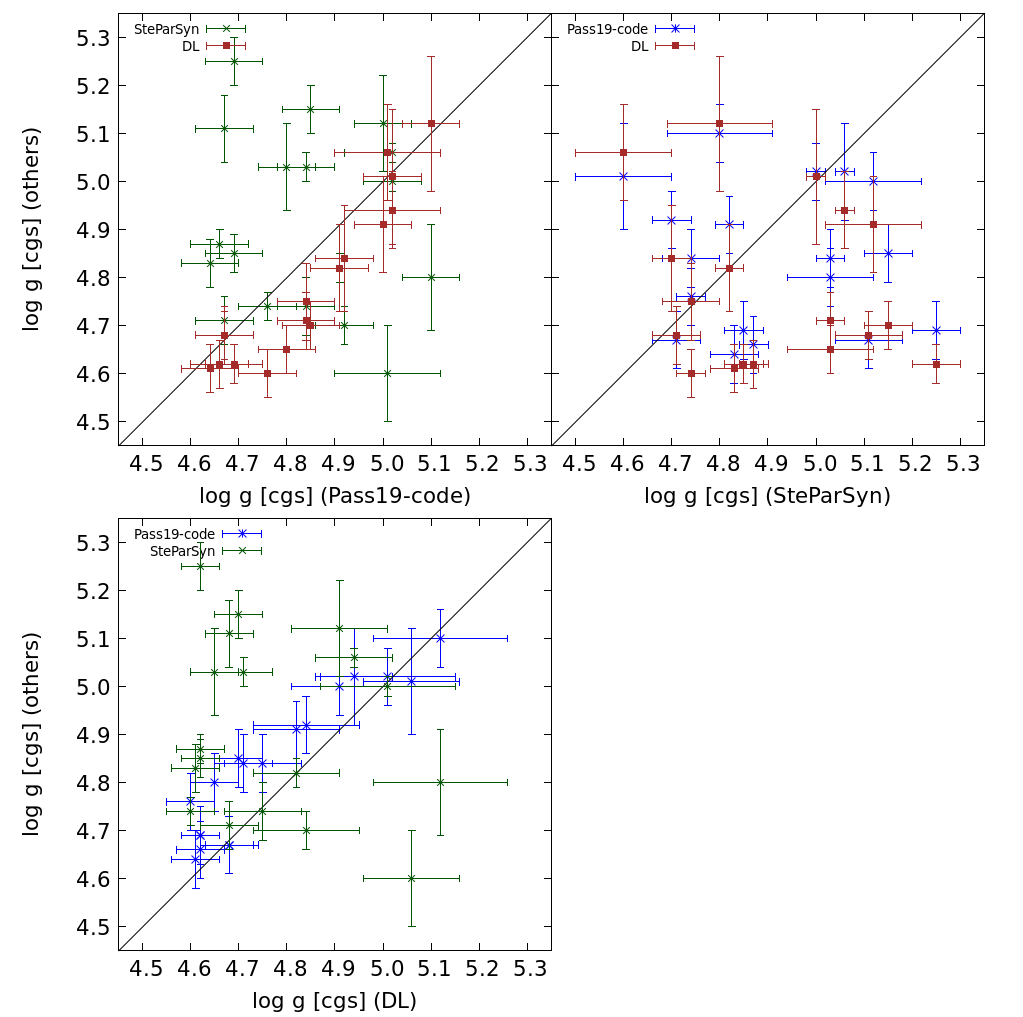}
\caption{Comparison between our methods, showing the derived $\log{g}$ in Run A. Each method is indicated by a different color and symbol. Each panel compares one method (denoted by the $x$-axis label) to all other methods. 
{\tt ODUSSEAS} did not derive $\log{g}$.}
\label{fig:logg_all_A}
\end{figure*}
\clearpage

\begin{figure*}[!ht]
  \centering
  \includegraphics[width=0.60\linewidth]{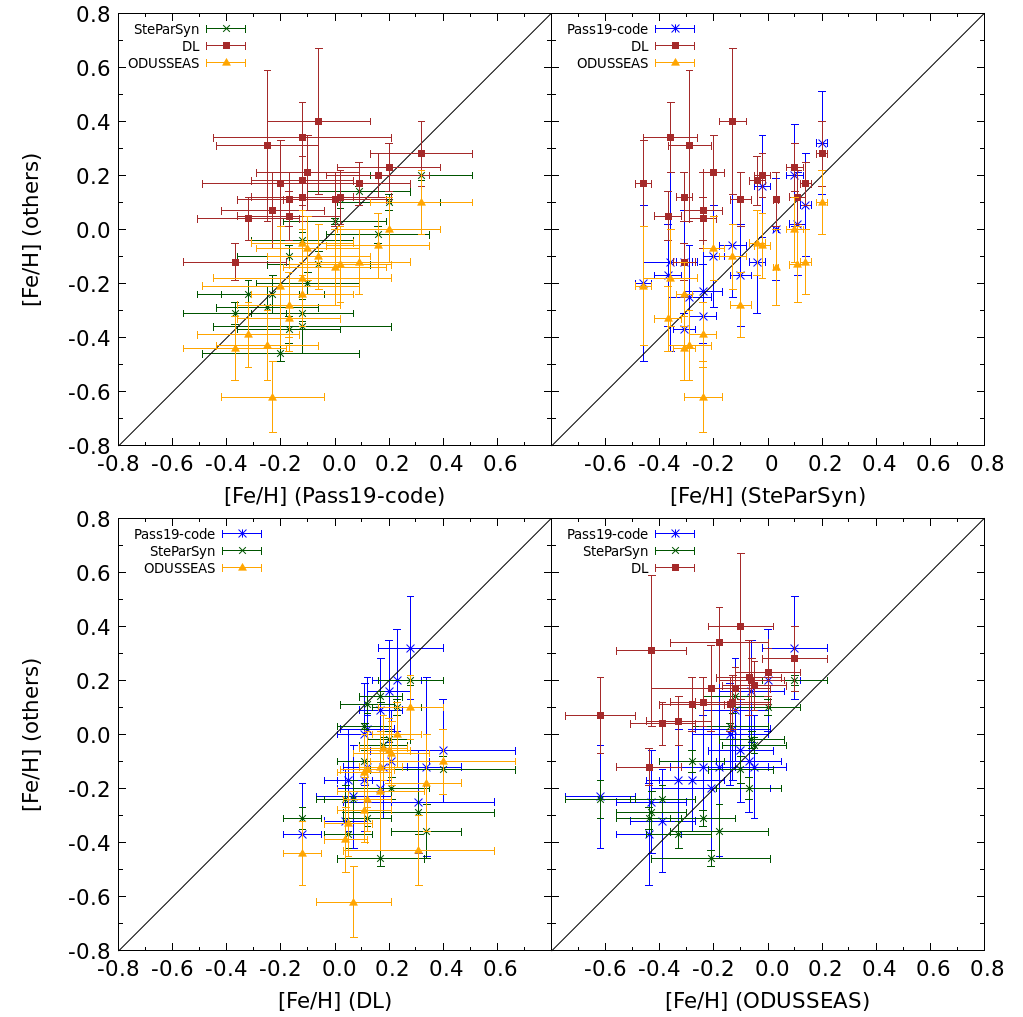}
\caption{Comparison between our methods, showing the derived [Fe/H] in Run A. Each method is indicated by a different color and symbol. Each panel compares one method (denoted by the $x$-axis label) to all other methods.}
\label{fig:metal_all_A}
\end{figure*}

\begin{figure*}[!ht]
  \centering
  \includegraphics[width=0.60\linewidth]{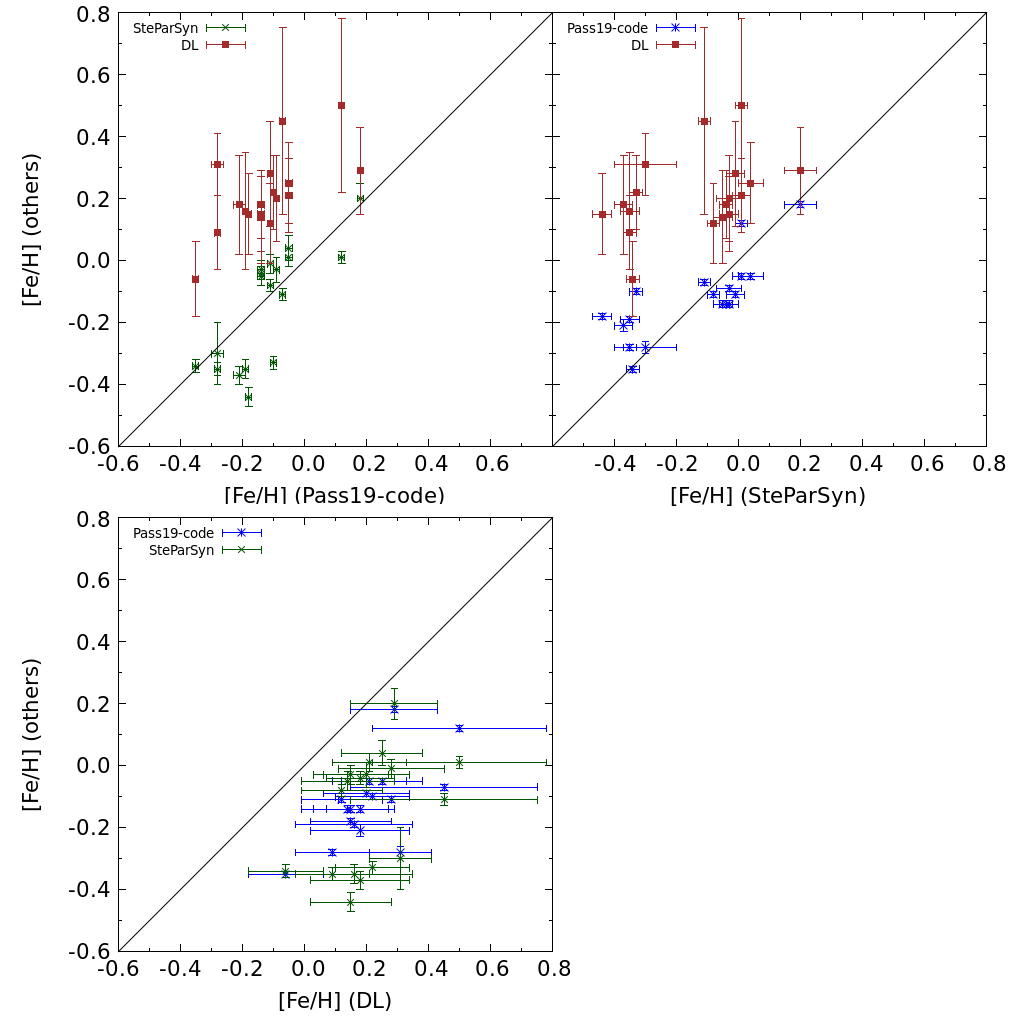}
\caption{Comparison between our methods, showing the derived [Fe/H] in Run B. Each method is indicated by a different color and symbol. Each panel compares one method (denoted by the $x$-axis label) to all other methods.
{\tt ODUSSEAS} did not participate in Run B.}
\label{fig:metal_all_B}
\end{figure*}
\clearpage

\begin{figure*}[!ht]
  \centering
  \includegraphics[width=0.60\linewidth]{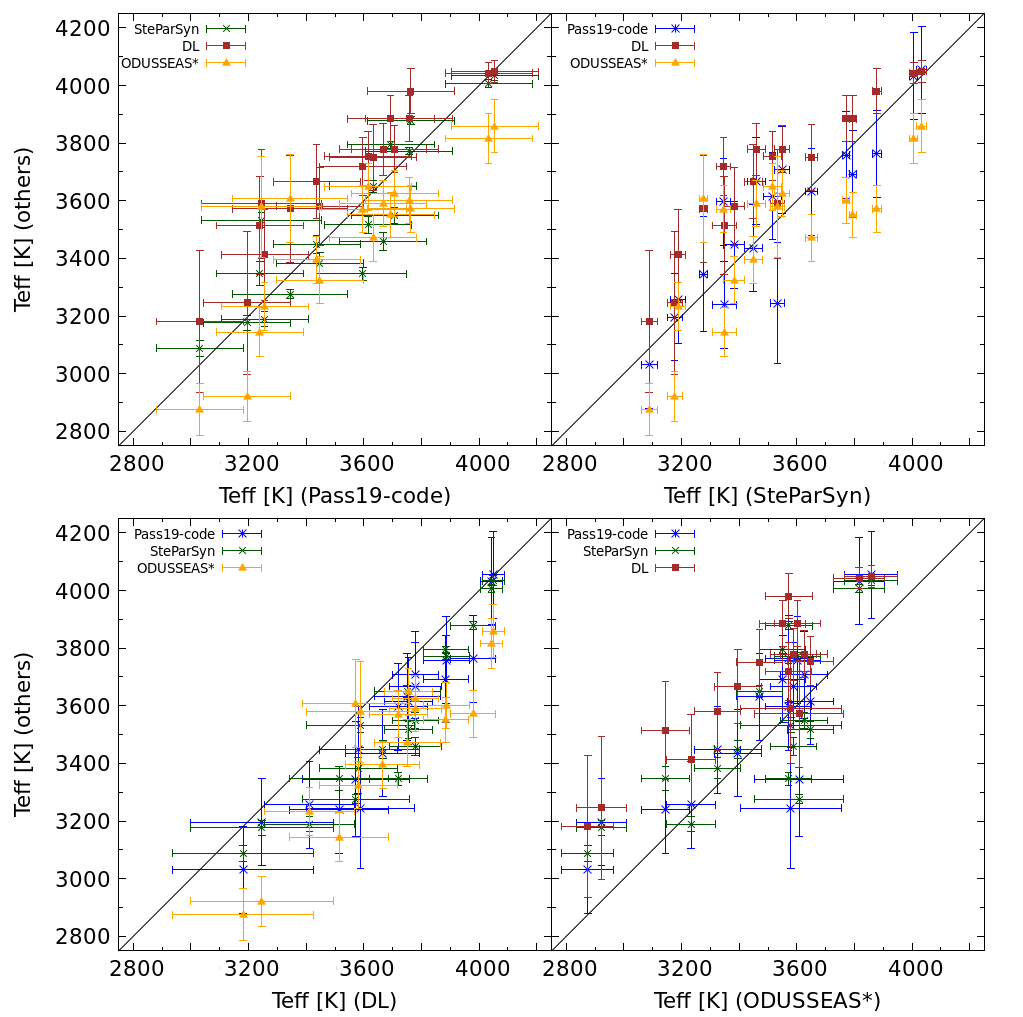}
\caption{Comparison between our methods, showing the derived $T_{\rm eff}$ in Run C. Each method is indicated by a different color and symbol. Each panel compares one method (denoted by the $x$-axis label) to all other methods.
The values from {\tt ODUSSEAS} correspond to Run C*.}
\label{fig:Teff_all_C}
\end{figure*}

\begin{figure*}[!ht]
  \centering
  \includegraphics[width=0.60\linewidth]{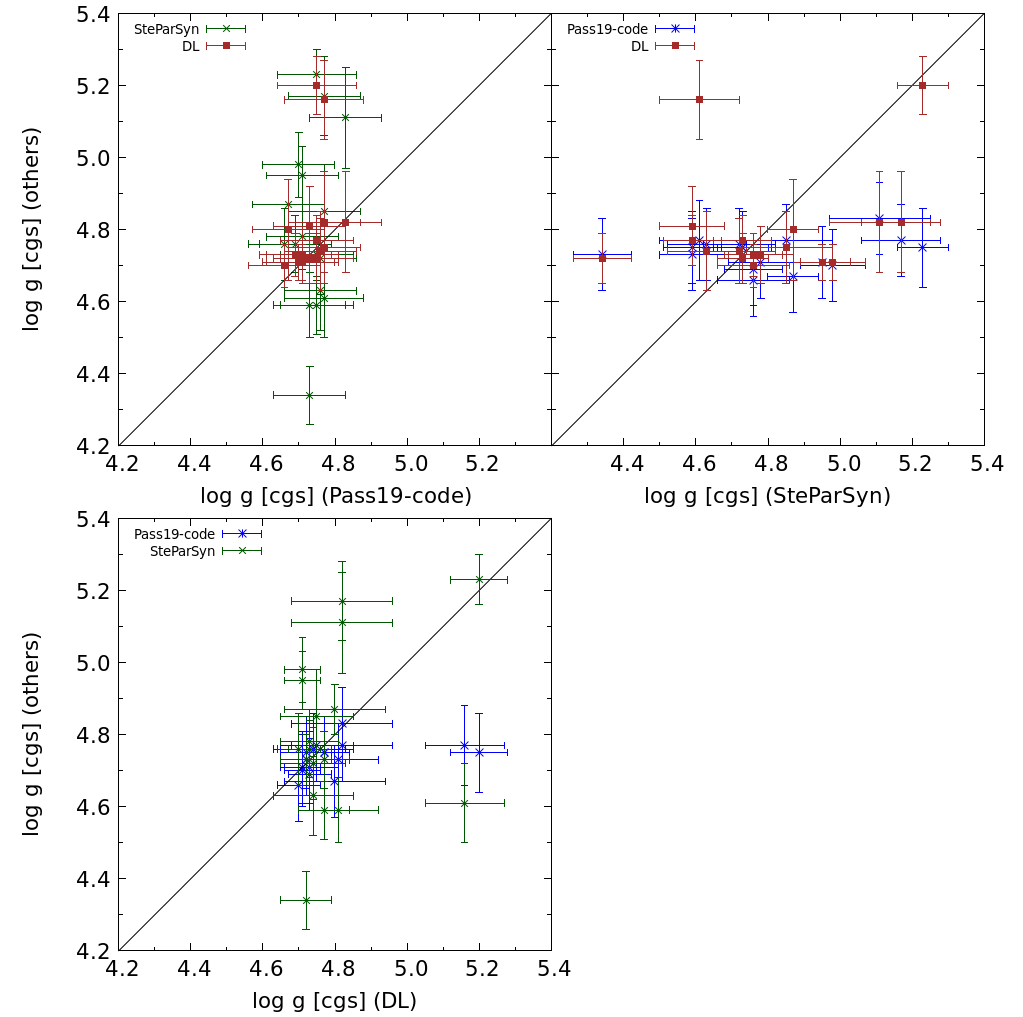}
\caption{Comparison between our methods, showing the derived $\log{g}$ in Run C. Each method is indicated by a different color and symbol. Each panel compares one method (denoted by the $x$-axis label) to all other methods. 
{\tt ODUSSEAS} did not derive $\log{g}$.}
\label{fig:logg_all_C}
\end{figure*}
\clearpage

\begin{figure*}[!ht]
  \centering
  \includegraphics[width=0.60\linewidth]{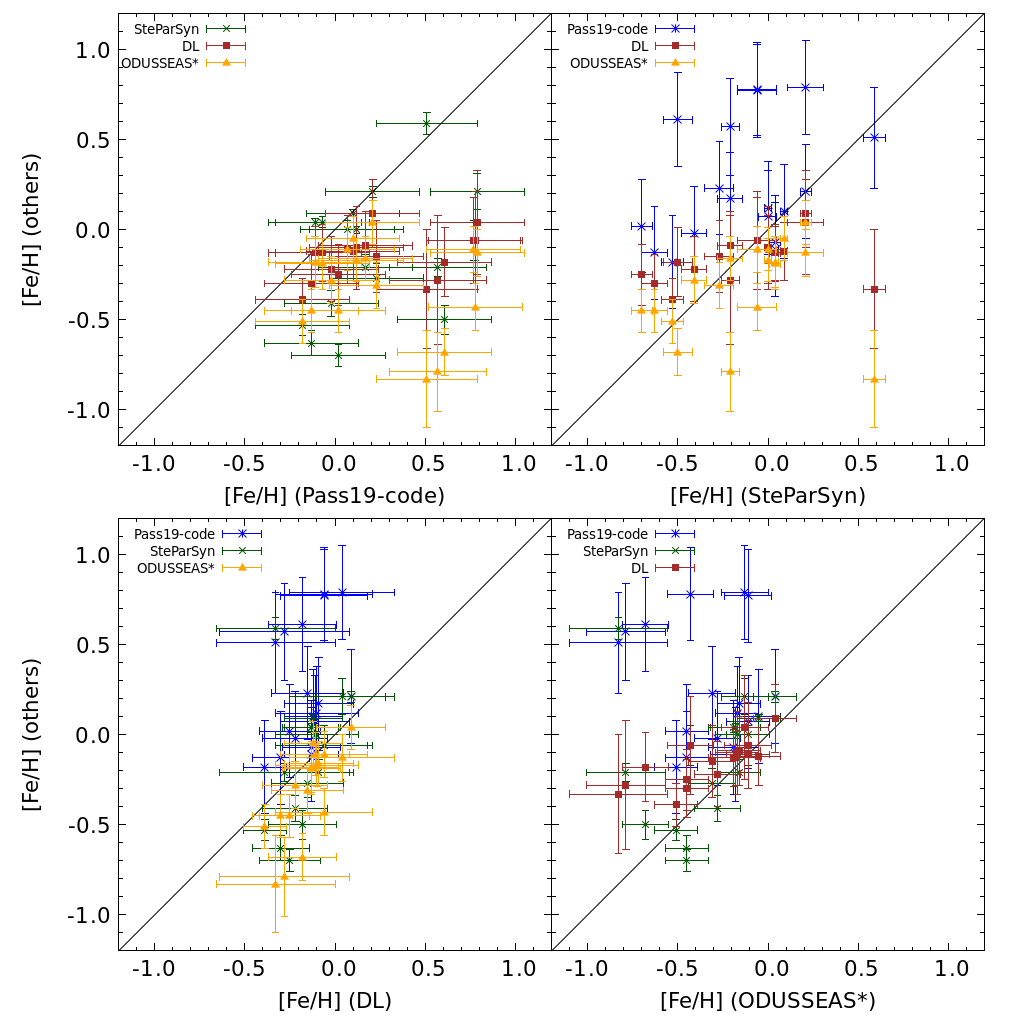}
\caption{Comparison between our methods, showing the derived [Fe/H] in Run C. Each method is indicated by a different color and symbol. Each panel compares one method (denoted by the $x$-axis label) to all other methods. The values from {\tt ODUSSEAS} correspond to Run C*.}
\label{fig:metal_all_C}
\end{figure*}

\begin{figure*}[!ht]
  \centering
  \includegraphics[width=0.60\linewidth]{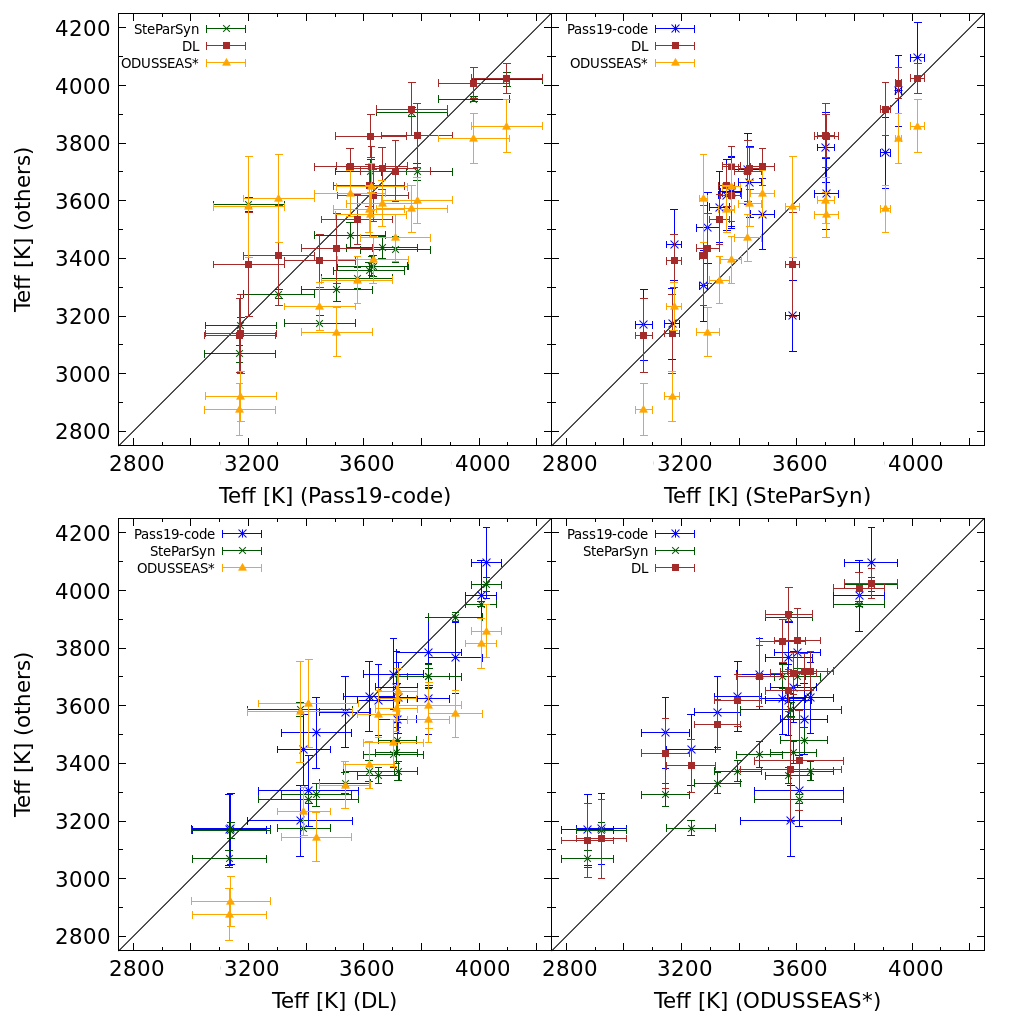}
\caption{Comparison between our methods, showing the derived $T_{\rm eff}$ in Run C2. Each method is indicated by a different color and symbol. Each panel compares one method (denoted by the $x$-axis label) to all other methods.
The values from {\tt ODUSSEAS} correspond to Run C*.}
\label{fig:Teff_all_C2}
\end{figure*}
\clearpage

\begin{figure*}[!ht]
  \centering
  \includegraphics[width=0.60\linewidth]{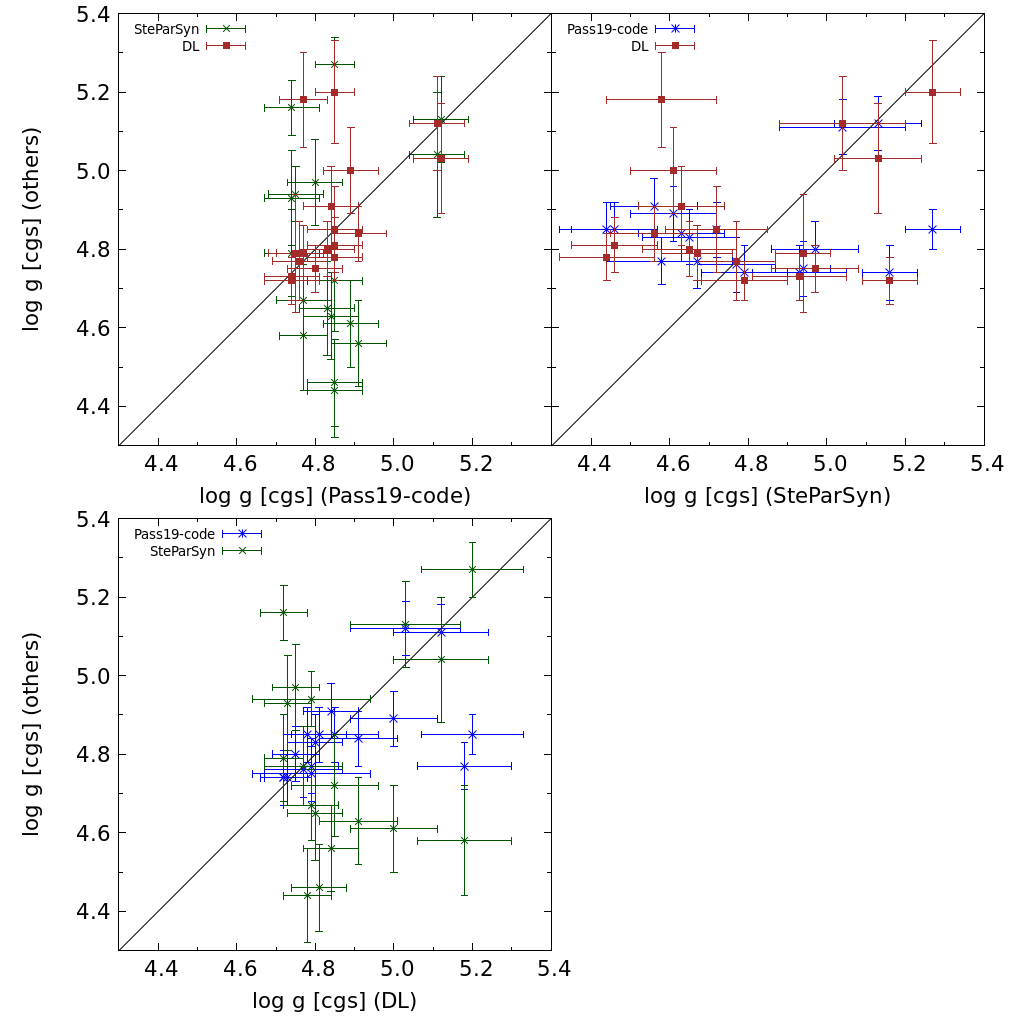}
\caption{Comparison between our methods, showing the derived $\log{g}$ in Run C2. Each method is indicated by a different color and symbol. Each panel compares one method (denoted by the $x$-axis label) to all other methods. 
{\tt ODUSSEAS} did not derive $\log{g}$.}
\label{fig:logg_all_C2}
\end{figure*}

\begin{figure*}[!ht]
  \centering
  \includegraphics[width=0.60\linewidth]{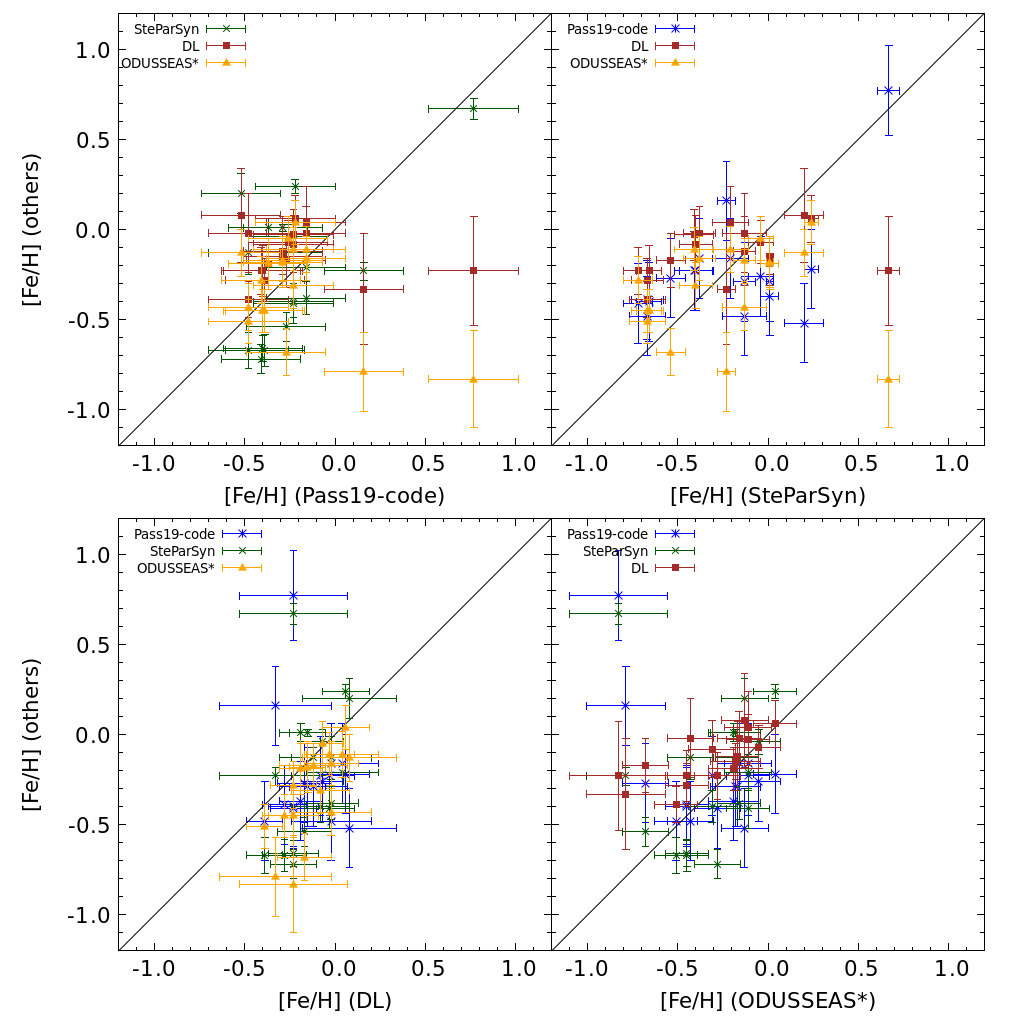}
\caption{Comparison between our methods, showing the derived [Fe/H] in Run C2. Each method is indicated by a different color and symbol. Each panel compares one method (denoted by the $x$-axis label) to all other methods.
The values from {\tt ODUSSEAS} correspond to Run C*.}
\label{fig:metal_all_C2}
\end{figure*}
\clearpage

\section{Results}

Tables C.1 and C.2 are available in their entirety in a machine-
readable form at the CDS. An excerpt is shown here for guidance
regarding their form and content.

\onecolumn
\begin{longtable}{lccccccc}
\label{tab:results1}\\
\caption[]{Stellar parameters for each method from Runs A and B.}\\
\hline\hline\noalign{\smallskip}
       &         & 
\multicolumn{3}{c}{Run A} & 
\multicolumn{3}{c}{Run B} \\
Karmn  &  Method & 
$T_{\rm eff}$\,[K] & $\log{g}$\,[dex] & [Fe/H]\,[dex] & 
$T_{\rm eff}$\,[K] & $\log{g}$\,[dex] & [Fe/H]\,[dex] \\ 
\noalign{\smallskip}\hline\noalign{\smallskip}          
\endfirsthead

\caption[]{continued.}\\ 
\hline\hline\noalign{\smallskip}                
       &        & 
\multicolumn{3}{c}{Run A} & 
\multicolumn{3}{c}{Run B} \\
Karmn  & Method & 
$T_{\rm eff}$\,[K] & $\log{g}$\,[dex] & [Fe/H]\,[dex] & 
$T_{\rm eff}$\,[K] & $\log{g}$\,[dex] & [Fe/H]\,[dex] \\ 
\noalign{\smallskip}\hline\noalign{\smallskip}
\endhead

\noalign{\smallskip}\hline
\endfoot
 
J00067$-$075 & {\tt Pass19-code} & $3024\pm54$     & $5.10\pm0.06$   & $-0.25\pm0.19$ 
                                 & $2960\pm103$     & $5.10\pm0.11$   & $-0.07\pm0.01$ \\
             & {\tt SteParSyn}   & $3023\pm22$     & $4.80\pm0.11$   & $-0.29\pm0.08$ 
                                 & $2960\pm103$     & $5.10\pm0.11$   & $-0.11\pm0.02$ \\
             & DL                & $2951\pm94$     & $5.12\pm0.14$   &  $+0.31\pm0.28$ 
                                 & $2960\pm103$     & $5.10\pm0.11$   &  $+0.45\pm0.30$ \\
             & {\tt ODUSSEAS}            & $2748\pm85$     & \ldots          & $-0.43\pm0.13$ 
                                 & \ldots              & \ldots          & \ldots        
                                 \\
\noalign{\smallskip}
J00183$+$440 & {\tt Pass19-code} & $3576\pm54$     & $4.84\pm0.06$   & $-0.32\pm0.19$ 
                                 & $3603\pm84$     & $4.85\pm0.04$   & $-0.28\pm0.01$ \\
             & {\tt SteParSyn}   & $3549\pm13$     & $5.03\pm0.03$   & $-0.24\pm0.05$
                                 & $3603\pm84$     & $4.85\pm0.04$   & $-0.35\pm0.02$ \\
              & DL               & $3672\pm45$     & $4.71\pm0.06$   &  $+0.04\pm0.08$
                                 & $3603\pm84$     & $4.85\pm0.04$   &  $+0.09\pm0.12$ \\
             & {\tt ODUSSEAS}          & $3539\pm78$     & \ldots          & $-0.39\pm0.12$ 
                                 & \ldots          & \ldots          & \ldots \\
\noalign{\smallskip}
J04429$+$189 & {\tt Pass19-code} & $3716\pm54$     & $4.67\pm0.06$   &  $+0.16\pm0.19$ 
                                 & $3592\pm85$     & $4.76\pm0.05$   & $-0.05\pm0.01$ \\
             & {\tt SteParSyn}   & $3528\pm15$     & $4.71\pm0.05$   & $-0.02\pm0.03$ 
                                 & $3592\pm85 $    & $4.76\pm0.05$   &  $+0.01\pm0.03$ \\
             & DL                & 3630 $\pm$ 44     & 4.68 $\pm$ 0.06   &  $+0.20\pm0.08$ 
                                 & 3592 $\pm$ 85     & 4.76 $\pm$ 0.05   &  $+0.21\pm0.12$\\
             & {\tt ODUSSEAS}          & 3376 $\pm$ 78     & \ldots          & $-0.06\pm0.12$ 
                                 & \ldots          & \ldots          & \ldots \\
\noalign{\smallskip}
J05314$-$036 & {\tt Pass19-code} & 3930 $\pm$ 54     & 4.64 $\pm$ 0.06   &  $+0.32\pm0.19$ 
                                 & 3801 $\pm$ 112    & 4.70 $\pm$ 0.05   &  $+0.18\pm0.01$ \\
             & {\tt SteParSyn}   & 3726 $\pm$ 10     & 4.83 $\pm$ 0.05   &  $+0.20\pm0.02$ 
                                 & 3801 $\pm$ 112    & 4.70 $\pm$ 0.05   &  $+0.20\pm0.05$ \\
             & DL                & 3809 $\pm$ 57     & 4.61 $\pm$ 0.05   &  $+0.28\pm0.12$ 
                                 & 3801 $\pm$ 112    & 4.70 $\pm$ 0.05   &  $+0.29\pm0.14$ \\
             & {\tt ODUSSEAS}          & 3527 $\pm$ 78     & \ldots          &  $+0.10\pm0.12$ 
                                 & \ldots          & \ldots          & \ldots \\
\noalign{\smallskip}
J07558$+$833 & {\tt Pass19-code} & 3191 $\pm$ 131     & 5.02 $\pm$ 0.10   & $-0.20\pm0.29$ 
                                 & 3220 $\pm$ 87     & 5.01 $\pm$ 0.07   & $-0.21\pm0.02$ \\
             & {\tt SteParSyn}   & 3355 $\pm$ 7      & 5.06 $\pm$ 0.02   & $-0.46\pm0.03$ 
                                 & 3220 $\pm$ 87     & 5.01 $\pm$ 0.07   & $-0.37\pm0.03$ \\
             & DL                & 3175 $\pm$ 29     & 4.94 $\pm$ 0.08   &  $+0.17\pm0.16$ 
                                 & 3220 $\pm$ 87     & 5.01 $\pm$ 0.07   &  $+0.18\pm0.16$ \\
             & {\tt ODUSSEAS}          & 3439 $\pm$ 134     & \ldots          & $-0.21\pm0.22$ 
                                 & \ldots          & \ldots          & \ldots \\
\noalign{\smallskip}
J09143$+$526 & {\tt Pass19-code} & 4045 $\pm$ 54     & 4.69 $\pm$ 0.06   &  $+0.00\pm0.19$ 
                                 & 3954 $\pm$ 88     & 4.69 $\pm$ 0.04   & $-0.11\pm0.01$ \\
             & {\tt SteParSyn}   & 3901 $\pm$ 9      & 4.85 $\pm$ 0.04   &  $+0.03\pm0.01$
                                 & 3954 $\pm$ 88     & 4.69 $\pm$ 0.04   & $-0.08\pm0.02$ \\
             & DL                & 4076 $\pm$ 62     & 4.62 $\pm$ 0.04   &  $+0.11\pm0.10$
                                 & 3954 $\pm$ 88     & 4.69 $\pm$ 0.04   &  $+0.12\pm0.13$ \\
             & {\tt ODUSSEAS}          & 3830 $\pm$ 85     & \ldots          & $-0.14\pm0.14$ 
                                 & \ldots          & \ldots          & \ldots \\
\noalign{\smallskip}
J09144$+$526 & {\tt Pass19-code} & 4021 $\pm$ 54     & 4.69 $\pm$ 0.06   &  $+0.02\pm0.19$
                                 & 3891 $\pm$ 89     & 4.71 $\pm$ 0.04   & $-0.14\pm0.01$ \\
             & {\tt SteParSyn}   & 3891 $\pm$ 18     & 5.25 $\pm$ 0.05   &  $+0.11\pm0.03$ 
                                 & 3891 $\pm$ 89     & 4.71 $\pm$ 0.04   & $-0.03\pm0.03$ \\
             & DL                & 4032 $\pm$ 60     & 4.62 $\pm$ 0.04   &  $+0.12\pm0.10$ 
                                 & 3891 $\pm$ 89     & 4.71 $\pm$ 0.04   &  $+0.15\pm0.12$ \\
             & {\tt ODUSSEAS}          & 3805 $\pm$ 84     & \ldots          & $-0.13\pm0.14$ 
                                 & \ldots          & \ldots          & \ldots \\
\noalign{\smallskip}
J10508$+$068 & {\tt Pass19-code} & 3284 $\pm$ 54     & 4.92 $\pm$ 0.06   & $-0.10\pm0.19$ 
                                 & 3235 $\pm$ 82     & 4.92 $\pm$ 0.06   & $-0.11\pm0.01$\\
             & {\tt SteParSyn}   & 3232 $\pm$ 11     & 4.70 $\pm$ 0.04   & $-0.20\pm0.04$ 
                                 & 3235 $\pm$ 82     & 4.92 $\pm$ 0.06   & $-0.01\pm0.02$ \\
             & DL                & 3281 $\pm$ 54     & 4.84 $\pm$ 0.11   &  $+0.21\pm0.14$ 
                                 & 3235 $\pm$ 82     & 4.92 $\pm$ 0.06   &  $+0.28\pm0.17$\\
             & {\tt ODUSSEAS}          & 3090 $\pm$ 79     & \ldots          & $-0.07\pm0.12$ 
                                 & \ldots          & \ldots          & \ldots \\
\noalign{\smallskip}
J11033$+$359 & {\tt Pass19-code} & 3555 $\pm$ 54     & 4.80 $\pm$ 0.06   & $-0.17\pm0.19 $
                                 & 3555 $\pm$ 76     & 4.81 $\pm$ 0.05   & $-0.18\pm0.01$\\
             & {\tt SteParSyn}   & 3550 $\pm$ 12     & 5.03 $\pm$ 0.09   & $-0.37\pm0.05 $
                                 & 3555 $\pm$ 76     & 4.81 $\pm$ 0.05   & $-0.44\pm0.03$ \\
             & DL                & 3766 $\pm$ 57     & 4.65 $\pm$ 0.05   &  $+0.05\pm0.09 $
                                 & 3555 $\pm$ 76     & 4.81 $\pm$ 0.05   &  $+0.15\pm0.13$\\
             & {\tt ODUSSEAS}          & 3469 $\pm$ 78     & \ldots          & $-0.33\pm0.12 $
                                 & \ldots          & \ldots          & \ldots \\
\noalign{\smallskip}
J11054$+$435 & {\tt Pass19-code} & 3609 $\pm$ 54     & 4.85 $\pm$ 0.06   & $-0.37\pm0.19$ 
                                 & 3619 $\pm$ 95     & 4.81 $\pm$ 0.04   & $-0.35\pm0.01$\\
             & {\tt SteParSyn}   & 3566 $\pm$ 14     & 5.15 $\pm$ 0.05   & $-0.31\pm0.04$ 
                                 & 3619 $\pm$ 95     & 4.81 $\pm$ 0.04   & $-0.34\pm0.02$ \\
             & DL                & 3774 $\pm$ 44     & 4.70 $\pm$ 0.05   & $-0.12\pm0.07$ 
                                & 3619 $\pm$ 95     & 4.81 $\pm$ 0.04   & $-0.06\pm0.12$\\
             & {\tt ODUSSEAS}          & 3581 $\pm$ 80     & \ldots          & $-0.44\pm0.12$ 
                                 & \ldots          & \ldots          & \ldots \\
\noalign{\smallskip}
J11421$+$267 & {\tt Pass19-code} & 3455 $\pm$ 54     & 4.84 $\pm$ 0.06   & $-0.12\pm0.19$ 
                                 & 3478 $\pm$ 81     & 4.83 $\pm$ 0.05   & $-0.09\pm0.01$\\
             & {\tt SteParSyn}   & 3492 $\pm$ 17     & 4.74 $\pm$ 0.06   & $-0.04\pm0.03$ 
                                 & 3478 $\pm$ 81     & 4.83 $\pm$ 0.05   & $-0.03\pm0.04$ \\
             & DL                & 3514 $\pm$ 47     & 4.75 $\pm$ 0.08   &  $+0.18\pm0.09$ 
                                 & 3478 $\pm$ 81     & $4.83\pm0.05$   &  $+0.20\pm0.14$\\
             & {\tt ODUSSEAS}          & 3314 $\pm$ 78     & \ldots          & $-0.05\pm0.12$ 
                                 & \ldots          & \ldots          & \ldots \\
\noalign{\smallskip}
J13005$+$056 & {\tt Pass19-code} & 3142 $\pm$ 134     & 5.01 $\pm$ 0.11   & $-0.12\pm0.33$ 
                                 & 3140 $\pm$ 100     & 4.81 $\pm$ 0.10   & $-0.28\pm0.02$\\
                                 
             & {\tt SteParSyn}   & 3148 $\pm$ 40     & 4.60 $\pm$ 0.10   & $-0.36\pm0.10$
                                 & 3140 $\pm$ 100     & 4.81 $\pm$ 0.10   & $-0.30\pm0.10$\\
                                 
             & DL                & 3071 $\pm$ 46     & 5.06 $\pm$ 0.10   &  $+0.34\pm0.13$ 
                                 & 3140 $\pm$ 100    & 4.81 $\pm$ 0.10   &  $+0.31\pm0.10$\\
                                 
             & {\tt ODUSSEAS}          & 3417 $\pm$ 117     & \ldots          & $-0.18\pm0.18$
                                 & \ldots          & \ldots          & \ldots \\
\noalign{\smallskip}
J13457$+$148 & {\tt Pass19-code} & $3628\pm54$     & $4.76\pm0.06$   & $-0.12\pm0.19$ 
                                 & $3648\pm88$     & $4.75\pm0.04$   & $-0.10\pm0.01$\\
             & {\tt SteParSyn}   & $3569\pm23$     & $4.74\pm0.03$   & $-0.31\pm0.03$ 
                                 & $3648\pm88$     & $4.75\pm0.04$   & $-0.33\pm0.02$ \\
             & DL                & $3975\pm70$     & $4.60\pm0.05$   &  $+0.12\pm0.09$ 
                                 & $3648\pm88$     & $4.75\pm0.04$   &  $+0.22\pm0.12$\\
             & {\tt ODUSSEAS}          & $3590\pm78$     & \ldots          & $-0.24\pm0.12$ 
                                 & \ldots          & \ldots          & \ldots \\
\noalign{\smallskip}
J15194$-$077 & {\tt Pass19-code} & $3390\pm54$     & $4.91\pm0.06$   & $-0.17\pm0.19$ 
                                 & $3404\pm82$     & $4.91\pm0.06$   & $-0.14\pm0.01$\\
             & {\tt SteParSyn}   & $3422\pm10$      & $4.82\pm0.03$   & $-0.10\pm0.04$ 
                                 & $3404\pm82$     & $4.91\pm0.06$   & $-0.05\pm0.03$ \\
             & DL                & $3385\pm47$     & $4.82\pm0.09$   &  $+0.11\pm0.10$ 
                                 & $3404\pm82$     & $4.91\pm0.06$   &  $+0.14\pm0.15$\\
             & {\tt ODUSSEAS}          & $3280\pm79$     & \ldots          & $-0.28\pm0.12$ 
                                 & \ldots          & \ldots          & \ldots \\
\noalign{\smallskip}
J16581$+$257 & {\tt Pass19-code} & $3825\pm54$     & $4.67\pm0.06$   &  $+0.09\pm0.19$ 
                                 & $3683\pm78$     & $4.72\pm0.06$   & $-0.14\pm0.01$\\
             & {\tt SteParSyn}   & $3673\pm12$     & $5.11\pm0.07$   &  $+0.14\pm0.02$ 
                                 & $3683\pm78$     & $4.72\pm0.06$   & $-0.04\pm0.02$ \\
             & DL                & $3748\pm43$     & $4.68\pm0.05$   & $+0.17\pm0.08$ 
                                 & $3683\pm78$     & $4.72\pm0.06$   & $+0.18\pm0.11$\\
             & {\tt ODUSSEAS}          & $3561\pm79$     & \ldots          & $-0.12\pm0.12$ 
                                 & \ldots          & \ldots          & \ldots \\
\noalign{\smallskip}
J17578$+$046 & {\tt Pass19-code} & $3231\pm54$     & $5.00\pm0.06$   & $-0.23\pm0.19$ 
                                 & $3243\pm75$     & $5.05\pm0.07$   & $-0.19\pm0.01$\\
             & {\tt SteParSyn}   & $3282\pm14$     & $5.12\pm0.10$   & $-0.24\pm0.07$ 
                                 & $3243\pm75$     & $5.05\pm0.07$   & $-0.35\pm0.03$ \\
             & DL                & $3352\pm55$     & $4.91\pm0.10$   &  $+0.07\pm0.14$ 
                                 & $3243\pm75$     & $5.05\pm0.07$   &  $+0.16\pm0.19$ \\
             & {\tt ODUSSEAS}          & $3172\pm80$     & \ldots          & $-0.62\pm0.13$ 
                                 & \ldots          & \ldots          & \ldots \\
\noalign{\smallskip}
J22565$+$165 & {\tt Pass19-code} & $3842\pm54$     & $4.66\pm0.06$   &  $+0.20\pm0.19$ 
                                 & $3714\pm79$     & $4.71\pm0.04$   & $-0.05\pm0.01$ \\
             & {\tt SteParSyn}   & $3714\pm9$      & $4.87\pm0.03$   &  $+0.10\pm0.03$ 
                                 & $3714\pm79$     & $4.71\pm0.04$   &  $+0.04\pm0.04$ \\
             & DL                & $3765\pm49$     & $4.62\pm0.05$   &  $+0.23\pm0.09$ 
                                 & $3714\pm79$     & $4.71\pm0.04$   &  $+0.25\pm0.13$ \\
             & {\tt ODUSSEAS}          & $3509\pm79$     & \ldots          &  $+0.00\pm0.12$ 
                                 & \ldots          & \ldots          & \ldots \\
\noalign{\smallskip}
J23419$+$441 & {\tt Pass19-code} & $3069\pm54$     & $5.02\pm0.06$   & $-0.06\pm0.19$ 
                                 & $3058\pm80$     & $5.02\pm0.10$   &  $+0.12\pm0.01$ \\
             & {\tt SteParSyn}   & $3140\pm7$      & $5.00\pm0.02$   & $-0.13\pm0.05$ 
                                 & $3058\pm80$     & $5.02\pm0.10$   &  $+0.01\pm0.02$ \\
             & DL                & $2995\pm81$     & $5.01\pm0.14$   & $+0.40\pm0.27$ 
                                 & $3058\pm80$     & $5.02\pm0.10$   & $+0.50\pm0.28$ \\
             & {\tt ODUSSEAS}          & $2831\pm79$     & \ldots          & $-0.10\pm0.12$ 
                                 & \ldots          & \ldots          & \ldots \\

\end{longtable}

\begin{flushleft}
\end{flushleft}

\begin{longtable}{lccccccc}
\label{tab:results2}\\
\caption[]{Stellar parameters for each method from Runs C and C2.}\\
\hline\hline\noalign{\smallskip}
       &         & 
\multicolumn{3}{c}{Run C} & 
\multicolumn{3}{c}{Run C2} \\
Karmn  &  Method & 
$T_{\rm eff}$\,[K] & $\log{g}$\,[dex] & [Fe/H]\,[dex] & 
$T_{\rm eff}$\,[K] & $\log{g}$\,[dex] & [Fe/H]\,[dex] \\ 
\noalign{\smallskip}\hline\noalign{\smallskip}          
\endfirsthead

\caption[]{continued.}\\ 
\hline\hline\noalign{\smallskip}                
       &        & 
\multicolumn{3}{c}{Run C} & 
\multicolumn{3}{c}{Run C2} \\
Karmn  & Method & 
$T_{\rm eff}$\,[K] & $\log{g}$\,[dex] & [Fe/H]\,[dex] & 
$T_{\rm eff}$\,[K] & $\log{g}$\,[dex] & [Fe/H]\,[dex] \\ 
\noalign{\smallskip}\hline\noalign{\smallskip}
\endhead

\noalign{\smallskip}\hline
\endfoot
 
J00067$-$075 & {\tt Pass19-code} & $3031\pm151$ & $4.83\pm0.10$ &  $+0.78\pm0.26$ 
                                 & $3169\pm123$ & $5.11\pm0.07$ & $-0.48\pm0.22$ \\
             & {\tt SteParSyn}   & $3088\pm28$     & $5.11\pm0.14$   & $-0.06\pm0.11$  
                                 & $3069\pm29$ & $5.04\pm0.16$ & $-0.13\pm0.12$ \\
             & DL                & $3181\pm246$    & $4.82\pm0.14$   & $-0.06\pm0.27$  
                                 & $3133\pm129$ & $5.12\pm0.12$ & $-0.02\pm0.22$ \\
             & {\tt ODUSSEAS}$^{*}$              & $2875\pm90$  & \ldots & $-0.43\pm0.13$        
                                 & \ldots          & \ldots          & \ldots \\
             
\noalign{\smallskip}
J00183$+$440 & {\tt Pass19-code} & $3667\pm151$ & $4.75\pm0.10$ & $-0.13\pm0.26$  
                                 & $3664\pm123$ & $4.83\pm0.07$ & $-0.39\pm0.22$ \\
             & {\tt SteParSyn}   & $3459\pm31$     & $4.59\pm0.08$   & $-0.63\pm0.07$  
                                 & $3437\pm39$ & $4.65\pm0.12$ & $-0.67\pm0.09$ \\
              & DL               & $3779\pm90$     & $4.77\pm0.07$   & $-0.30\pm0.16$  
                                 & $3713\pm73$ & $4.80\pm0.07$ & $-0.28\pm0.12$ \\
             & {\tt ODUSSEAS}$^{*}$          & $3589\pm80$ & \ldots & $-0.45\pm0.12$ 
                                 & \ldots          & \ldots          & \ldots \\
              
\noalign{\smallskip}
J04429$+$189 & {\tt Pass19-code} & 3632 $\pm$ 151      & 4.71 $\pm$ 0.10      &  $+0.07\pm0.26$  
                                 & $3710\pm123$ & $4.76\pm0.07$ & $-0.23\pm0.22$ \\
             & {\tt SteParSyn}   & 3651 $\pm$ 21     & 4.78 $\pm$ 0.09   &  $+0.00\pm0.05$  
                                 & $3430\pm46$ & $4.77\pm0.10$ & $-0.41\pm0.11$ \\
             & DL                & 3751 $\pm$ 114    & 4.73 $\pm$ 0.08   & $-0.11\pm0.19$ 
                                 & $3703\pm105$ & $4.77\pm0.10$ & $-0.03\pm0.14$ \\
             & {\tt ODUSSEAS}$^{*}$         & $3471\pm81$ & \ldots & $-0.11\pm0.12$    
                                 & \ldots          & \ldots          & \ldots \\
             
\noalign{\smallskip}
J05314$-$036 & {\tt Pass19-code} & 3763$\pm151$ & 4.66$\pm0.10$ &  $+0.21\pm0.26$  
                                 & $3766\pm123$ & $4.74\pm0.07$ & $-0.22\pm0.22$ \\
             & {\tt SteParSyn}   & 3878 $\pm$ 15     & 4.76 $\pm$ 0.10   &  $+0.21\pm0.03$  
                                 & $3908\pm17$ & $4.79\pm0.11$ & $+0.24\pm0.04$ \\
             & DL                & 3980 $\pm$ 78     & 4.70 $\pm$ 0.06   &  $+0.09\pm0.19$  
                                 & $3918\pm93$ & $4.72\pm0.05$ & $+0.06\pm0.13$ \\
             & {\tt ODUSSEAS}$^{*}$          & $3572\pm81$ & \ldots & $+0.04\pm0.12$
                                 & \ldots          & \ldots          & \ldots \\
\noalign{\smallskip}
J07558$+$833 & {\tt Pass19-code} & 3345 $\pm$ 199    & 4.75 $\pm$ 0.11   &  $+0.57\pm0.27$  
                                 & $3305\pm123$ & $4.85\pm0.05$ & $+0.16\pm0.22$ \\
             & {\tt SteParSyn}   & 3276 $\pm$ 15     & 5.23 $\pm$ 0.07   & $-0.21\pm0.05$  
                                 & $3276\pm15$ & $5.27\pm0.07$ & $-0.23\pm0.05$ \\
             & DL                & 3572 $\pm$ 186    & 5.20 $\pm$ 0.08   & $-0.28\pm0.36$  
                                 & $3409\pm173$ & $5.20\pm0.13$ & $-0.33\pm0.31$ \\
             & {\tt ODUSSEAS}$^{*}$          & $3608\pm154$ & \ldots & $-0.79\pm0.22$ 
                                 & \ldots          & \ldots          & \ldots \\
\noalign{\smallskip}
J09143$+$526 & {\tt Pass19-code} & 4054 $\pm$ 151    & 4.70 $\pm$ 0.10   & $-0.07\pm0.26$  
                                 & $4096\pm151$ & $4.74\pm0.07$ & $-0.37\pm0.22$ \\
             & {\tt SteParSyn}   & 4034 $\pm$ 17     & 4.98 $\pm$ 0.09   & $+0.04\pm0.03$  
                                 & $4020\pm24$ & $4.93\pm0.12$ & $+0.01\pm0.05$ \\
             & DL                & 4049 $\pm$ 38     & 4.71 $\pm$ 0.05   & $-0.13\pm0.16$  
                                 & $4026\pm52$ & $4.73\pm0.06$ & $-0.19\pm0.12$ \\
             & {\tt ODUSSEAS}$^{*}$          & $3859\pm92$ & \ldots & $-0.19\pm0.14$ 
                                 & \ldots          & \ldots          & \ldots \\
\noalign{\smallskip}
J09144$+$526 & {\tt Pass19-code} & 4033 $\pm$ 151    & 4.71 $\pm$ 0.10   &  $-0.11\pm0.26$  
                                 & $3982\pm123$ & $4.74\pm0.07$ & $-0.29\pm0.22$ \\
             & {\tt SteParSyn}   & 4006 $\pm$ 14     & 4.95 $\pm$ 0.08   & $+0.04\pm0.02$  
                                 & $3953\pm9$ & $5.16\pm0.07$ & $+0.01\pm0.02$ \\
             & DL                & 4043 $\pm$ 37     & 4.71 $\pm$ 0.05   & $-0.13\pm0.15$  
                                 & $4008\pm54$ & $4.72\pm0.06$ & $-0.15\pm0.10$ \\
             & {\tt ODUSSEAS}$^{*}$          & $3816\pm87$ & \ldots & $-0.18\pm0.14$ 
                                 & \ldots          & \ldots          & \ldots \\
\noalign{\smallskip}
J10508$+$068 & {\tt Pass19-code} & 3239 $\pm$ 151    & 4.77 $\pm$ 0.10   & $+0.77\pm0.26$ 
                                 & $3506\pm123$ & $4.85\pm0.07$ & $-0.16\pm0.22$ \\
             & {\tt SteParSyn}   & 3348 $\pm$ 42     & 4.85 $\pm$ 0.13   & $-0.06\pm0.11$  
                                 & $3291\pm40$ & $4.72\pm0.13$ & $-0.21\pm0.10$ \\
             & DL                & 3514 $\pm$ 171    & 4.75 $\pm$ 0.10   & $-0.06\pm0.24$ 
                                 & $3435\pm121$ & $4.85\pm0.11$ & $+0.04\pm0.20$ \\
             & {\tt ODUSSEAS}$^{*}$          & $3144\pm84$ & \ldots & $-0.11\pm0.13$ 
                                 & \ldots          & \ldots          & \ldots \\
\noalign{\smallskip}
J11033$+$359 & {\tt Pass19-code} & 3597 $\pm$ 151    & 4.73 $\pm$ 0.10   & $+0.02\pm0.26$ 
                                 & $3619\pm123$ & $4.85\pm0.07$ & $-0.40\pm0.22$ \\
             & {\tt SteParSyn}   & 3346 $\pm$ 24     & 4.34 $\pm$ 0.08   & $-0.70\pm0.06$  
                                 & $3357\pm28$ & $4.46\pm0.11$ & $-0.66\pm0.07$ \\
             & DL                & 3719 $\pm$ 101    & 4.72 $\pm$ 0.07   & $-0.25\pm0.17$ 
                                 & $3652\pm73$ & $4.81\pm0.07$ & $-0.23\pm0.14$ \\
             & {\tt ODUSSEAS}$^{*}$          & $3571\pm81$ & \ldots & $-0.45\pm0.12$
                                 & \ldots          & \ldots          & \ldots \\
\noalign{\smallskip}
J11054$+$435 & {\tt Pass19-code} & 3707 $\pm$ 151    & 4.75 $\pm$ 0.10   & $-0.18\pm0.26 $
                                 & $3553\pm123$ & $4.91\pm0.07$ & $-0.48\pm0.22$ \\
             & {\tt SteParSyn}   & 3549 $\pm$ 27     & 4.73 $\pm$ 0.08   & $-0.53\pm0.06$  
                                 & $3481\pm42$ & $4.56\pm0.11$ & $-0.67\pm0.10$ \\
             & DL                & 3779 $\pm$ 81     & 4.77 $\pm$ 0.07   & $-0.39\pm0.12$ 
                                 & $3717\pm65$ & $4.84\pm0.07$ & $-0.39\pm0.10$ \\
             & {\tt ODUSSEAS}$^{*}$          & $3625\pm81$ & \ldots & $-0.51\pm0.12$
                                 & \ldots          & \ldots          & \ldots \\
\noalign{\smallskip}
J11421$+$267 & {\tt Pass19-code} & 3436 $\pm$ 151    & 4.75 $\pm$ 0.10   & $+0.17\pm0.26$ 
                                 & $3632\pm123$ & $4.77\pm0.07$ & $-0.16\pm0.22$ \\
             & {\tt SteParSyn}   & 3449 $\pm$ 32     & 4.73 $\pm$ 0.07   & $-0.21\pm0.07$  
                                 & $3373\pm36$ & $4.67\pm0.09$ & $-0.38\pm0.09$ \\
             & DL                & 3666 $\pm$ 128    & 4.72 $\pm$ 0.07   & $-0.09\pm0.19$ 
                                 & $3618\pm89$ & $4.79\pm0.07$ & $-0.02\pm0.15$ \\
             & {\tt ODUSSEAS}$^{*}$          & $3395\pm81$ & \ldots & $-0.16\pm0.12$ 
                                 & \ldots          & \ldots          & \ldots \\
\noalign{\smallskip}
J13005$+$056 & {\tt Pass19-code} & 3245 $\pm$ 210    & 4.77 $\pm$ 0.11  &  $+0.51\pm0.28 $
                                 & $3201\pm123$ & $4.77\pm0.06$ & $+0.77\pm0.25$ \\
             & {\tt SteParSyn}   & 3533 $\pm$ 5     & 4.61 $\pm$ 0.11   & $+0.59\pm0.06$  
                                 & $3586\pm24$ & $4.58\pm0.14$ & $+0.67\pm0.06$ \\
                                 
             & DL                & 3589 $\pm$ 188    & 5.16 $\pm$ 0.11   & $-0.33\pm0.33 $
                                 & $3379\pm181$ & $5.18\pm0.12$ & $-0.23\pm0.30$ \\
                                 
             & {\tt ODUSSEAS}$^{*}$          & $3579\pm175$ & \ldots & $-0.83\pm0.27$ 
                                 & \ldots          & \ldots         & \ldots \\
                                 
\noalign{\smallskip}
J13457$+$148 & {\tt Pass19-code} & 3615 $\pm$ 151    & 4.73 $\pm$ 0.10   & $-0.02\pm0.26 $
                                 & $3627\pm123$ & $4.85\pm0.07$ & $-0.41\pm0.22$ \\
             & {\tt SteParSyn}   & 3516 $\pm$ 31     & 4.59 $\pm$ 0.09   & $-0.41\pm0.07$  
                                 & $3373\pm32$ & $4.44\pm0.12$ & $-0.72\pm0.08$ \\
             & DL                & 3755 $\pm$ 84     & 4.81 $\pm$ 0.11   & $-0.22\pm0.18$ 
                                 & $3720\pm68$ & $4.78\pm0.06$ & $-0.23\pm0.13$ \\
             & {\tt ODUSSEAS}$^{*}$          & $3648\pm81$ & \ldots & $-0.28\pm0.13$ 
                                 & \ldots          & \ldots          & \ldots \\
\noalign{\smallskip}
J15194$-$077 & {\tt Pass19-code} & 3447 $\pm$ 151    & 4.76 $\pm$ 0.10   & $+0.23\pm0.26 $
                                 & $3578\pm123$ & $4.84\pm0.07$ & $-0.23\pm0.22$ \\
             & {\tt SteParSyn}   & 3383 $\pm$ 36     & 4.72 $\pm$ 0.10   & $-0.27\pm0.08 $ 
                                 & $3332\pm35$ & $4.63\pm0.11$ & $-0.40\pm0.09$ \\
             & DL                & 3581 $\pm$ 135    & 4.74 $\pm$ 0.09   & $-0.15\pm0.20 $
                                 & $3535\pm87$ & $4.91\pm0.10$ & $-0.08\pm0.16$ \\
             & {\tt ODUSSEAS}$^{*}$          & $3325\pm81$ & \ldots & $-0.31\pm0.13$
                                 & \ldots          & \ldots          & \ldots \\
\noalign{\smallskip}
J16581$+$257 & {\tt Pass19-code} & 3758 $\pm$ 151    & 4.67 $\pm$ 0.10   & $+0.12\pm0.26 $
                                 & $3785\pm123$ & $4.75\pm0.07$ & $-0.29\pm0.22$ \\
             & {\tt SteParSyn}   & 3772 $\pm$ 12     & 4.87 $\pm$ 0.07   & $+0.00\pm0.02$  
                                 & $3701\pm29$ & $4.94\pm0.07$ & $-0.13\pm0.06$ \\
             & DL                & 3886 $\pm$ 80     & 4.80 $\pm$ 0.14   & $-0.10\pm0.23 $
                                 & $3826\pm112$ & $4.79\pm0.15$ & $-0.12\pm0.19$ \\
             & {\tt ODUSSEAS}$^{*}$          & $3602\pm80$ & \ldots & $-0.17\pm0.12$ 
                                 & \ldots          & \ldots          & \ldots \\
\noalign{\smallskip}
J17578$+$046 & {\tt Pass19-code} & 3256 $\pm$ 151    & 4.76 $\pm$ 0.10   & $+0.61\pm0.26$ 
                                 & $3448\pm123$ & $4.89\pm0.07$ & $-0.27\pm0.22$ \\
             & {\tt SteParSyn}   & 3189 $\pm$ 26     & 4.63 $\pm$ 0.11   & $-0.50\pm0.08$  
                                 & $3175\pm26$ & $4.61\pm0.11$ & $-0.54\pm0.08$ \\
             & DL                & 3412 $\pm$ 157    & 4.74 $\pm$ 0.11   & $-0.18\pm0.19$  
                                 & $3392\pm92$ & $5.00\pm0.11$ & $-0.17\pm0.15$ \\
             & {\tt ODUSSEAS}$^{*}$          & $3233\pm84$ & \ldots & $-0.68\pm0.13$ 
                                 & \ldots          & \ldots          & \ldots \\
\noalign{\smallskip}
J22565$+$165 & {\tt Pass19-code} & $3693\pm151$    & $4.69\pm0.10$   & $+0.10\pm0.26$  
                                 & $3624\pm123$ & $4.80\pm0.07$ & $-0.26\pm0.22$ \\
             & {\tt SteParSyn}   & $3795\pm12$     & $4.76\pm0.08$   &  $+0.09\pm0.02$  
                                 & $3702\pm41$ & $4.97\pm0.11$ & $-0.04\pm0.07$ \\
             & DL                & $3885\pm79$     & $4.73\pm0.06$   & $-0.12\pm0.16$  
                                 & $3824\pm74$ & $4.75\pm0.06$ & $-0.07\pm0.12$ \\
             & {\tt ODUSSEAS}$^{*}$          & $3551\pm79$ & \ldots & $-0.05\pm0.12$ 
                                 & \ldots          & \ldots          & \ldots \\
\noalign{\smallskip}
J23419$+$441 & {\tt Pass19-code} & $3195\pm151$    & $4.77\pm0.10$   & $+0.79\pm0.26$  
                                 & $3173\pm123$ & $5.12\pm0.07$ & $-0.52\pm0.22$ \\
             & {\tt SteParSyn}   & $3176\pm26$     & $5.17\pm0.11$   &  $+0.21\pm0.10$  
                                 & $3167\pm27$ & $5.13\pm0.11$ & $+0.20\pm0.11$ \\
             & DL                & $3246\pm248$    & $4.82\pm0.14$   & $+0.04\pm0.29$  
                                 & $3139\pm137$ & $5.03\pm0.14$ & $+0.08\pm0.26$ \\
             & {\tt ODUSSEAS}$^{*}$          & $2922\pm87$ & \ldots & $-0.83\pm0.27$
                                 & \ldots          & \ldots          & \ldots \\

\end{longtable}
\tablefoot{$^{(*)}$ Corresponding to Run C*.}
\begin{flushleft}
\end{flushleft}


\twocolumn

\clearpage

\end{document}